\documentclass[12pt,letterpaper]{article}
\pdfoutput=1
\usepackage[colorlinks=true,
linkcolor=BlueViolet, 
citecolor=BlueViolet,
filecolor=BlueViolet,
urlcolor=BlueViolet,
linktoc=page, 
pdfstartview=FitV,
bookmarksopen=true]{hyperref}
\usepackage[left=2cm,top=1cm,right=3cm,nohead]{geometry}
\usepackage[dvipsnames]{xcolor}
\usepackage{colortbl}
\usepackage[utf8]{inputenc} 
\usepackage{color}
\usepackage{epsfig,graphics,latexsym,amsfonts,mathtools,amsthm,amssymb,amsbsy,multirow,slashed,mathrsfs,wasysym,textcomp,subfigure,wrapfig,comment,bbold,array,longtable,multirow,setspace,amsmath,pifont, enumitem}
\usepackage{cleveref}
\usepackage{tabularx}
\usepackage{graphicx}
\usepackage{graphbox}
\usepackage{setspace}
\usepackage{multirow}
\usepackage{rotate}	
\usepackage{slashed}
\usepackage{accents}
\usepackage[bf]{caption}
\usepackage{braket}
\usepackage{hyperref}
\usepackage{fancyhdr}
\usepackage{comment}
\usepackage{float}
\usepackage{tikz}
\usetikzlibrary{mindmap,trees,shadows}
\colorlet{color1}{gray!25}
\usetikzlibrary{positioning}
\usetikzlibrary{intersections} 

\newlength{\PicScale}
\usepackage[UKenglish]{babel}
\usepackage[toc,page]{appendix}
\usepackage{hhline}
\usepackage{geometry}
\usepackage{cite}
\usepackage{datetime}
\usepackage{bbm} 
\usepackage{bm} 
\hypersetup{
	pdftitle={},
	pdfauthor={},
	pdfsubject={}
} 
\usepackage{subfiles}
\definecolor{Gray}{gray}{0.94}
\usepackage{simpler-wick}

\newlength{\dhatheight}

\numberwithin{equation}{section}

\newcommand{\bmat}{\left(\begin{array}}
\newcommand{\emat}{\end{array}\right)}

\def\yzero{\smash{\hbox{$y\kern-4pt\raise1pt\hbox{${}^\circ$}$}}}

\def\beq{\begin{equation}}
\def\eeq{\end{equation}}
\def\beqa{\begin{eqnarray}}
\def\eeqa{\end{eqnarray}}

\def\-{\hphantom{-}}

\def\s2{\frac{1}{\sqrt2}}

\def\beq{\begin{equation}}
\def\eeq{\end{equation}}
\def\beqa{\begin{eqnarray}}
\def\eeqa{\end{eqnarray}}

\def\IF{\relax{\rm I\kern-.18em F}}
\def\II{\relax{\rm I\kern-.18em I}}
\def\IP{\relax{\rm I\kern-.18em P}}
\def\IC{\relax\hbox{\kern.25em$\inbar\kern-.3em{\rm C}$}}
\def\IR{\relax{\rm I\kern-.18em R}}

\def\Dsl{\,\raise.15ex\hbox{/}\mkern-13.5mu D} 
\def\IZ{Z\kern-.4em  Z}

\def\h31{h^{3,1}}

\def\va{\vec{a}}
\def\vb{\vec{b}}
\def\vc{\vec{c}}

\catcode`\@=11   
\newdimen\@rotdimen
\newbox\@rotbox  

\def\@vspec#1{\special{ps:#1}}
\def\@rotstart#1{\@vspec{gsave currentpoint currentpoint translate
   #1 neg exch neg exch translate}}
\def\@rotfinish{\@vspec{currentpoint grestore moveto}}
\def\@rotr#1{\@rotdimen=\ht#1\advance\@rotdimen by\dp#1%
   \hbox to\@rotdimen{\hskip\ht#1\vbox to\wd#1{\@rotstart{90 rotate}%
   \box#1\vss}\hss}\@rotfinish}
\def\@rotl#1{\@rotdimen=\ht#1\advance\@rotdimen by\dp#1%
   \hbox to\@rotdimen{\vbox to\wd#1{\vskip\wd#1\@rotstart{270 rotate}%
   \box#1\vss}\hss}\@rotfinish}%
\def\@rotu#1{\@rotdimen=\ht#1\advance\@rotdimen by\dp#1%
   \hbox to\wd#1{\hskip\wd#1\vbox to\@rotdimen{\vskip\@rotdimen
   \@rotstart{-1 dup scale}\box#1\vss}\hss}\@rotfinish}%
\def\@rotf#1{\hbox to\wd#1{\hskip\wd#1\@rotstart{-1 1 scale}%
   \box#1\hss}\@rotfinish}%
\def\rotate{\@ifnextchar[{\@rotate}{\@rotate[l]}}
\def\@rotate[#1]#2{\setbox\@rotbox=\hbox{#2}\@nameuse{@rot#1}\@rotbox}

\catcode`\@=12

\usepackage{xcolor}

\begin{document}

\numberwithin{equation}{section}

\begin{flushright}
\end{flushright}
\pagestyle{empty}

\vspace{20mm}

 \begin{center}
{\LARGE \textbf{{Affine algebras at infinite distance limits \\ \vspace{0.3cm} in the Heterotic String}}}

\vspace{10mm} {\Large Veronica Collazuol, Mariana Gra\~na,\\  
\vspace{0.3cm}
Alvaro Herr\'aez and H\'ector Parra De Freitas}\\

\vspace{5mm}

Institut de Physique Th\'eorique, \\
Universit\'e Paris Saclay, CEA, CNRS, \\
Orme des Merisiers,  F-91191 Gif sur Yvette, France \\

\vspace{.5mm} {\small\upshape\ttfamily veronica.collazuol, mariana.grana, alvaro.herraezescudero, hector.parradefreitas \\   @ ipht.fr} \\

\vspace{15mm}

\textbf{Abstract}
\end{center}

We analyze the boundaries of the moduli spaces of compactifications of the heterotic string on $T^d$, making particular emphasis on $d=2$  and its F-theory dual.  We compute the OPE algebras as we approach all the infinite distance limits that correspond to (possibly partial) decompactification limits in some dual frame. When decompactifying $k$ directions, we find  infinite towers of states becoming light that enhance the algebra arising at a given point in the moduli space of the $T^{d-k}$ compactification to its $k$-loop version, where the central extensions are given by the $k$ KK vectors. For $T^2$ compactifications, we reproduce all the affine algebras that arise in the F-theory dual, and show all the towers explicitly, including some that are not manifest in the F-theory counterparts. Furthermore, we construct the affine $SO(32)$ algebra arising in the full decompactification limit, both in the heterotic and in the F-theory sides, showing that not only affine algebras of exceptional type arise in the latter. 

\vspace{3mm}

\newpage
\setcounter{page}{0}
	\pagestyle{empty}
	
\tableofcontents
\newpage
\section{Introduction}
\label{intro}
\setcounter{page}{1}
\pagestyle{plain}
In the recent years, there has been an increasing interest in the so-called Swampland program \cite{Vafa:2005ui} (reviewed in \cite{Brennan:2017rbf,Palti:2019pca,vanBeest:2021lhn,Grana:2021zvf,Harlow:2022gzl}), which aims at drawing the boundaries between the low energy effective theories than can  be consistently coupled to Quantum Gravity and those that cannot. These boundaries are typically formulated as a series of conjectures that constrain the low energy effective theories, whose arguments range from general black hole physics to explicit String Theory constructions. One of the best established conjectures at the moment is the Swampland Distance Conjecture \cite{Ooguri:2006in}, which predicts the presence of an infinite tower of states becoming exponentially light with the distance as one moves far away in the moduli space of the theory. The nature of the possible massless towers has been analysed in \cite{Lee:2019wij}, where it was conjectured that it must always come from either a KK tower (or dual versions), or from the excitations of an emergent critical string becoming tensionless (this is known as the Emergent String Conjecture).

On the other hand, massless vectors at a given point in moduli space are a signature of a larger gauge group, and all massless states  arrange into  representations of this group.  Thus, exploring specific points (or more generally, hypersufaces) in moduli space allows to examine the symmetry enhancements associated to this phenomenon. In particular, a detailed knowledge of all the possible enhancements is essential to determine whether all consistent low energy vacua in Quantum Gravity can be obtained from String Theory (i.e. string universality).  Progress along this direction has been recently made by systematically studying toroidal compactifications of the heterotic/CHL string \cite{Fraiman:2018ebo,Font:2020rsk,Font:2021uyw,Fraiman:2021soq} and F-theory on K3 surfaces \cite{https://doi.org/10.48550/arxiv.math/0007171}, as well as by applying general Swampland arguments to supergravity vacua \cite{Hamada:2021bbz, Montero:2020icj, Tarazi:2021duw, Long:2021jlv}  . In this context, a natural question arises, namely what is the interplay between these symmetry enhancements and  infinite distance boundaries of the moduli space, specially in relation to the Distance Conjecture. Such points have been recently studied from this perspective in the circle compactification of the heterotic string \cite{Collazuol:2022jiy} and in 8d theories realised from F-theory on K3 surfaces \cite{Lee:2021qkx,Lee:2021usk,Cvetic:2022uuu}, where it was shown that in these limits one gets an infinite tower of massless states that realize an affine algebra. Particularly, in the decompactification limit of the heterotic string on the circle the KK tower combines with the  $E_8 \times E_8$ or $Spin(32)/\mathbb{Z}_2$ gauge group to give rise to the affine version of their algebras, respectively, with the $U(1)$ of the circle direction playing the role of imaginary root. 

In this paper, we study the boundaries of the moduli space   of  toroidal compactifications of the heterotic string, which correspond to (duals of) different decompactification limits. 
Despite being a very simple playground, the toroidally-compactified heterotic theory proves to be rich enough to describe all the possible decompactifications and the related affine algebras that are realised in the corresponding moduli space. 
We show that when decompactifying $k$ dimensions from $10-d$ to a point in the $10-d+k$ dimensional theory with gauge algebra $\mathcal{A}_L \times U(1)^{d-k}_R$, we obtain the $k$-loop version of $\mathcal{A}_L \times U(1)^{d-k}_R$. The central extension for the left/right moving algebra is the holomorphic/antiholomorphic  part of the current associated to the $k$ directions that are decompactified.  

We show that for $T^2$ compactifications, the heterotic results match those obtained in the dual F theory on K3, in the context of Kulikov models and K3 complex structure degenerations \cite{Lee:2021qkx,Lee:2021usk}. The heterotic framework allows however for a clearer characterisation of the decompactification limits and corresponding algebras in all the corners of moduli space, and moreover can be (straightforwardly) generalised to lower dimensions.

The paper is organised as follows. After setting our conventions and reviewing the affinisation process in the decompactification of the heterotic theory on $S^1$ in Section \ref{sect0}, we discuss in Section \ref{8ddecompactification} its generalisations,  first to $T^2$ and then to a generic $T^d$ compactification, showing the explicit world-sheet realisation of the algebras in Appendix \ref{appendix}. In Section \ref{Ftheory} we review the string junction picture, which we explicitly use in Section \ref{Het-Ftheory} in order to make contact with the F-theory dual description of \cite{Lee:2021qkx,Lee:2021usk}, reviewed in Appendix \ref{appendix2}. We also build the explicit realisation of the Kulikov Weierstrass model for $\widehat{\widehat{D}}_{16}$, which from the heterotic dual is predicted to be, together with $(\widehat{E}_9\oplus \widehat{E}_9)/\sim$, the only algebra in the full decompactification limit. Finally, we summarise and make some concluding remarks  in Section \ref{conclusions}.

\section{Decompactification limits of the Heterotic String}
\label{sect0}
To set our notation, in this section we summarize the main features of heterotic compactification on a $d$ dimensional torus $T^d$, and review the behaviour of the theory on $S^1$ when going to infinite distance in the moduli space.

Let $T^d$ be associated with a lattice $\Lambda$ generated by the vectors $e_i, \, i=1,...,d$, and dual lattice $\Tilde{\Lambda}$ generated by the dual vectors $e^{*i}$. The background of the theory compactified on this space is specified by\footnote{One should also include the dilaton, but we are taking it to be fixed (and small).} 
\begin{itemize}
    \item the metric on the torus $G_{ij}=e^{a}_i \delta_{ab} e^{b}_j, \, \,  a=1,...,d$\,, 
    \item the internal B-field $B_{ij}$ and
    \item the Wilson lines $A^{I}_i$, \, $I=1,...,16$\,.
\end{itemize}
Restricting the worldsheet action to the bosonic sector and setting $\alpha'=1$, we have
\begin{equation}
\label{eqn:action}
\begin{split}
     S = - \frac{1}{4\pi}  \int d^2 \sigma \Big[ \eta^{\alpha \beta} \big( G_{MN} \partial_{\alpha}X^{M} &\partial_{\beta}X^{N} + \frac{1}{2} \delta_{IJ}  \partial_{\alpha}X^{I} \partial_{\beta}X^{J} \big)  \\
     &+  \epsilon^{\alpha \beta} \big( B_{MN}\partial_{\alpha}X^{M}\partial_{\beta}X^{N}+
     A^{I}_i\partial_{\alpha}Y^i\partial_{\beta}X^{I} \big) \Big] \, .
\end{split}
\end{equation}
In the following we split the 10 dimensional index $M=0,...,9$ as $M=(\mu, i)$, with $\mu=0,...9-d$ and $i = 10-d,...,9$. We write $X^M=(X^{\mu}, Y^i)$ with $Y^i$ the compact spacetime bosons.

The heterotic states are characterized by the winding and momentum numbers associated to the $T^d$ directions, respectively $w^i, \, n_i \in \mathbb{Z}$, and by the heterotic momenta $\pi^I$ along the 16 dimensional heterotic torus. These can be arranged in a charge vector 
\beq \label{Zdef}
Z=(\pi^I, \, w^i, \, n_i ) \ . 
\eeq
The vector $\pi$ lies either in the root lattice of $E_8 \times E_8$, denoted $\Gamma_8 \times \Gamma_8$, or in the weight lattice of $Spin(32)/\mathbb{Z}_2$, denoted $\Gamma_{16}$. Moreover, the states have the following left ($L$) and right ($R$) internal momenta along the $T^d$ directions,
\begin{equation}
\label{eqn:tdmomenta}
    \begin{split}
    p_{R,i} &= \frac{1}{\sqrt{2}} (n_i - E_{ij} w^j - \pi \cdot A_i) \, , \\
    p_{L,i} &= \frac{1}{\sqrt{2}} (n_i + (2G_{ij} - E_{ij}) w^j - \pi \cdot A_i) \, ,
  \end{split}
\end{equation}
where
\begin{equation}
    E_{ij} = G_{ij} + \frac{1}{2} A_i \cdot A_j + B_{ij} \, , 
\end{equation}
and along the heterotic torus, 
\begin{equation}
\label{eqn:hetmom}
    p^I = \pi^I+ A^I_i w^i \, .
\end{equation}

The vector $\boldsymbol{p}= (p_{R,i} e^{*i}_{a}; p_{L,i} e^{*i}_{a},p^I) = (p_{R,a};p_{L,a},p^I) \equiv (\boldsymbol{p}_R; \boldsymbol{p}_L)$ lies in an even and self-dual lattice $\Gamma_{d,d+16}$ with Lorentzian signature $(-_d,+_{d+16})$, as seen from the expression
\begin{equation}
    \boldsymbol{p}^2 = - \boldsymbol{p}_R^2 + \boldsymbol{p}_L^2 = 2 n_i w^i + |\pi|^2 \in 2\, \mathbb{Z} \, .
\end{equation}
The mass formula and the level-matching condition (LMC) for the NS sector of the spectrum then read
\begin{equation}
\label{eqn:spectrum}
    \begin{split}
             M^2&= \boldsymbol{p}_L^2 + p_R^2 + 2 \Big( N + \tilde{N} -\frac{3}{2} \Big) \, , \\
             0&= \boldsymbol{p}_L^2 - p_R^2 + 2 \Big( N - \tilde{N} -\frac{1}{2} \Big)  \, ,
    \end{split}         
\end{equation}
where $N$ and $\tilde N$ are respectively the left and right-moving oscillator numbers. These formulas determine the group of symmetries of the spectrum to be $O(d,d+16,\mathbb{Z})$, which we refer to as the T-duality group.

In order to read the symmetry algebra at a given point in moduli space one should look at the set of massless vectors. For generic values of the moduli, these are the $U(1)^d_R\times U(1)^{16+d}_L$  vector bosons, characterized by $N=1, \tilde{N}=\frac{1}{2}, \boldsymbol{p}_L=0, \boldsymbol{p}_R=0$:
\begin{equation}
\label{eqn:12}
          \alpha^{\mu}_{-1} \tilde{\psi}^{i}_{-\frac{1}{2}} \ket{0}_{NS} \, ,  \alpha^i_{-1} \tilde{\psi}^{\mu}_{-\frac{1}{2}} \ket{0}_{NS}\, , \,    \alpha^{I}_{-1} \tilde{\psi}^{\mu}_{-\frac{1}{2}} \ket{0}_{NS} \longrightarrow \, (g_{\mu i}\mp B_{\mu i}), \,  A_{\mu}{}^{I} \, .
\end{equation}
Due to the moduli dependence of \eqref{eqn:spectrum} it turns out that at fixed points of the T-duality group there can be additional massless vectors with $N=0, \bar{N}=\frac{1}{2}, |\boldsymbol{p}_L|^2=2,~ \boldsymbol{p}_R=0$, of the form
\begin{equation} \label{eq:roots}
     \tilde{\psi}_{-\frac{1}{2}}^{\mu} \ket{\pi_{\alpha}, w^i, n_i}_{NS}\, \, \longrightarrow \, A_{\mu}^{\alpha} \, 
\end{equation}
for some particular values of $w^i, n_i$. These enhance the gauge group in the left sector to a rank $(16+d)$ semisimple group of ADE type. 
\subsection{Decompactification limit of the $E_8 \times E_8$ Heterotic String on $S^1$}
We now restrict to $d=1$ (compactifications on a circle of radius $R$) and the $E_8 \times E_8$ Heterotic String, and review the results of \cite{Collazuol:2022jiy} showing that in the limits $R \to 0, \, \infty$ and by appropriately tuning the Wilson line, one finds a symmetry enhancement to the affine algebra $(E_9 \oplus E_9) / \sim$ (where $\sim$ refers to the identification of the central extensions of the two $E_8$ factors) or $\widehat{D}_{16}$. Let us start with the case $R \to \infty$. For any finite value of the Wilson line, the massless vectors are the momentum towers of the $E_8 \times E_8 \times U(1)$ gauge bosons, namely  
\begin{equation}
    Z=(\pi_{\alpha},0,n), \quad n \in \mathbb{Z}, \, \pi_{\alpha} \in \Gamma_8 \times \Gamma_8 \, , |\pi_{\alpha}|^2=0, 2 \ ,
\end{equation}
where $|\pi_{\alpha}|^2=0$ correspond to the states in \eqref{eqn:12}, and the $|\pi_{\alpha}|^2=2$ states to those in \eqref{eq:roots}. From these we can build the following asymptotically conserved currents
\begin{equation}
\label{eqn:e9current}
\begin{split}
    J^{I}_n(z)  &=  i (\partial X^{I}(z) - A^{I} \partial Y(z)) e^{i n Y(z)} \, , \\
     J^{\alpha}_{n}(z) &=c_{\alpha} e^{i \pi^{I}_{\alpha} (X^{I}(z) - A^{I} Y(z)) }  e^{inY(z)} \, , \\
\end{split}
\end{equation}
where $c_{\alpha}$ are the cocycle factors
\begin{equation}
    c_{\alpha} c_{\beta} \, = \,  \epsilon(\alpha, \beta) \, c_{\alpha + \beta}\, , \quad \epsilon(\alpha,\beta) = \pm 1 \quad \text{for} \, \alpha+\beta \, \text{root} 
\end{equation}
and zero otherwise, with   $c_{0}\equiv 1$.
The zero modes of the currents \eqref{eqn:e9current}
\begin{equation}
\label{eqn:zeromode}
    (J^{a}_n)_0 \equiv \oint \frac{dz}{2 \pi i} J^{a}_n(z) \, , \quad a=\{I, \alpha\}
\end{equation}
satisfy the following commutation relations \cite{Collazuol:2022jiy}
\begin{equation}
\label{eqn:e9commutator}
\begin{split}
   & [(J^{I}_n)_0, (J^{J}_m)_0] = i n \delta^{IJ} \delta_{n+m, 0} (\partial Y)_0 \, ,\\
    &[(J^{I}_n)_0, (J^{\alpha}_m)_0] = \pi_{\alpha}^{I} (J^{\alpha}_{n+m})_0 \, ,\\
    &[(J^{\alpha}_n)_0, (J^{\beta}_m)_0] = \begin{cases}
          \epsilon(\alpha, \beta) (J^{\alpha + \beta}_{n+m})_0  \qquad \quad \qquad \quad \, \,    \,  \alpha+\beta \, \, {\rm root}, \\
          \pi^{I}_{\alpha} (J^{I}_{n+m})_0 + i n \delta_{n+m, 0} (\partial Y)_0 \quad   \alpha=-\beta\, ,\\
         0 \qquad \qquad \qquad \qquad \qquad \qquad \, \, \,  {\rm otherwise,} 
    \end{cases}
    \end{split}
\end{equation}
which correspond to an $(E_9 \oplus E_9) / \sim$ algebra, where both the $E_8$ factors are made affine by the current associated to the circle direction, $\partial Y$. 

We see that at infinite radius the nine-dimensional theory decompactifies to the corresponding ten-dimensional one, and it displays the affine version of the ten-dimensional gauge algebra $E_8 \oplus E_8$, realized by the KK tower becoming massless at this infinite distance limit. Not surprisingly, this algebra can also be obtained in the T-dual frame at $R\to 0$ with zero Wilson line considering the winding tower.\footnote{Note that for $R\to 0$ the only infinite distance point within the fundamental domain that includes the $R\to \infty$ boundary has Wilson line $(0_7,1,0_7,1)$ \cite{Keurentjes:2006cw}.}  

\subsection{Decompactification limit of the $SO(32)$ Heterotic String on $S^1$}
\label{sec:decomSO32}

Same as for the $E_8 \times E_8$ theory, there are different ways to reach the affine $SO(32)$ algebra. Here we show how to get it by starting from the  $E_8 \times E_8$ theory, with a Wilson line $A=(0_7, 1, 0_7, 1)$ and $R\to 0$.
This Wilson line  breaks the $E_8 \times E_8 \times U(1)$ to $SO(16) \times SO(16) \times U(1)$. In the limit $R \to 0$, the massless states and their associated holomorphic currents are
\begin{align}
\label{eqn:so32c}
    \alpha_{-1}^{I} \Tilde{\psi}_{-\frac{1}{2}}^{\mu} \ket{Z}_{NS} &\to J^{I}_m(z)  =  i \partial X^{I}(z) e^{i m Y(z)} \, , \\
\label{eqn:so32kk}    
    \alpha_{-1}^{9} \Tilde{\psi}_{-\frac{1}{2}}^{\mu} \ket{Z}_{NS} &\to J^{9}_m(z)  =  i \partial Y(z) e^{i m Y(z)} \, , \\
\label{eqn:so32r}    
    \Tilde{\psi}_{-\frac{1}{2}}^{\mu}\ket{Z}_{NS} &\to J^{\alpha}_{m}(z) =c_{\alpha} e^{i \pi^{I}_{\alpha}X^{I}(z)}  e^{i(m+\pi_{\alpha}^I \Lambda^I) Y(z)} \, ,
\end{align}
where $\pi_{\alpha} \in \Gamma_{16}$, $\Lambda=\big(\big(\frac{1}{2} \big)_8,0_8 \big)$ is the Wilson line that in the $SO(32)$ embedding breaks the gauge group to $SO(16) \times SO(16) \times U(1)$ and the charge vector is 
\begin{equation}
\label{currentsSO}
    Z= (\pi_{\alpha},w,n) \quad \text{with} \quad
    \begin{cases}
    w=-2m + 2 \, \pi_{\beta} \cdot \Lambda\, , \\
    n=2m + \pi_{\beta} \cdot (A - 2 \Lambda) \, , \\
    \pi_{\alpha}=2m A+ \pi_{\beta} - 2(\Lambda \cdot \pi_{\beta})A\, , 
    \end{cases} \forall m \in \mathbb{Z} \, , \pi_{\beta} \in \Gamma_{16} \, , |\pi_{\beta}|^2=0, 2 \, .
\end{equation}
The holomorphic currents one uses to compute the OPEs are 
\begin{equation}
\label{eqn:so32currents}
\begin{split}
    J^{I}_m(z)  &=  i (\partial X^{I}(z) + \Lambda^{I} \partial Y(z)) e^{i m Y(z)} \, , \\
    J^{\alpha}_{m}(z) &=c_{\alpha} e^{i \pi^{I}_{\alpha} (X^{I}(z) - \Lambda^{I} Y(z)) }  e^{imY(z)} \, . \\
\end{split}
\end{equation}
These can be shown to satisfy the commutation relations \eqref{eqn:e9commutator} with $n\to m$ and $\pi_{\alpha}$ being $SO(32)$ roots. This algebra is thus the $\widehat{D}_{16}$ affine algebra.  

We can also decompactify the $SO(32)$ to get the $(E_9\oplus E_9)/\sim$ taking the $R\to 0$ limit and Wilson line $\Lambda$.

\section{Decompactifications of the Heterotic String on tori}
\label{8ddecompactification}
The decompactification pattern from nine to ten dimensions of the heterotic theory on $S^1$ can be generalized to the decompactification of several dimensions of the theory on $T^d$. 

In the next section we show the possible decompactification patterns in  the case $d=2$, which is particularly interesting in light of its dual realisation in F-theory on K3, analyzed in \cite{Lee:2021qkx, Lee:2021usk}. In section \ref{Dddecompactification}, following the intuition in 8 dimensions, we extend to generic $d$.

\subsection{Decompactification limits of the Heterotic String on $T^2$} \label{8d}
Consider a generic background given by 
\begin{equation}
\label{eqn:8dbackground}
    G_{ij}= \begin{pmatrix}
        R_8^2 & R_8 R_9 \cos \alpha \\
        R_8 R_9 \cos \alpha & R_9^2
        \end{pmatrix} , \quad B_{ij}= \begin{pmatrix}
        0 & b \\
        -b & 0
        \end{pmatrix}, \quad A^I_i, \quad i=8,9.
\end{equation}
where $R_8$ and $R_9$ are the lengths of the two dimensional vectors $e_{1,2}$ generating the $T^2$ lattice and $\alpha$ is the angle between them.
In what follows we will also use the inverse metric, given by
\begin{equation}
\label{Ginvese}
    G^{ij}= \frac{1}{\sin^2 \alpha}\begin{pmatrix}
        \frac{1}{ R_8^2} & -\frac{\cos \alpha}{R_8 R_9} \\
        -\frac{\cos \alpha}{R_8 R_9} & \frac{1}{R_9^2}
        \end{pmatrix} \, .
\end{equation}

One can distinguish two types of decompactification limits\footnote{T-dual versions are considered as equivalent. For example, in the case of diagonal torus metric $\alpha=\frac{\pi}{2}$ to which we will restrict in the following, T duality acts as a shift of the Wilson lines by a vector in $\Gamma_8 \times \Gamma_8$ or as $A_i'= - \frac{
A_i}{R_ i^2
+ \frac{A_i^2}{2}}, \, R_ i'=\frac{R_i}{R_ i^2 + \frac{A_i^2}{2}}, \, w^i \leftrightarrow n_i, \, i=8,9$.},  according to the values of $R_8,\,R_9 \, \cos \alpha$: 
\begin{itemize}
    \item only one radius diverges:    decompactification from eight to nine dimensions; 
    \item both $R_8, \, R_9  \to \infty$, independently of their ratio ($\frac{R_8}{R_9} \to 0, \, \text{const}, \,  \infty)$: decompactification from eight to ten dimensions.
\end{itemize}
The possible interesting profiles for $\cos \alpha$ along the path taken in moduli space depend on the type of decompactification, and they will be specified case by case.

\subsubsection{$8d\to9d$ decompactification}
\label{89}
Let us focus on the first situation, where only one radius diverges, $R_8\to \infty$.  
The massless states can only come from states with no winding in the eighth direction, $w^8=0$. The massless and the level-matching conditions reduce then to 
\begin{equation}
\label{eqn:lowdimmc}
\begin{split}
    0 &=  \frac{1}{\sin^2 \alpha} \left(\frac{1}{R_9^2} (p_{
    R,9}^2 +p_{L,9}^2)  - (w^9)^2 R_9^2 \cos^2 \alpha \right) + p_I^2 + 2 \left( N + \Tilde{N} - \frac{3}{2} \right) \, ,\\
    0 &= 2n_9 w^9 + |\pi_{\alpha}|^2 + 2 \left( N - \Tilde{N} - \frac{1}{2} \right) \, ,
\end{split}
\end{equation}
where we have used the  inverse metric in \eqref{Ginvese}, and 
\begin{equation}
\label{eqn:9dmom}
   \begin{split}
    p_{R,9} &= \frac{1}{\sqrt{2}} \left(n_9 -\left (R_9^2 + \frac{1}{2} A_9 \cdot A_9 \right) w^9 - \pi_{\alpha} \cdot A_9\right) \, , \\
    p_{L,9} &= \frac{1}{\sqrt{2}} \left(n_9 + \left(R_{9}^2 - \frac{1}{2} A_9 \cdot A_9\right) w^9 - \pi_{\alpha} \cdot A_9 \right) \, , \\
    p^I &= \pi_{\alpha}^I + A^I_9 w^9 \, .
  \end{split}
\end{equation}
Note that \eqref{eqn:lowdimmc} with the expressions \eqref{eqn:9dmom} for the internal momenta correspond, in the limit $\cos \alpha \to 0$, to the massless and the level-matching conditions of the nine-dimensional theory. On the other hand, when 
$\cos \alpha \to c \ne 0$, which corresponds to the asymptotic behaviour $G_{89}  \to \infty$, nine-dimensional Lorentz invariance is broken. We therefore restrict to $\cos \alpha =0$. Similarly, nontrivial values of the Kalb-Ramond field break Lorentz invariance, so we also restrict to $b = 0$.

If we choose the moduli such that $R_9$ and $A^{I}_9$ correspond to a point with gauge algebra $\mathcal{A}$ in the left sector of the nine-dimensional theory, with roots of the form $Z_{\cal A}=(\pi_{\alpha},w^9,n_9)$ satisfying $p_{L,9} = 2$ and $p_{R,9} = 0$, the massless spectrum of the eight-dimensional theory will contain the momentum tower of the nine-dimensional gauge vectors. This is given by the following left-moving vectors together with their associated currents, where  
\begin{itemize}
    \item Cartan sector: 
    \begin{equation}
    \label{eqn:8dcartan}
    \begin{split}
        \alpha^{\hat{I}}_{-1} \Tilde{\psi}^{\mu}_{-\frac{1}{2}} \ket{0,n_8}_{NS} &\to J^{\hat{I}}_{n_8}(z) = i \partial X^{\hat{I}}(z) e^{i n_8 Y^8(z)} \, , \\
        \alpha^{8}_{-1} \Tilde{\psi}^{\mu}_{-\frac{1}{2}} \ket{0,n_8}_{NS} &\to J^{8}_{n_8}(z) = i \sqrt{2} \partial Y^{8}(z) e^{i n_8 Y^8(z)} \, ;
    \end{split}
    \end{equation}
    \item root sector:
    \begin{equation}
    \label{eqn:8droots}
        \tilde{\psi}_{-\frac{1}{2}}^{\mu} \ket{Z_{\cal A}, n_8}_{NS} \to J^{\alpha}_{n_8}(z) = c_{\alpha} e^{i p_{\alpha;\hat{I}}X^{\hat{I}}(z)} e^{in_8 Y^8(z)} \, ,
    \end{equation}
\end{itemize}
where the first number in the kets represents the nine dimensional quantum numbers, $n_8 \in {\mathbb Z}$, and $w^8 = 0$ is implied. We have defined the index $\hat{I}=\{9,I\}$, the left moving $X^{\hat{I}}(z) = \left(\sqrt{2} e_{\,\, 9}^9 Y^9(z), X^{I}(z)\right)$,  and $p_{\alpha;\hat{I}}=(e_9^{*9} p_{L,9}, p_{I})$. We have also restricted to $A^I_8 = 0$, which together with the choices $\cos \alpha,~b = 0$ implies in particular that $E_{89},~ E_{98} = 0$; the case $A^I_8 \ne 0$ is  discussed in Appendix \ref{appa1}, but in both cases the resulting BPS algebra is the same and so we present the former for simplicity. 

As derived in Appendix \ref{appa1}, for $R_8 \to \infty$ the OPEs among the asymptotic currents of equations \eqref{eqn:8dcartan} and \eqref{eqn:8droots} are 
\begin{align}
\label{eqn:cc}
    J^{\hat{I}}_{n_8}(z) J^{\hat{J}}_{m_8}(w) &\sim  \delta^{\hat{I}\hat J} \frac{:e^{i(n_8+m_8)Y^8(w)}:}{(z-w)^2} + i \delta^{\hat I \hat J} n_8 \frac{:\partial Y^8(w) e^{i (n_8+m_8)Y^8(w)}: }{z-w} + \mathcal{O}(1)  \, , \\
    \label{eqn:cr}
    J^{\hat{I}}_{n_8}(z) J^{\alpha}_{m_8}(w) &\sim \frac{p_{\alpha}^{\hat I} J^{\alpha}_{n_8+m_8}(w)}{z-w} + \mathcal{O}(1) \, , \\
    \label{eqn:rr1}
    J^{\alpha}_{n_8}(z) J^{\beta}_{m_8}(w) &\sim \begin{cases}
        \frac{\epsilon(\alpha, \beta) J^{\alpha+\beta}_{n_8+m_8}(w)}{z-w} + \mathcal{O}(1)\qquad \qquad \qquad \qquad \qquad \qquad \qquad \qquad \quad  \alpha+\beta \,  {\rm root,} \\
        \frac{:e^{i (n_8+m_8) Y^8(w)}:}{(z-w)^2} + \frac{p_{\alpha;\hat{I}} J^{\hat{I}}_{n_8+m_8}(w)+in_8:\partial Y^8(w) e^{i (n_8+m_8)Y^8(w)}:}{z-w} +\mathcal{O}(1) \, \quad \alpha=-\beta\, , \\
        0 \quad \qquad \qquad \qquad \qquad \qquad \qquad \qquad \qquad\qquad \qquad \qquad \quad \quad \, \, \,  {\rm \, otherwise.} 
    \end{cases}
\end{align}
All the OPEs involving $J^{8}_{n_8}(z)$ are trivial in the limit $R \to \infty$.

The asymptotic algebra is obtained from the commutation relations of the zero modes of the currents \eqref{eqn:8dcartan}-\eqref{eqn:8droots}, defined as in \eqref{eqn:zeromode}. We find (see  Appendix \ref{appa1} for the detailed computation) 
\begin{equation}
\label{eqn:8dcommutator}
\begin{split}
    &[(J^{\hat{I}}_n)_0, (J^{\hat{J}}_m)_0] = i n \delta^{\hat{I}\hat{J}} \delta_{n+m, 0} (\partial Y^8)_0 \, ,\\
    &[(J^{\hat{I}}_n)_0, (J^{\alpha}_m)_0] = p_{\alpha}^{\hat{I}} (J^{\alpha}_{n+m})_0 \, ,\\
    &[(J^{\alpha}_n)_0, (J^{\beta}_m)_0] = \begin{cases}
          \epsilon(\alpha, \beta) (J^{\alpha + \beta}_{n+m})_0  \qquad \quad \qquad \quad\quad \, \,   \,  \alpha+\beta \,  {\rm root}, \\
          p_{\alpha;\hat{I}} (J^{\hat{I}}_{n+m})_0 + i n \delta_{n+m, 0} (\partial Y^8)_0 \quad  \alpha=-\beta\, ,\\
         0 \qquad \qquad \quad \qquad \qquad \qquad \qquad \, \,  {\rm otherwise,} 
    \end{cases}
\end{split}
\end{equation}
which is precisely the algebra of $\hat{\mathcal{A}}$, with central extension $(\partial Y^8)_0$.
Importantly, if the algebra $\mathcal{A}$ is semisimple, $\mathcal{A}=\mathcal{A}_1 \oplus ... \oplus \mathcal{A}_n$, \eqref{eqn:8dcommutator} describes the algebra $\hat{\mathcal{A}}=(\widehat{\mathcal{A}}_1 \oplus ... \oplus \widehat{\mathcal{A}}_n) / \sim$, where $\sim$ means that the central extension of all the factors (including $U(1)$'s) is identified, namely as $(\partial Y^8)_0$. This signifies that all the factors are made affine by the universal presence of the momentum tower.

As an example, let us show explicitly the case of $(\widehat{E}_8 \oplus \widehat{E}_8 \oplus \widehat{A}_1) / \sim$ (with $A_1$ the algebra of $SU(2)$, not to be confused with the aforementioned generic algebra $\mathcal{A}_1$) in the decompactification of the $E_8 \times E_8$ theory from 8 to 9 dimensions.
We start from the background 
\begin{equation}
    G_{ij}= \begin{pmatrix}
        R_8^2 & 0 \\
        0 & 1
        \end{pmatrix}, \quad b = 0, \quad A_8^{I}, \quad A_9^I=0\,,
\end{equation}
 and take the limit $R_8 \to \infty$. As we argued above, without loss of generality we can take $A_8^I$ to vanish. The choice  $G_{99}=1$ and $A_9^I=0$ is instead dictated by the fact that we want to decompactify to a theory with gauge algebra $E_8 \oplus E_8 \oplus A_1$ in 9 dimensions.  The massless states here are the momentum towers along $y^8$ of the $E_8 \oplus E_8$ states, 
\begin{align}
\label{eqn:e8c}
    \alpha^I_{-1} \tilde{\psi}^{\mu}_{-\frac{1}{2}}\ket{0, n_8}_{NS} \, &\to J_{n_8}^I(z) = i \partial X^I(z) e^{i n_8 Y^8(z)} \, , \\
\label{eqn:e8r}
    \tilde{\psi}^{\mu}_{-\frac{1}{2}}\ket{Z_{E_8 \oplus E_8}, n_8}_{NS} \,  &\to J^{\alpha}_{n_8}(z) = c_{\alpha} e^{i \pi_{\alpha;I} X^I(z)} e^{i n_8 Y^8(z)}\, ,
\end{align}
where $Z_{E_8 \oplus E_8}=(\pi_{\alpha}, 0, 0), \pi_{\alpha}$ are the roots of $E_8 \oplus E_8$, and in addition we have the corresponding momentum towers of the $A_1$ states
\begin{align}
\label{eqn:a1c}
    \alpha^9_{-1} \tilde{\psi}^{\mu}_{-\frac{1}{2}}\ket{0, n_8}_{NS} \, &\to J^9_{n_8}(z) = i \sqrt{2} \partial Y^9(z) e^{i n_8 Y^8(z)}, \, \\
\label{eqn:a1r}    
    \tilde{\psi}^{\mu}_{-\frac{1}{2}}\ket{Z_{A_1}, n_8}_{NS} \, &\to J^{\pm}_{n_8}(z) = e^{\pm i 2  Y^9(z)} e^{i n_8 Y^8(z)} \, ,
\end{align}
with $Z_{A_1}\equiv (\pi,w_9,n_9)=(0,\pm 1 ,\pm 1)$. Here  \eqref{eqn:e8c} and \eqref{eqn:a1c} are associated to the Cartan currents and \eqref{eqn:e8r} and \eqref{eqn:a1r} to the root generators. The affinization of the $E_8 \oplus E_8$ part of the algebra works exactly  as in the decompactification from 9 to 10 dimensions presented in Section \ref{sect0}, with the replacements $n \to n_8$ and $Y \to Y^8$. 
The affinisation of the $A_1$ factor is derived from the general structure of the OPEs \eqref{eqn:survive}, \eqref{eqn:rrgeneral} and \eqref{eqn:ialpha} among the affine currents \eqref{eqn:a1c} and \eqref{eqn:a1r}. These give the subalgebra
\begin{equation}
\label{eqn:a1commutator}
\begin{split}
    [(J^{9}_{n_8})_0, (J^{9}_{m_8})_0] &= i n_8 \delta_{n_8+m_8, 0} (\partial Y^8)_0 \, ,\\
    [(J^{9}_{n_8})_0, (J^{\pm}_{m_8})_0] &= \pm \sqrt{2} (J^{\pm}_{n_8+m_8})_0 = \pi_{A_1, \pm} (J^{\pm}_{n_8+m_8})_0 \, ,\\
    [(J^{\pm}_{n_8})_0, (J^{\pm}_{m_8})_0] &= 0 \, , \\
    [(J^{\pm}_{n_8})_0, (J^{\mp}_{m_8})_0] &= \pi_{A_1,\pm} 
    (J^{\pm}_{n_8+m_8})_0 + i n_8 \delta_{n_8+m_8,0} (\partial Y^8)_0
\end{split}
\end{equation}
among the current zero modes, which corresponds to $\widehat{A}_1$ with central extension $(\partial Y^8)_0$, the commutators between $\widehat{E}_8 \oplus \widehat{E}_8$ modes and $\widehat{A}_1$ ones vanishing in the limit. This is indeed a $(\widehat{E}_8 \oplus \widehat{E}_8 \oplus \widehat{A}_1) / \sim$ algebra for the current zero modes.

We emphasize that the Cartans corresponding to the algebra of the decompactified theory get affinized independently of them being enhanced or not to nonabelian algebras. In fact, the same is true for those Cartans lying in the gravity multiplet, which do not admit enhancements. These get affinized by the antiholomorphic counterpart $(\bar \partial \tilde Y^8)_0$ of the left-moving central extension (see Appendix \ref{appa1} for details).

\subsubsection{$8d\to10d$ decompactification}
\label{sec:doubleloop}
In this limit both  $R_8 \text{ and } R_9$ diverge. They can go to infinity with different speeds $\frac{R_8}{R_9} \to 0, \, \infty$ or at the same rate, $\frac{R_8}{R_9} \to \text{const}$. The first can effectively capture a decompactification by steps, $8 \to 9 \to 10$ dimensions, while the second is always a one-step process $8 \to 10$ dimensions. 
Again, in the limit $R_8, \, R_9 \to \infty$, the massless spectrum is characterised by $w^8=w^9=0$, with no restriction on $n_8, \, n_9$. 

The non-trivial contributions to the massless and level-matching conditions are then
\begin{equation}
\label{eqn:onlyhet}
\begin{split}
    0 &= | \pi_{\alpha}|^2  + 2 \left( N + \Tilde{N} - \frac{3}{2} \right) \, ,\\
    0 &=  |\pi_{\alpha}|^2 + 2 \left( N - \Tilde{N} - \frac{1}{2} \right) \, ,
\end{split}
\end{equation}
regardless of the values of $\cos \alpha$, $b$, $A^I_8$ and $A^I_9$. The massless left-moving vectors are then the momentum towers of the $E_8 \oplus E_8 \oplus U(1)^2$ vectors or $D_{16}\oplus U(1)^2$ vectors, depending on which theory one starts with. 
From the algebra point of view turning on $B_{ij}$ or the Wilson lines is equivalent to setting them to zero, as shown in Appendix \ref{appa2}, so we take for simplicity $B_{ij}=0$, $A^I_8=0$, $A^I_9=0$ and $G_{89}=0$ in the following expressions for the currents. The two KK Cartans are associated to ($i=8,9$) 
\begin{equation}
\label{eqn:kkc}
    \alpha^{i}_{-1} \Tilde{\psi}^{\mu}_{-\frac{1}{2}} \ket{0, n_8, n_9}_{NS} \to J^{i}_{n_8, n_9}(z) = i \sqrt{2} \partial Y^{i}(z) e^{in_8 Y^8(z)} e^{in_9 Y^9(z)} \, ,
\end{equation}
the $E_8 \oplus E_8$ Cartans to
\begin{equation}
\label{eqn:e8cartanss}
    \alpha^{I}_{-1} \Tilde{\psi}^{\mu}_{-\frac{1}{2}} \ket{0, n_8,n_9}_{NS} \to J^{I}_{n_8, n_9}(z) = i  \partial X^{I}(z) e^{in_8 Y^8(z)} e^{in_9 Y^9(z)} \, ,
\end{equation}
and the $E_8 \oplus E_8$ roots to
\begin{equation}
\label{eqn:allroots}
    \Tilde{\psi}_{-\frac{1}{2}}^{\mu} \ket{Z_{E_8 \oplus E_8}, n_8, n_9}_{NS} \to J^{\alpha}_{n_8, n_9}(z) = c_{\alpha} e^{i\pi_{\alpha;I} X^{I}(z)} e^{in_8 Y^8(z)} e^{in_9 Y^9(z)} \, ,
\end{equation}
As shown in detail in Appendix \ref{appa2}, in the decompactification limit the OPEs among these currents are, regardless of $\frac{R_8}{R_9}$,
\begin{align}
    J^{I}_{n_8, n_9}(z) J^{J}_{m_8, m_9}(w) &= \frac{\delta^{IJ}:e^{i (n_i+m_i) Y^i(w)}:}{(z-w)^2} + i \delta^{IJ} n_i \frac{:\partial Y^i(w) e^{i (n_j+m_j) Y^j(w)}:}{z-w} + \mathcal{O}(1) \, , \\
    J^{I}_{n_8, n_9}(z) J^{\alpha}_{m_8, m_9}(w)  &=  \frac{\pi^{I}_{\alpha} J^{\alpha}_{n_8+m_8,n_9+m_9}(w)}{z-w} + \mathcal{O}(1)  \, , \\
    J^{\alpha}_{n_8, n_9}(z) J^{\beta}_{m_8, m_9}(w)  &=  
    \begin{cases}
        \frac{\epsilon(\alpha, \beta) J^{\alpha+\beta}_{n_8+m_8,n_9+m_9}(w)}{z-w} + \mathcal{O}(1)\qquad \qquad \qquad \qquad \qquad \qquad \qquad \, \, \, \, \quad   \alpha+\beta \,  {\rm root,} \\
        \frac{:e^{i (n_i+m_i) Y^i(w)}:}{(z-w)^2} + \frac{\pi_{\alpha;I} J^{I}_{n_8+m_8,n_9+m_9}(w)+i n_i:\partial Y^i(w) e^{i (n_j+m_j)Y^j(w)}:}{z-w} +\mathcal{O}(1) \, \, \alpha=-\beta\, , \\
        0 \qquad \qquad \qquad \qquad \qquad \qquad \qquad\qquad \qquad \qquad \qquad \qquad  \qquad \, \, \, {\rm \, otherwise,} 
    \end{cases}
\end{align}
All the other OPEs are either finite or vanishing as the inverse metric components.
These have the structure of the double loop versions of the ten dimensional algebra ($\mathcal{A}=E_8 \oplus E_8, \, D_{16}$) with the addition of two central extensions, which can be identified from the structure of the single pole in the OPEs and are the zero modes $(\partial Y^i)_0$.  The algebra of the zero modes \eqref{eqn:zeromode} is
\begin{equation}
\label{eqn:loope9commutator}
\begin{split}
   & [(J^{I}_{n_8,n_9})_0, (J^{J}_{m_8,m_9})_0] = i \delta^{IJ} (n_8  \delta_{n_8+m_8, 0} (\partial Y^8_{0,n_9+m_9})_0 + n_9  \delta_{n_9+m_9, 0} (\partial Y^9_{n_8+m_8,0})_0 ) \, ,\\
   &[(J^{I}_{n_8,n_9})_0, (J^{\alpha}_{m_8,m_9})_0] = \pi_{\alpha}^{I} (J^{\alpha}_{n_8+m_8, n_9+m_9})_0 \, ,\\
    &[(J^{\alpha}_{n_8,n_9})_0, (J^{\beta}_{m_8,m_9})_0] = \begin{cases}
          \epsilon(\alpha, \beta) (J^{\alpha + \beta}_{n_8+m_8,n_9+m_9})_0  \qquad \qquad \qquad \qquad \qquad \qquad \qquad \,\,\, \, \,    \,  \alpha+\beta \,  {\rm root}, \\
          \pi_{\alpha;I} (J^{I}_{n_8+m_8,n_9+m_9})_0 + \\
          \quad + i (n_8  \delta_{n_8+m_8, 0} (\partial Y^8_{0, n_9+m_9})_0 + n_9  \delta_{n_9+m_9, 0} (\partial Y^9_{n_8+m_8,0})_0 )  \quad  \, \alpha=-\beta\, ,\\
         0 \qquad \qquad \qquad \qquad \qquad \qquad \qquad \qquad \qquad \qquad \qquad \qquad \,   {\rm otherwise,} 
    \end{cases}
    \end{split}
\end{equation}
 where
\begin{equation}
    (\partial Y^i_{n_j+m_j,0})_0 = \oint \frac{dz}{2\pi i} \partial Y^i(z) e^{i (n_j+m_j)Y^j(z)} \, , \quad i \ne j 
\end{equation}
are the momentum modes of the two central extensions of the algebra, $(\partial Y^8)_0$ and $(\partial Y^9)_0$, along the $y^9$ and $y^8$ directions, respectively. This gives the double loop version $(\widehat{E}_9 \oplus \widehat{E}_9)/ \sim$ or ${\widehat{\widehat{D}}}_{16}$, and we see that indeed the operators associated to the two torus directions are good central extensions.
Let us notice that in the limit $R_{8,9}\to \infty$ the only possible decompactification is to the ten dimensional theory one started with, independently of their ratio.  Since the toroidally compactified $E_8 \times E_8 \text{ and } SO(32)$  theories are T dual,  it is also possible to reach the 10 dimensional $E_8 \times E_8$ heterotic theory from the eight dimensional $SO(32)$ one and viceversa.\footnote{For this we can take the path discussed in section \ref{sec:decomSO32}, namely $R_9 \to 0$ and Wilson line $\Lambda=\big(\big(\frac{1}{2} \big)_8,0_8 \big)$ and then take $R_8 \to \infty$ with any Wilson line, or equivalently $8\leftrightarrow 9$.}

\subsection{Decompactification limits of the Heterotic String on $T^d$}
\label{Dddecompactification}

One can easily generalize the above discussion to the decompactification of an arbitrary number $k$ of dimensions starting from the heterotic theory compactified on $T^d$, $k \leq d$. 

We start from a fixed point in the moduli space of $T^{d-k}$, specified by a given
$G_{\hat{\imath} \hat{\jmath}}$, $B_{\hat{\imath} \hat{\jmath}}$ and $A^I_{\hat{\imath}}$ $(\hat{\imath}=1,..,d-k)$, giving an enhanced gauge algebra $\mathcal{A}$. As explained in Section \ref{89}, in order to decompactify to a Lorentz invariant vacuum of a higher dimensional toroidally compactified heterotic theory, the $T^d$ background must be such that $T^d=T^k \times T^{d-k}$ as follows 
\begin{equation}
\label{eqn:metric}
    G_{ij}= \begin{pmatrix}
        G_{\bar{\imath} \bar{\jmath}} & 0 \\
        0 & G_{\hat{\imath}\hat{\jmath}}
        \end{pmatrix} = e_{i} \cdot e_{j},  \quad B_{ij}= \begin{pmatrix}
        B_{\bar{\imath} \bar{\jmath}} & 0 \\
       0 & B_{\hat{\imath}\hat{\jmath}}
        \end{pmatrix}, \quad A^I_{i}=(A^I_{\bar{\imath}},A^I_{\hat{\imath}})
\end{equation}
with $i =1,...,d$, $i=(\bar{\imath},\hat{\imath})$.  
$G_{\bar{\imath} \bar{\jmath}}$ is parametrised by $k$ radii $R_{\bar{\imath}}$ and $\frac{k(k-1)}{2}$ angles. The decompactification limit is realized by taking  $R_{\bar{\imath}} \to \infty$, and a generic finite value for  $A_{\bar{\imath}}^I$. Again, T dual backgrounds can be obtained via an $O(d,d+16, \mathbb{Z})$ transformation to the one described above.

The general pattern is that, under these assumptions, the $10-d$ dimensional theory  displays the $k$-times loop version of the gauge algebra in $10-(d-k)$ dimensions at the point in moduli space given by $\{g_{\hat{\imath}\hat{\jmath}}, \, B_{\hat{\imath}\hat{\jmath}} \, A^I_{\hat{\imath}}\}$, with central extensions $(\partial Y^{\bar{\imath}})_0$. In the case of a semi-simple algebra $\mathcal{A}=\mathcal{A}_1 \oplus \mathcal{A}_2 \oplus ... \oplus \mathcal{A}_n$ all the factors  are made affine 

For concreteness, let us introduce the index $\hat{I}=\{\hat{\imath},I\}$, $X^{\hat{I}}(z)=(\sqrt{2} e_{\, \, \hat{\imath}}^{\hat{a}} Y^{\hat{\imath}}(z),X^I(z))$, and call and call $p_{\alpha;\hat{I}}$ the roots of the algebra $\mathcal{A}$ in dimension $10-(d-k)$, corresponding to the $16+d-k$ momenta $p_{\alpha;\hat{I}} = (e^{*\hat{\imath}}_{\hat{a}}p_{L,\hat{\imath}},p_I)$.

 The left-moving massless states and their associated holomorphic conserved currents, related to the Cartan and ladder generators respectively, are then
\begin{equation}
\begin{split}
    \alpha^{\hat{I}}_{-1} \bar{\psi}^{\mu}_{-\frac{1}{2}} \ket{0}_{NS} &\longrightarrow J^{\hat{I}}
    (z) = i \partial X^{\hat{I}}(z) \, ,\\        \bar{\psi}_{-\frac{1}{2}}^{\mu} \ket{Z_{\mathcal{A}}}_{NS} &\longrightarrow J^{\alpha}(z) = c_{\alpha} e^{i p_{\alpha; \hat{I}} \cdot X^{\hat{I}}(z)} \, .
\end{split}
\end{equation}
and they satisfy the massless and level-matching conditions 
\begin{equation}
    \label{eqn:lowdim}
    \begin{cases}
    \begin{split}
             0&= G^{\hat{\imath}\hat{\jmath}} (p_{L,\hat{\imath}} p_{L,\hat{\jmath}} + p_{R,\hat{\imath}} p_{R,\hat{\jmath}}) +  p_I^2  + 2 \Big( N + \Tilde{N} -\frac{3}{2} \Big) \, , \\
             0&= 2 n_{\hat{\imath}} w^{\hat{\imath}} + 2 \Big( N - \Tilde{N} -\frac{1}{2} \Big)  \, ,
    \end{split}         
     \end{cases}
\end{equation}
Additionally compactifying $k$ directions as in \eqref{eqn:metric}, the internal momenta $P$ of the $10-d$ dimensional theory in terms of the ones in the $10-(d-k)$ dimensional ones, $p$, are 
\begin{equation}
\label{eqn:Pp}
\begin{split}
    P_{R,\hat{\imath}} 
    &= 
     p_{R,\hat{\imath}} - \frac{1}{\sqrt{2}} E_{\hat{\imath}\bar{\jmath}} w^{\bar{\jmath}}\, , \\
    P_{L,\hat{\imath}} &= p_{L,\hat{\imath}} - \frac{1}{\sqrt{2}} E_{\hat{\imath}\bar{\jmath}} w^{\bar{\jmath}} \, , \\
    P^{I} &= p^{I} + A^I_{\bar{\jmath}} w^{\bar{\jmath}} \, ,
\end{split}
\end{equation}
while $P_{R/L,\bar{\imath}}$ are independent from $p_{R/L,\hat{\imath}}$. 

The inverse internal metric components satisfy $G^{\bar{\imath} \bar{\jmath}} \sim \frac{1}{R_{\bar{\imath}} R_{\bar{\jmath}}} \to 0$. In order to have massless states one should set to zero the possibly divergent term in the $10-d$ dimensional mass formula 
\begin{equation} 
\begin{split}
     G^{\bar{\imath} \bar{\jmath}} \left((2g_{\bar{\imath} \bar{k}}-E_{\bar{\imath} \bar{k}})(2 g_{\bar{\jmath} \bar{h}} - E_{\bar{\jmath} \bar{h}} )+ E_{\bar{\imath} \bar{h}} E_{\bar{\jmath} \bar{h}}  \right) w^{\bar{k}} w^{\bar{h}} \sim g_{\bar{k}\bar{h}} w^{\bar{k}} w^{\bar{h}} =0
\end{split}
\end{equation}
which is achieved by requiring  $w^{\bar{\imath}}=0$. The level-matching condition in $10-d$ dimensions reduces to
\begin{equation}
\label{LMC10-d+k}
    0= 2 n_{\hat{\imath}}w^{\hat{\imath}} + |\pi_{\alpha}|^2 +2 \left( N - \Tilde{N} - \frac{1}{2}  \right)
\end{equation}
which is exactly the same expression in $10-d+k$ dimensions with the only difference that the oscillators can be turned on also in the additional $k$ directions. This will provide  the states related to the central extensions. 
Moreover, having vanishing winding along the directions we are decompactifying in \eqref{eqn:Pp} gives asymptotically in moduli space
\begin{equation}
    P_{R,\hat{\imath}} = p_{R,\hat{\imath}} \, , \quad
    P_{L,\hat{\imath}} = p_{L,\hat{\imath}} \, , \quad
    P^{I} = p^I \, ,
\end{equation}
so that the massless condition is 
\begin{equation}
\label{eqn:0}
\begin{split}
    G^{\hat{\imath}\hat{\jmath}} \left( p_{L,\hat{\imath}} p_{L,\hat{\jmath}} +p_{R,\hat{\imath}} p_{R,\hat{\jmath}} \right) + G^{\bar{\imath}\bar{\jmath}} \left( P_{L,\bar{\imath}}  P_{L,\bar{\jmath}} + P_{R,\bar{\imath}} P_{R,\bar{\jmath}} \right) + \\ + p^2_I  + 2 \left( N + \Tilde{N} -  \frac{3}{2} \right) = 0 \, .
\end{split}
\end{equation}
where for $w^{\bar{\imath}}=0$
\begin{equation}
\begin{split}
    P_{R,\bar{\imath}} &= \frac{1}{\sqrt{2} } \left(n_{\bar{\imath}}-E_{\bar{\imath}\hat{\jmath}}w^{\hat{\jmath}} - \pi_{\alpha} \cdot A_{\bar{\imath}}\right) \, ,\\
    P_{L,\bar{\imath}} &= \frac{1}{\sqrt{2}} \left(n_{\bar{\imath}}-E_{\bar{\imath}\hat{\jmath}}w^{\hat{\jmath}}- \pi_{\alpha} \cdot A_{\bar{\imath}} \right) \, .
\end{split}
\end{equation}
Since at leading order $G^{\bar{\imath}\bar{\jmath}} \sim \frac{1}{R_{\bar{\imath}}R_{\bar{\jmath}}} \to 0$ and $G^{\hat{\imath}\hat{\jmath}} \sim \mathcal{O}(1)$,  \eqref{eqn:0} reduces to
\begin{equation}
    G^{\hat{\imath}\hat{\jmath}} \left( p_{L,\hat{\imath}} p_{L,\hat{\jmath}} +p_{R,\hat{\imath}} p_{R,\hat{\jmath}} \right) + p^2_I  + 2 \left( N + \Tilde{N} -  \frac{3}{2} \right) = 0 \, ,
\end{equation}
and together with \eqref{LMC10-d+k} this implies that the massless spectrum coincides with the momentum towers of the massless spectrum of the $10-(d-k)$ dimensional theory \eqref{eqn:lowdim}, with $k$ additional left KK vectors (with related momentum towers) to the Cartans in $10-d+k$ dimensions, so that the massless vectors are 
\begin{equation}
\label{eqn:masslessd}
\begin{split}
    \alpha^{\bar{I}}_{-1} \Tilde{\psi}^{\mu}_{-\frac{1}{2}} \ket{0,n_{\check{\jmath}}}_{NS} &\longrightarrow J^{\bar{I}}
    (z) = i \partial X^{\bar{I}}(z) e^{i n_{\check{\jmath}} Y^{\check{\jmath}}(z)} \, ,\\    
    \Tilde{\psi}_{-\frac{1}{2}}^{\mu} \ket{Z_{\mathcal{A}}, n_a}_{NS} &\longrightarrow J^{\alpha}(z) = c_{\alpha} e^{i p_{\alpha;\hat{I}}  X^{\hat{I}}(z)} e^{in_{\check{\jmath}} Y^{\check{\jmath}}(z)} \, .
\end{split}
\end{equation}
where $\bar{I}=(i,I)=(\bar{\imath},\hat{I})$, $X^{\bar{I}}=( \sqrt{2} Y^{\bar{\imath}},X^{\hat{I}})$.
The OPEs among these states are found in Appendix \ref{appa3}, where one can see that the algebra is the $k$-th loop version of ${\cal A}$ (with the corresponding $k$ central extensions).

\section{Review of string junctions}
\label{Ftheory}
In the previous section we argued that infinite distance points in the moduli space of the perturbative heterotic theory on $T^d$ are characterised by the presence of affine BPS algebras. In particular, the predictions in \ref{8d} for the case $d=2$ can be matched to the ones in the dual framework of F-theory on K3, worked out in \cite{Lee:2021usk}. The presence of affine algebras in these limits is locally detected through the intersection patterns of string junctions supported by stacks of 7-branes as argued already in \cite{DeWolfe:1998yf,DeWolfe:1998zf}.
In this section we give a brief overview of string junctions, their properties under monodromies and some key concepts for the following study of affine algebras. We follow mainly \cite{DeWolfe:1998zf} (see also \cite{Cvetic:2022uuu}).
The reader who is already familiar with these concepts can skip this section and go directly to Section \ref{Het-Ftheory}. 

\subsection{Basic concepts on String Junctions}
\begin{figure}[tb]
	\begin{center}
		\subfigure[]{
			\includegraphics[scale=0.23]{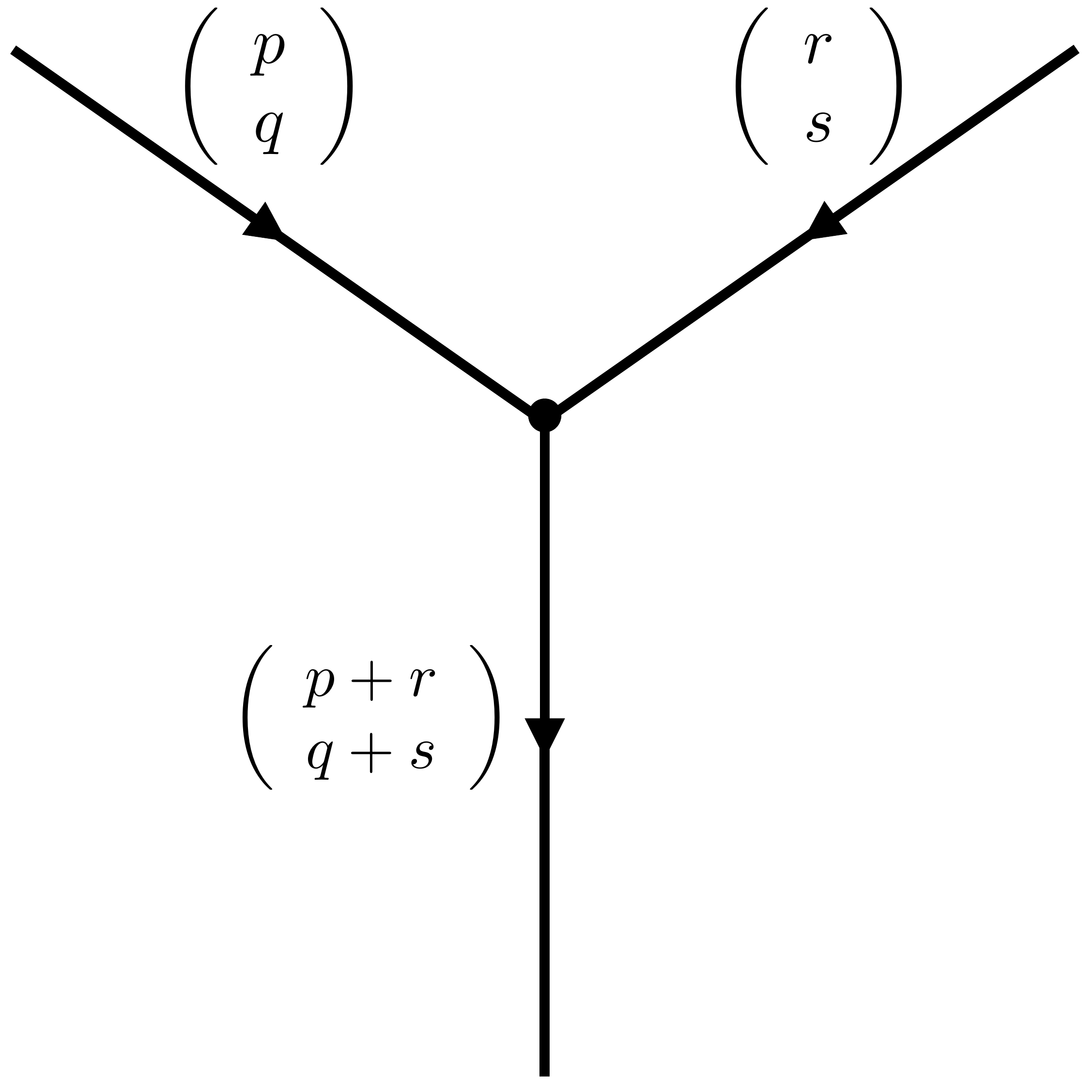} 
			\label{fig:sj1}
			}
		\subfigure[]{
			\includegraphics[scale=0.30]{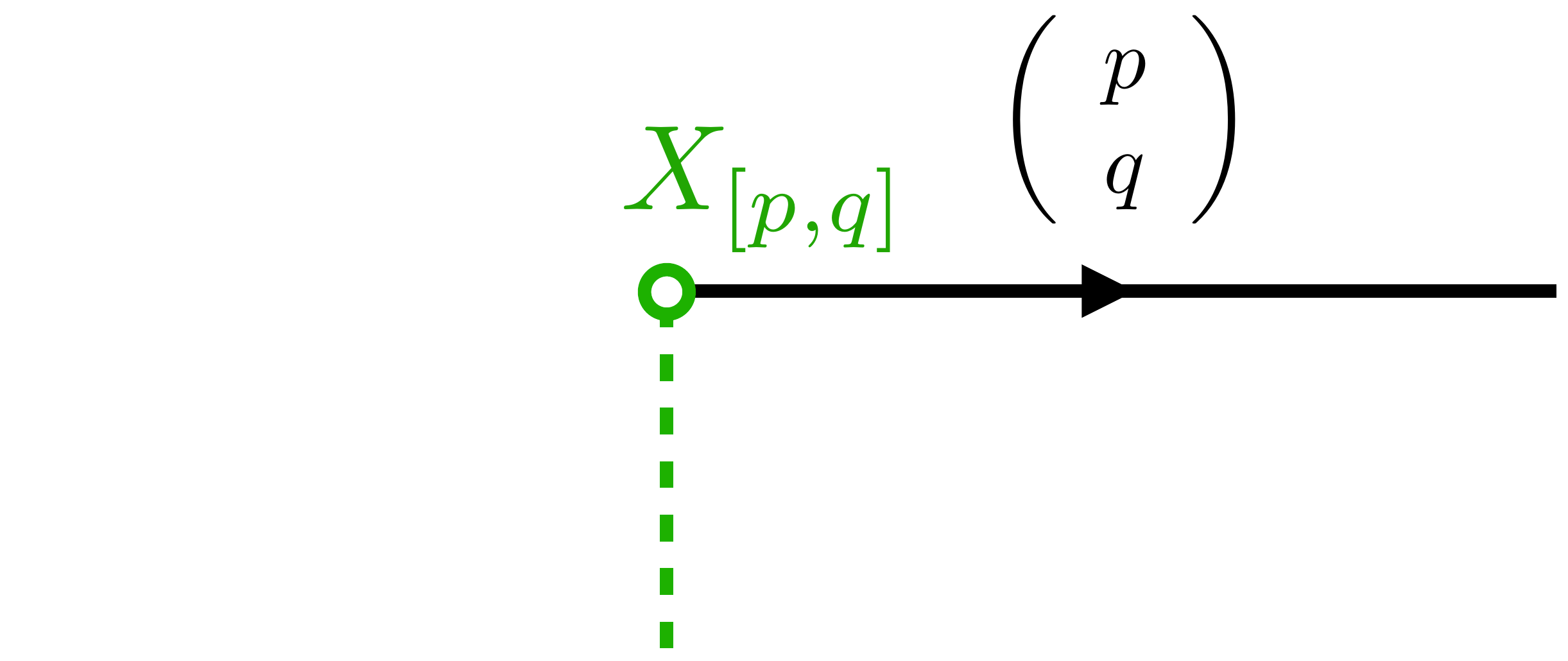}
			\label{fig:sj2}
			}
		\caption{\footnotesize \textbf{(a)} String junction from joining two strings with $(p,q)$ and $(r,s)$ charges at a point. \textbf{(b)} A $(p,q)$ prong. The dotted vertical line represents the branch cut of the 7-brane, which is not intersected by the string.}			
		    \label{fig:sj}
	\end{center}
\end{figure}

The spectrum of F-theory includes $(p,q)$-strings, bound states of $p$ F1 strings and $q$ D1 strings.
7-branes also carry $[p,q]$-indices, which indicate the kind of $(p,q)$-strings that end on them. A string junction is a web of $(p,q)$-strings that may join at a point\footnote{This is a pictorial way of seeing the intersection of two-dimensional worldsheets} in a way that conserves their $(p,q)$ charge, as shown in fig. \ref{fig:sj1}. A segment of $(p,q)$ string that starts or ends at a $[p,q]$ brane is called a prong, see fig. \ref{fig:sj2}. Junctions may have prongs at different $[p,q]$-branes and/or include  strings extending away from them that carry some asymptotic $(p,q)$ charge away from the branes. 

\begin{figure}[tb]
	\begin{center}
		\subfigure[]{
			\includegraphics[scale=0.25]{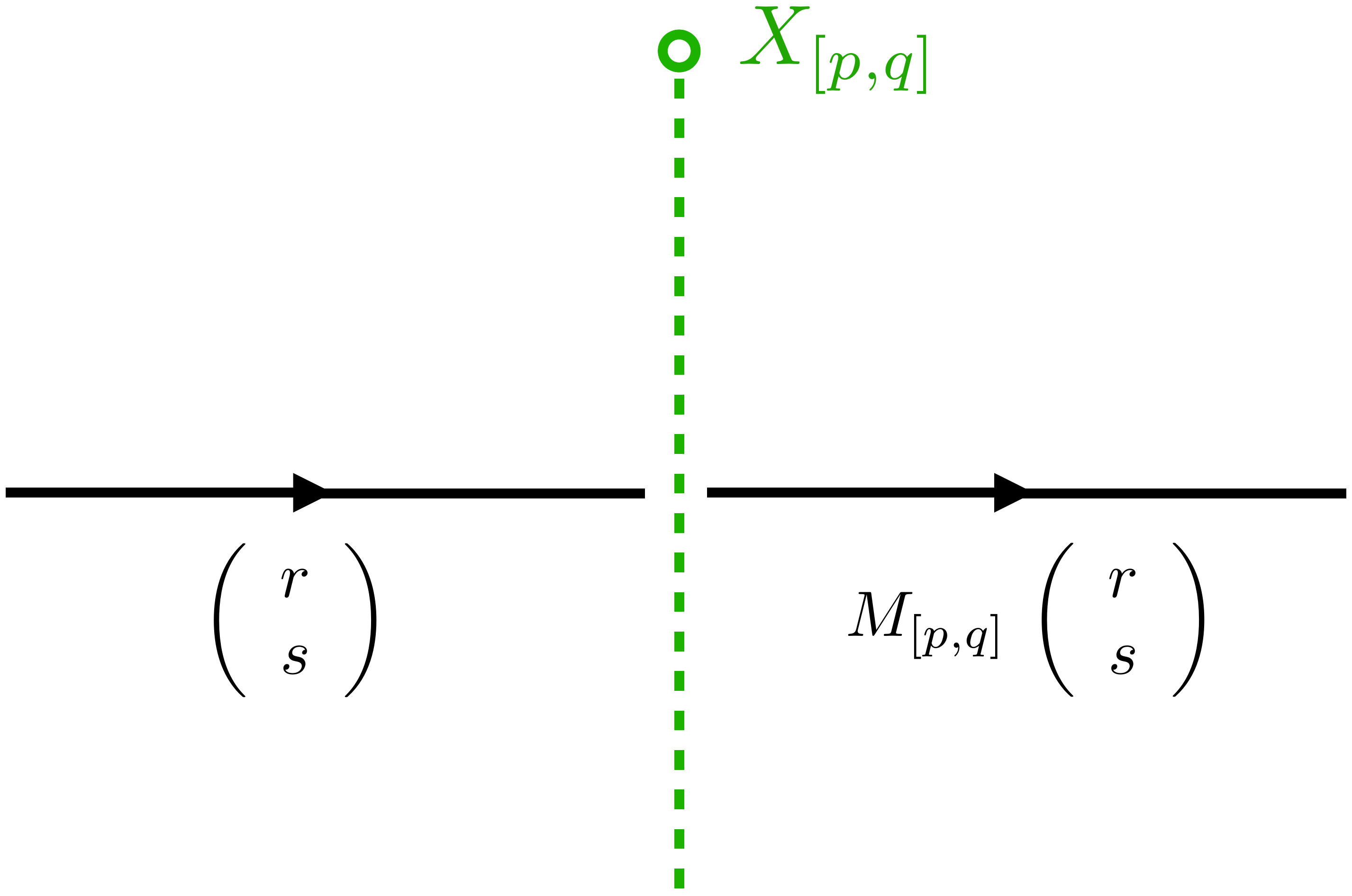}
			\label{fig:sjm1}
		}
		\subfigure[]{
			\includegraphics[scale=0.25]{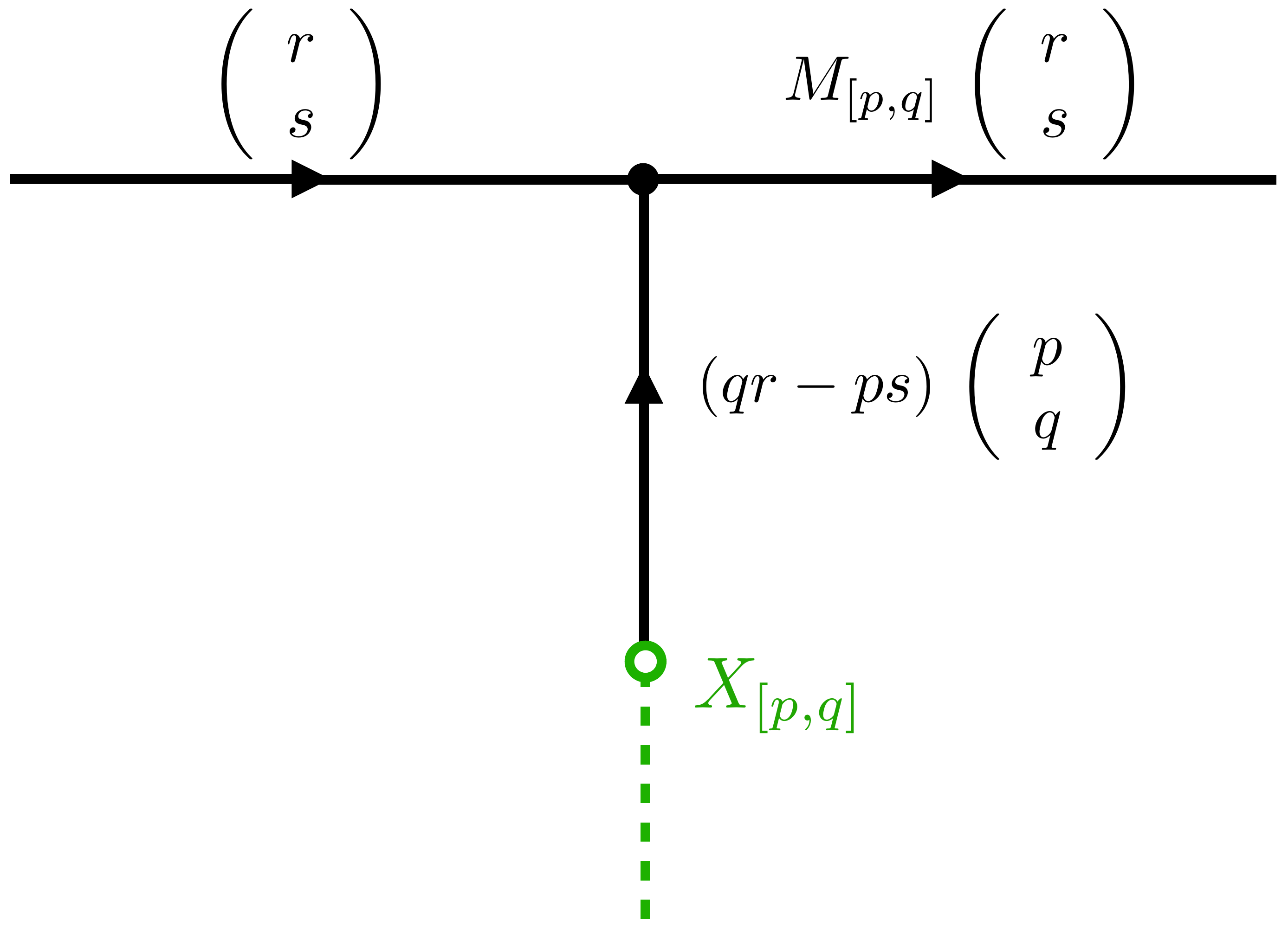}
			\label{fig:sjm2}
		}
		\caption{\footnotesize The Hanany-Witten effect in terms of string junctions  }
		\label{fig:sjm}
	\end{center}
\end{figure}

 Upon encircling a 7-brane $X_{[p,q]}$ (i.e. when crossing the branch cut), an $(r,s)$-string gets transformed by the corresponding monodromy as (see fig. \ref{fig:sjm1})
\begin{equation}
\label{eq:monodromy}
\left(\begin{array}{c}
r \\
s
\end{array}\right) \, \rightarrow \,
M_{[p, q]}\left(\begin{array}{c}
r \\
s
\end{array}\right) \, , \qquad M_{[p, q]} \, = \, \left(\begin{array}{c c}
1+pq & -p^2 \\
q^2 & 1-pq
\end{array}\right) \, = \, \mathbb{I}_{2x2} + \left(\begin{array}{c c}
pq & -p^2 \\
q^2 & -pq
\end{array}\right)\,
\end{equation}
Moreover, this segment of the junction can be moved across the 7-brane to a position in which it does not cross the corresponding branch-cut anymore. For this crossing to be consistent with the monodromy transformation above, there should be an extra prong leaving $X_{[p,q]}$ and joining the original segment of the junction (in analogy with the original Hanany-Witten effect \cite{Hanany:1996ie}), as shown in fig. \ref{fig:sjm2}. Note that the junction that is obtained after the monodromy transformation acts on the $(r,s)$-string is precisely
\begin{equation}
M_{[p, q]}\left(\begin{array}{c}
r \\
s
\end{array}\right) \, = \,  \left(\begin{array}{c}
r \\
s
\end{array}\right) + (rq-sp)\left(\begin{array}{c}
p \\
q
\end{array}\right) \, ,
\end{equation}
where the second term on the right hand side means there are $rq-sp$ strings of $(p,q)$ type. This shows that both configurations are indeed equivalent.

\begin{figure}[tb]
	\begin{center}
		
			\includegraphics[scale=0.20]{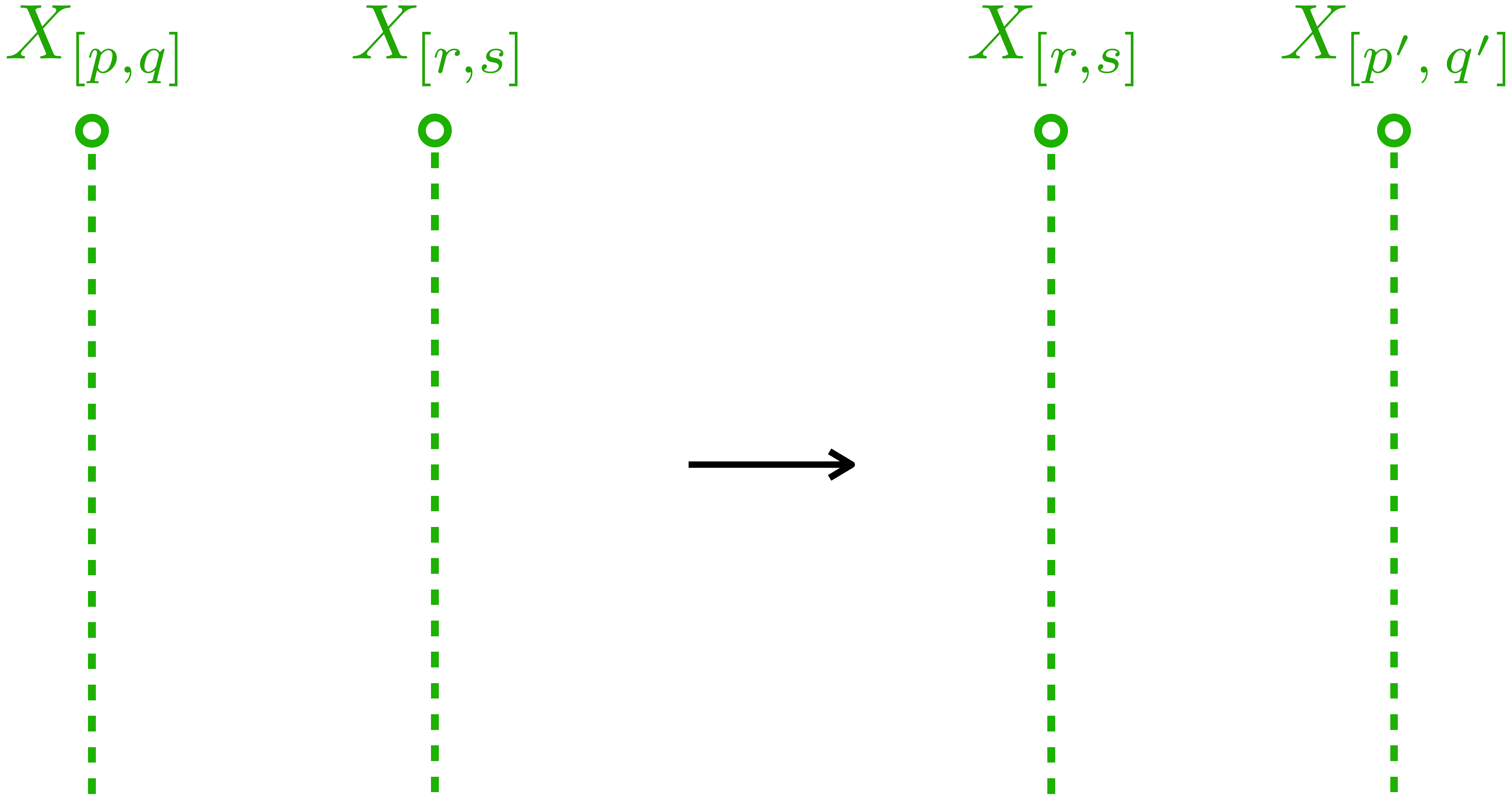} 
			
		\caption{\footnotesize Reordering of the branch-cuts. After displacing the $X_{[p,q]}$ brane from left to right across the branch cut of the $X_{[r,s]}$ brane it becomes a $X_{[p',q']}$ brane with $p'=p+(sp-rq)r$ and $q'=q+(sp-rq)s]$. }			
		    \label{fig:bcr}
	\end{center}
\end{figure}

7-brane configurations related by a global $SL(2,\mathbb{Z})$ transformation or by a relocation of branch-cuts (i.e. a reordering of the branes) are physically equivalent \cite{Gaberdiel:1998mv, DeWolfe:1998pr}. Moving an $X_{[p,q]}$ brane across the branch-cut of $X_{[r,s]}$ is equivalent to (see fig. \ref{fig:bcr})
\begin{equation}
X_{[p,q]}\,  X_{[r,s]} \, \rightarrow \, X_{[r,s]} \, X_{[p',q']} \qquad \text{with} \qquad p'=p-(rq-sp)r\,, \quad q'=q-(rq-sp)s
\label{eq:bcr}
\end{equation}

In a configuration with several 7-branes, one can always choose the branch-cuts to be vertical lines going downwards.  For such a stack of branes ordered from left to right, the full monodromy can be calculated by the product of the individual monodromies from right to left, that is
\begin{equation}
M_{X_{[p_1,q_1]}X_{[p_2,q_2]}\ldots X_{[p_n,q_n]}}\, = \, M_{[p_n,q_n]}\,  \ldots \, M_{[p_2,q_2]}\, M_{[p_1,q_1]} \, .
\end{equation}
Moreover, in eight dimensions, the tadpole cancellation condition requires a total of 24 $[p,q]$ 7-branes with total monodromy equal to the identity.

Any configuration of 7 branes can be decomposed into the combination of three particular 7 branes called $A$, $B$ and $C$
\begin{equation}
\begin{split}
&A\equiv X_{[1,0]} : \qquad \ \ M_A\, =\, M_{[1,0]}\, =\, \left( \begin{array}{cc}
1 & -1 \\
0 & 1
\end{array}\right)\, , \\
&B\equiv X_{[1,-1]} : \qquad M_B\, =\, M_{[1,-1]}\, =\, \left( \begin{array}{cc}
0 & -1 \\
1 & 2
\end{array}\right)\, , \\
&C\equiv X_{[1,1]} : \qquad \ \ M_C\, =\, M_{[1,1]}\, =\, \left( \begin{array}{cc}
2 & -1 \\
1 & 0
\end{array}\right)\, .
\end{split}
\end{equation}

Finally, let us mention that the $(p,q)$ strings have a natural interpretation in M-theory as M2-branes wrapping the $pA+qB$ cycle in the elliptic fiber, and the 1-cycle that characterizes the string  on the base. String junctions can then be understood as bound states of the corresponding M2-branes wrapping the relevant cycles in the fiber and the base. The intersection patterns of the relevant 2-cycles in homology, which are crucial for the identification of gauge groups, can be described effectively by the so called junction lattice \cite{DeWolfe:1998zf}, which we introduce now.

\subsection{The Junction Lattice}

Let us introduce a charge vector for a given string junction, whose entries are  the number of prongs starting/ending on a given 7-brane (i.e. these vectors should have 24 entries in a global eight dimensional compactification). By convention the charge is positive (negative) for a prong with outwards (inwards) orientation with respect to a brane. That is, if we introduce a basis for the charge space, by denoting $\vec{x}_{[p,q]}$ the unit vector associated to a particular $X_{[p,q]}$ brane in the configuration, the charge of $n$ $(p,q)$-strings leaving that brane would be $\vec{Q}\, =\, n\, \vec{x}_{[p,q]} $. In general, for a stack of branes ordered from left to right (with branch-cuts pointing downwards) $X_{[p_1,q_1]}X_{[p_2,q_2]}\ldots X_{[p_n,q_n]}$ we can characterize a string junction by the charge vector
\begin{equation}
\vec{Q}= \sum_ i Q^i \vec{x}_{[p_i,q_i]} \, , \qquad Q^i \in \mathbb{Z} \, .
\end{equation}
The lattice of all possible charges is the so-called junction lattice, and one can define a basis of strings as the ones that correspond to the basis vectors in the charge lattice. A junction can also carry some asymptotic $(p,q)$ charge away from the 7-branes, and it is given by the expression
\begin{equation}
\left( \begin{array}{c}
p\\
q
\end{array}\right)_{\text{asymptotic}}\, =\, \sum_ i Q^i \left( \begin{array}{c}
p_i\\
q_i
\end{array}\right)\, 
\end{equation} 

It is also possible to introduce a symmetric intersection bilinear pairing between two junctions, $( \cdot\, , \cdot)$  \cite{DeWolfe:1998zf}. This pairing captures the intersection properties of the corresponding 2-cycles in the M-theory uplift. By defining the self-intersection of any two basis strings to be 
\begin{equation}
\label{eq:xpqintequal}
(\vec{x}_{[p_i,q_i]}, \, \vec{x}_{[p_i,q_i]}) \, =\, -1\, ,
\end{equation}
exploiting bilinearity and the invariance under junction transformations (c.f. fig. \ref{fig:sjm}) one finds that the intersection between different basis elements takes the form
\begin{equation}
\label{eq:xpqintdiff}
(\vec{x}_{[p_i,q_i]}, \, \vec{x}_{[p_j,q_j]}) \, =\, \dfrac{1}{2}(p_i q_j - p_j q_i)\, ,
\end{equation}
when the position of the brane  $X_{[p_i,q_i]}$ is on the left of $X_{[p_j,q_j]}$.\footnote{Let us remark that the pairing is defined to be symmetric, but in order to compute it for a given configuration the ordering of the branes matters. That is, in this case one can define $(\vec{x}_{[p_j,q_j]}, \, \vec{x}_{[p_i,q_i]}) \, := \, (\vec{x}_{[p_i,q_i]}, \, \vec{x}_{[p_j,q_j]}) \, =\, \dfrac{1}{2}(p_i q_j - p_j q_i)\,$.}  For a set of branes given by $A_1 \ldots A_{n_a}\,  B_1 \ldots B_{n_b}\, C_1 \ldots C_{n_c}\,$, denoting the corresponding basis vectors by $\va_i$, $\vb_i$ and $\vc_i$, the relevant pairings are 
\begin{equation}
\begin{array}{l l l}
(\va_i, \va_j)= -\delta_{ij}\, , & (\vb_i, \vb_j)= -\delta_{ij}\, , & (\vc_i, \vc_j)= -\delta_{ij}\, , \\
(\va_i, \vb_j)= -1/2\, , &  (\va_i, \vc_j)= 1/2\, , & (\vb_i, \vc_j)= 1\, .
\end{array}
\label{eq:abcint}
\end{equation}
\begin{figure}[t]
	\begin{center}
		\subfigure[]{
			\includegraphics[width=190pt]{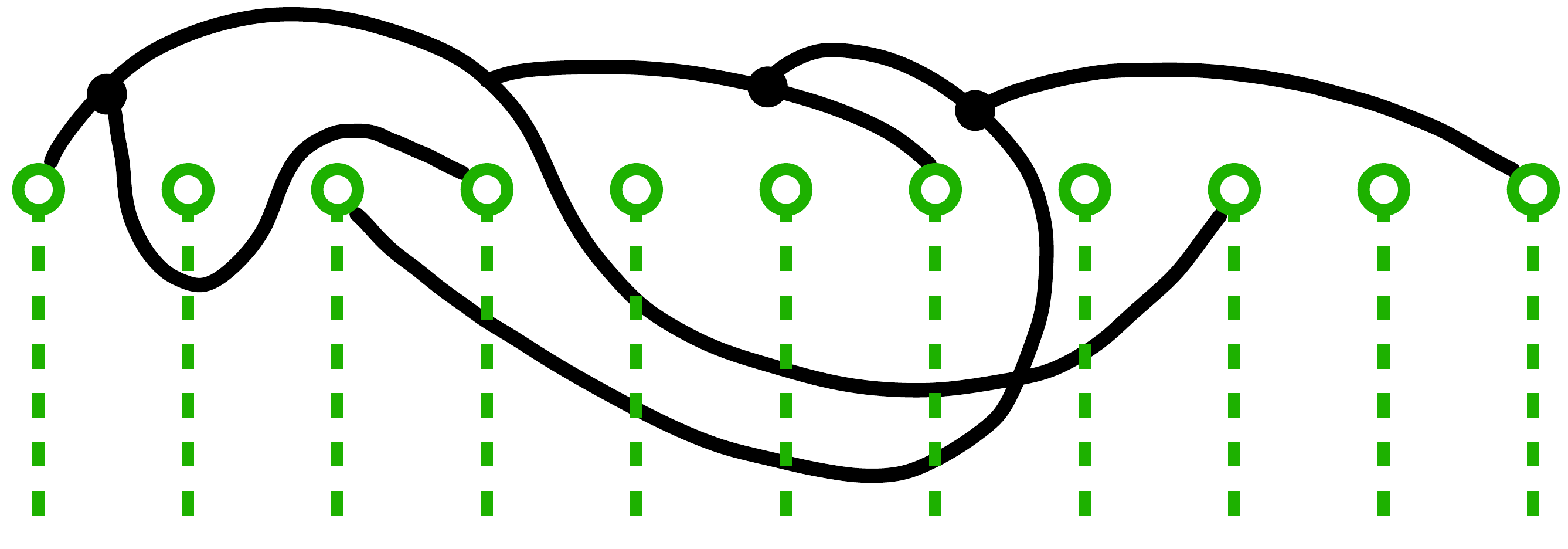}
			\label{fig:cj1}
		}
		\hspace{1.3cm}
		\subfigure[]{
			\includegraphics[width=190pt]{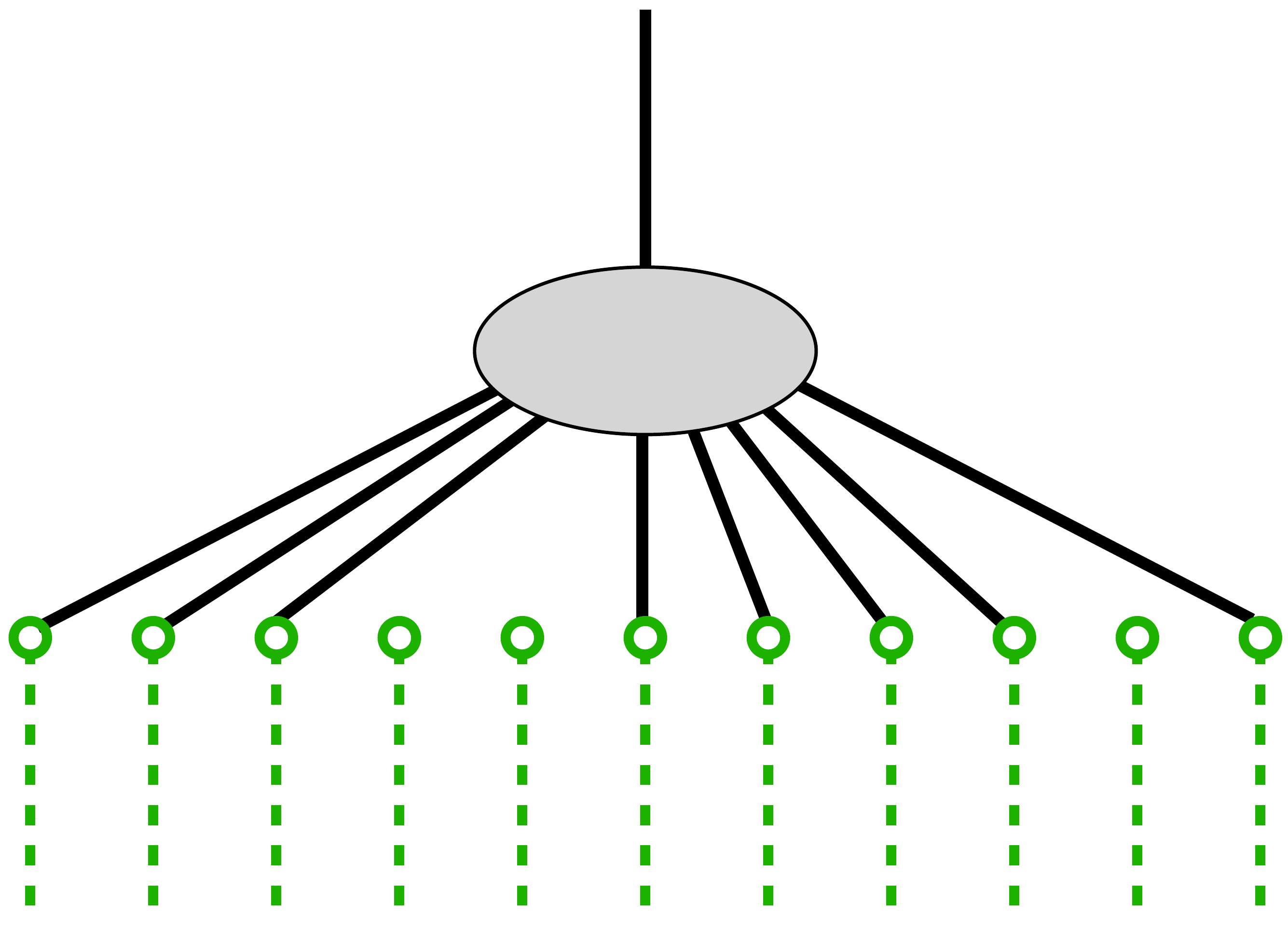}
			\label{fig:cj2}
		}
		\caption{\footnotesize \textbf{(a)} String junction in an arbityrary representation.  \textbf{(b)} The same string junction in the canonical representation, including the existence of new prongs after moving the original junction across the branes in order for it not to cross any branch-cut. }
		\label{fig:cj}
	\end{center}
\end{figure} 

Let us remark that, by construction, the charge and the pairing are invariant under the operations of moving a junction across a 7-brane  and reordering of the branes, as displayed in figs. \ref{fig:sjm} and \ref{fig:bcr}, respectively. The former operation can actually be used to define a \emph{canonical} representation for junctions, which is the one in which the junction itself does not cross any branch cut. That is, given a general string junction, we can construct its canonical representation by just moving any segment crossing a branch cut across the corresponding brane by adding the extra prongs that appear due to the Hanany-Witten effect, until no segment of the junctions crosses any branch cut. This is displayed schematichally in fig. \ref{fig:cj}. The latter operation, namely rearranging the branch-cuts and changing the order of the branes, corresponds to a base change that is taken into account by the transformation properties of the branes given in eq. \eqref{eq:bcr} \cite{Cvetic:2022uuu}.

\subsection{String Junctions and (Affine) Lie Algebras}

So far, we have introduced the string junctions, together with their allowed invariant charges, which form the junction lattice. This plays a crucial role in the discussion of  algebras, as certain string junctions turn out to represent the roots of the algebra (and their corresponding affine extensions), and their pairwise intersections (given by the pairing defined above) reproduce the negative of  the Cartan matrix of the associated algebra \cite{DeWolfe:1998zf, DeWolfe:1998bi,DeWolfe:1998yf,DeWolfe:1998eu,DeWolfe:1998pr}.

In the following, we will restrict to junctions with no asymptotic charge, as they will suffice to describe the root sector of the groups (and their corresponding affine or loop versions) that we  study here.\footnote{More general junctions including asymptotic charges are necessary to describe more general weight sectors of the corresponding groups \cite{DeWolfe:1998zf}.} 

Therefore, in order to characterize the algebra associated to a stack of 7-branes, one needs to specify the branes and find the particular string junctions that give rise to the (simple) roots. 

\subsubsection*{Example: The $A_n$ Algebra}
\begin{figure}[t]
	\begin{center}
		
			\includegraphics[scale=0.35]{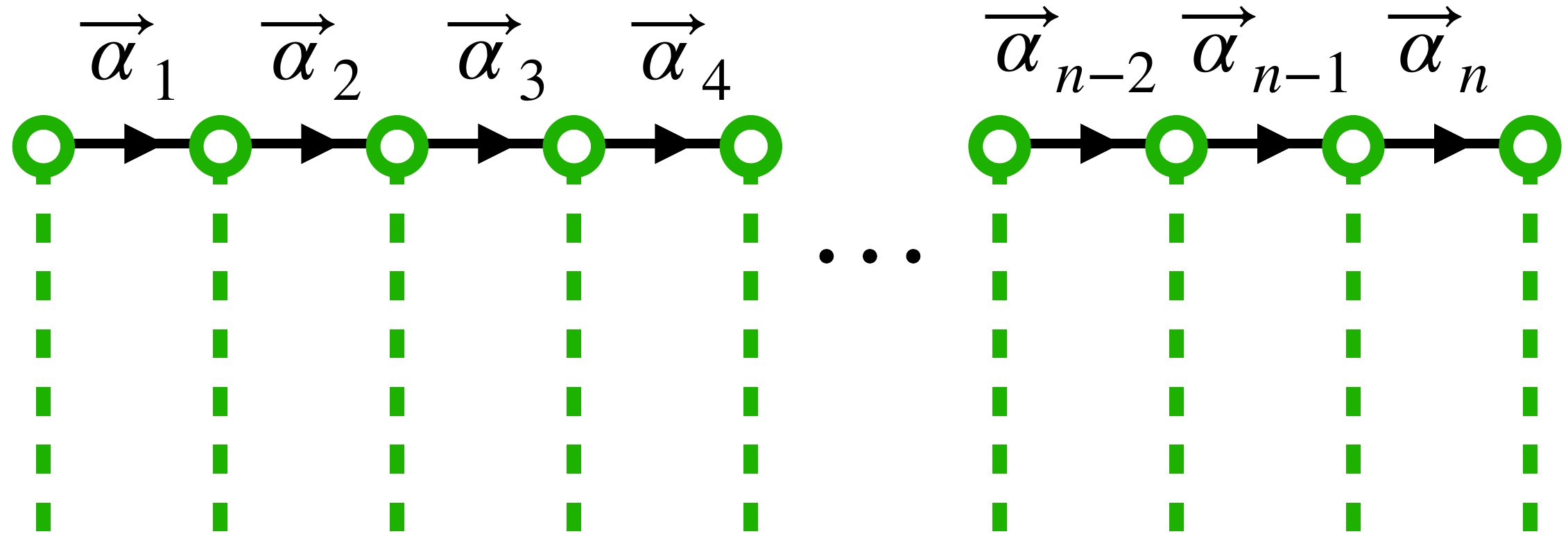} 
			
		\caption{\footnotesize String junctions that realize the simple roots $\vec{\alpha}_i$ of the $A_{n}$ algebra in a stack of $(n+1)$ A-branes. }			
		    \label{fig:su(n)}
	\end{center}
\end{figure}

Let us introduce the simple example of the $A_{n}=su(n+1)$ algebra, following \cite{DeWolfe:1998zf}. This algebra is realized by a stack of $(n+1)$ $A$ branes, with monodromy 
\begin{equation}
M_{A^{n+1}}\, = \, \left( \begin{array}{cc}
1 & -n-1 \\
0 & 1
\end{array}\right)\, ,
\label{eq:Monodromysu(n)}
\end{equation}
The junctions that represent the $n$ simple roots have charges
\begin{equation}
\vec{\alpha}_i= \vec{a}_i-\vec{a}_{i+1} \, 
\end{equation}
that is, the $i$-th entry is 1, the $(i+1)$-th entry is -1 and the rest are zero. From this charges we can conclude that the simple roots are strings starting at one brane and ending at the next, as displayed in fig. \ref{fig:su(n)}.

Using eq. \eqref{eq:abcint} one can calculate the intersection form, which yields
\begin{equation}
( \vec{\alpha}_i, \vec{\alpha}_j) \,  = \, \left( \begin{array}{ccccccccccc}
-2 & 1 &0 &0 &&\ldots && 0 & 0 & 0 &0 \\
1 & -2 &1 &0 &&\ldots && 0 & 0 & 0 &0\\
0 & 1 &-2 &1 &&\ldots && 0 & 0 & 0 &0\\
\vdots & \vdots &\vdots &\vdots && \ddots && \vdots & \vdots & \vdots & \vdots  \\
 0 & 0 & 0 &0 &&\ldots && 1 & -2 & 1 &0 \\
 0 & 0 & 0 &0 &&\ldots && 0 & 1 & -2 & 1 \\
 0 & 0 & 0 &0 &&\ldots && 0 & 0 & 1 &-2 
\end{array}\right)\, ,
\end{equation}
and indeed reproduces the negative of the Cartan Matrix for $A_n$.

\subsubsection{Loop Junctions and Affine/Loop Algebras}
\label{ss:JunctionsandAffine}

\begin{figure}[t]
	\begin{center}
		\subfigure[]{
			\includegraphics[scale=0.25]{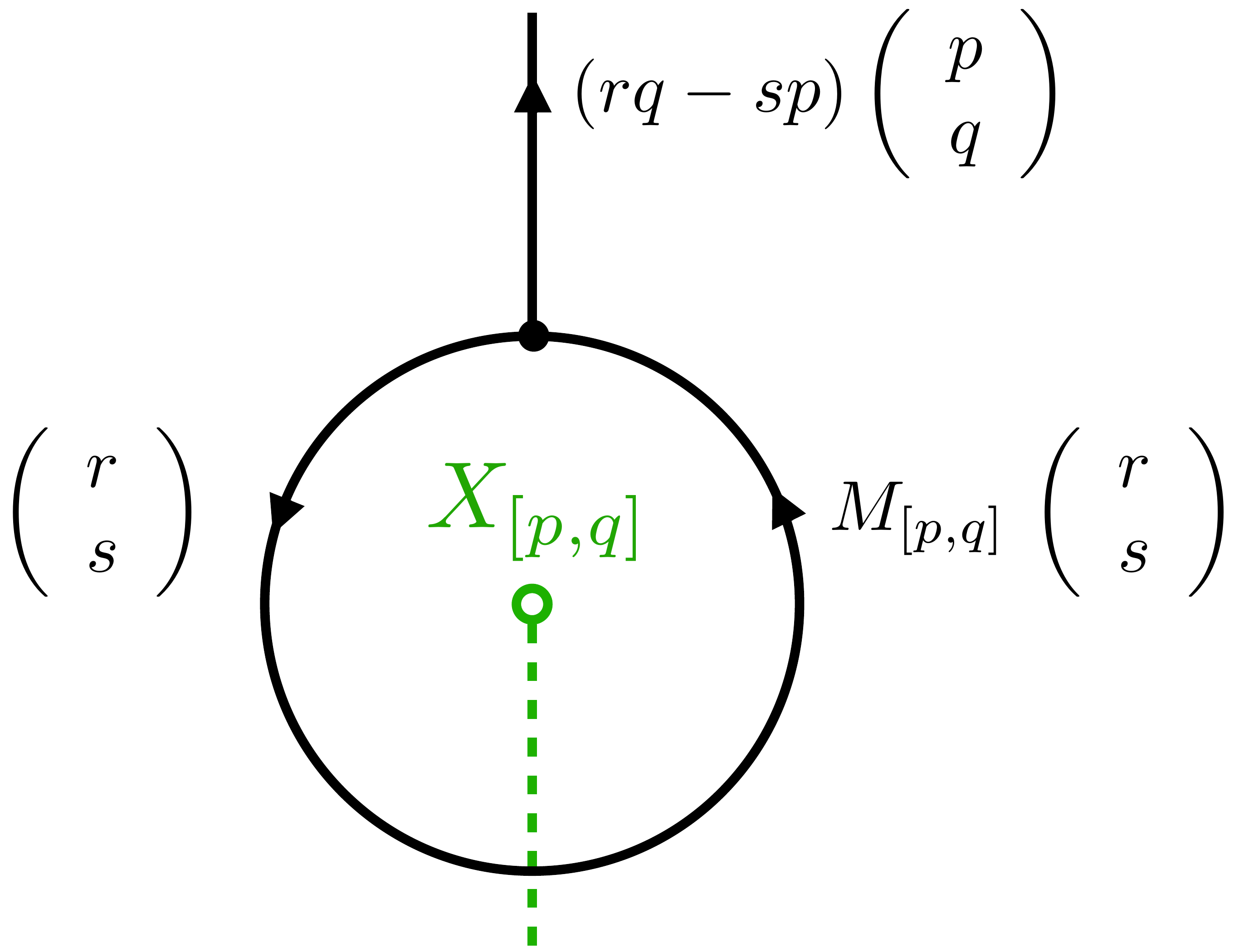}
			\label{fig:loop1}
		}
		\hspace{3cm}
		\subfigure[]{
			\includegraphics[scale=0.25]{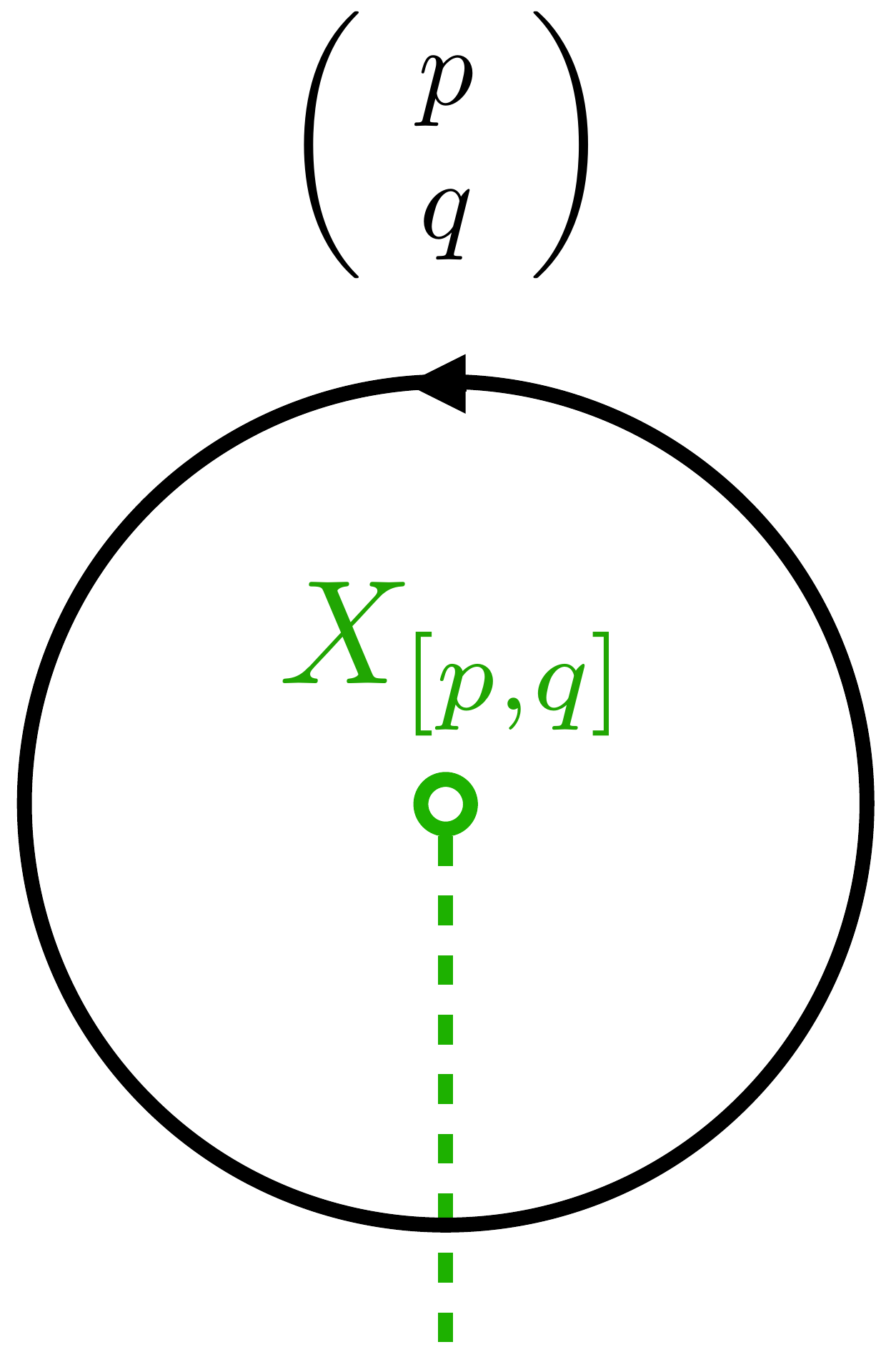}
			\label{fig:loop2}
		}
		\caption{\footnotesize Two loop junctions (in their non-canonical form) }
		\label{fig:loop}
	\end{center}
\end{figure} 

Consider now a junction that takes the form of a loop around a (stack) of 7-branes, and possibly an extra piece attached to it that may carry some asymptotic charge, as displayed in fig. \ref{fig:loop1}. The monodromy transformation associated to the (stack of) 7-brane(s) acts on the junction as it crosses the branch cut(s) such that the asymptotic charge is compatible with it. Furthermore, one can bring the junction into its canonical representation by moving it through the brane and taking into account the corresponding Hanany-Witten transition. In the case displayed in fig.  \ref{fig:loop1} this would correspond to a charge vector given by $\vec{Q}= (rq-sp) \vec{x}_{[p,q]}$, so the charge in that case is equivalent to the asymptotic charge under the brane surrounded by the loop junction. The situation becomes much richer when several different branes are encircled by the loop, as we will see in the following. Additionally, one can consider the simple case of a $(p,q)$ string surrounding a $X_{[p,q]}$ brane, as shown in fig. \ref{fig:loop2}, where the junction carries no asymptotic charge since the monodromy acts trivially on it. More generally, loop junctions surrounding a stack of branes and carrying no asymptotic charge can be built whenever the full monodromy action of the stack is such that it leaves the junction invariant. This situation will be particularly relevant in the identification of imaginary roots for the affine algebras below.

Let us now remark a couple of particularly interesting properties of loop junctions. First, it can be checked that loop junctions with no asymptotic charge have vanishing self-intersection, making them the perfect candidates for a junction realization of the imaginary root associated to the affine or loop version of an algebra. Second, recall that the roots of an algebra on a stack of branes are also realized  by string junctions with no asymptotic charge, so they always start and end on different branes. It is straightforward to see that a loop junction surrounding the whole configuration does not intersect any of roots, which is the other key condition that an imaginary root must fulfill. Note that these two properties can be properly checked by computing the intersection from the charges in the canonical representation.

We must stress here the importance of the loop junction not having an asymptotic charge, as it is the one that has vanishing self-intersection, and it is the one that indirectly indicates what must be added to a finite algebra in order to affinize it. As explained above, in order  for a loop junction with no asymptotic charge to be supported around a brane configuration, it must be left invariant under the total monodromy. That is, for a given stack of branes, the only allowed loop junctions are the ones that correspond to the eigenspaces that are left invariant by such monodromy. In many general stacks of branes (as e.g. the ones that realize the finite exceptional groups) this space is just empty.\footnote{To be precise, it cannot be realized by $(p,q)$-strings with integer charges.} Thus, the standard procedure in this case is to add one extra brane to the stack such that the new total monodromy allows for a loop junction in order to realize the affine version of the previous algebra. 

In general, one must also check that the imaginary root is non-trivial, meaning it has non-zero charge in the junction lattice. This might look like a meaningless check, but it turns out to be key in some cases. Consider for example the $su(n)$ algebra discussed above. Its total monodromy \eqref{eq:Monodromysu(n)} leaves invariant the space spanned by $(1,0)$ strings. However, a $(1,0)$ loop junction  surrounding the whole configuration does not make the algebra affine because it is actually contractible (one can just move the loop across all branes without generating any extra junction by the Hanany-Witten effect). This can be detected by the fact that its charge vector is $\vec{Q}=0$, as can be easily seen from its canonical representation. As we will see in section \ref{Het-Ftheory}, a different way to affinize such an algebra is by having some extra branes far away from their stack such that the $(1,0)$ loop junction has some non-trivial charge under them in the junction lattice. This is in contrast with the usual procedure (typically used for exceptional algebras) where one adds some extra brane to change the total monodromy in the right way.

\subsubsection*{Example: The $\widehat{A}_1$ Algebra}

\begin{figure}[t]
	\begin{center}
		\subfigure[]{
			\includegraphics[height=135pt]{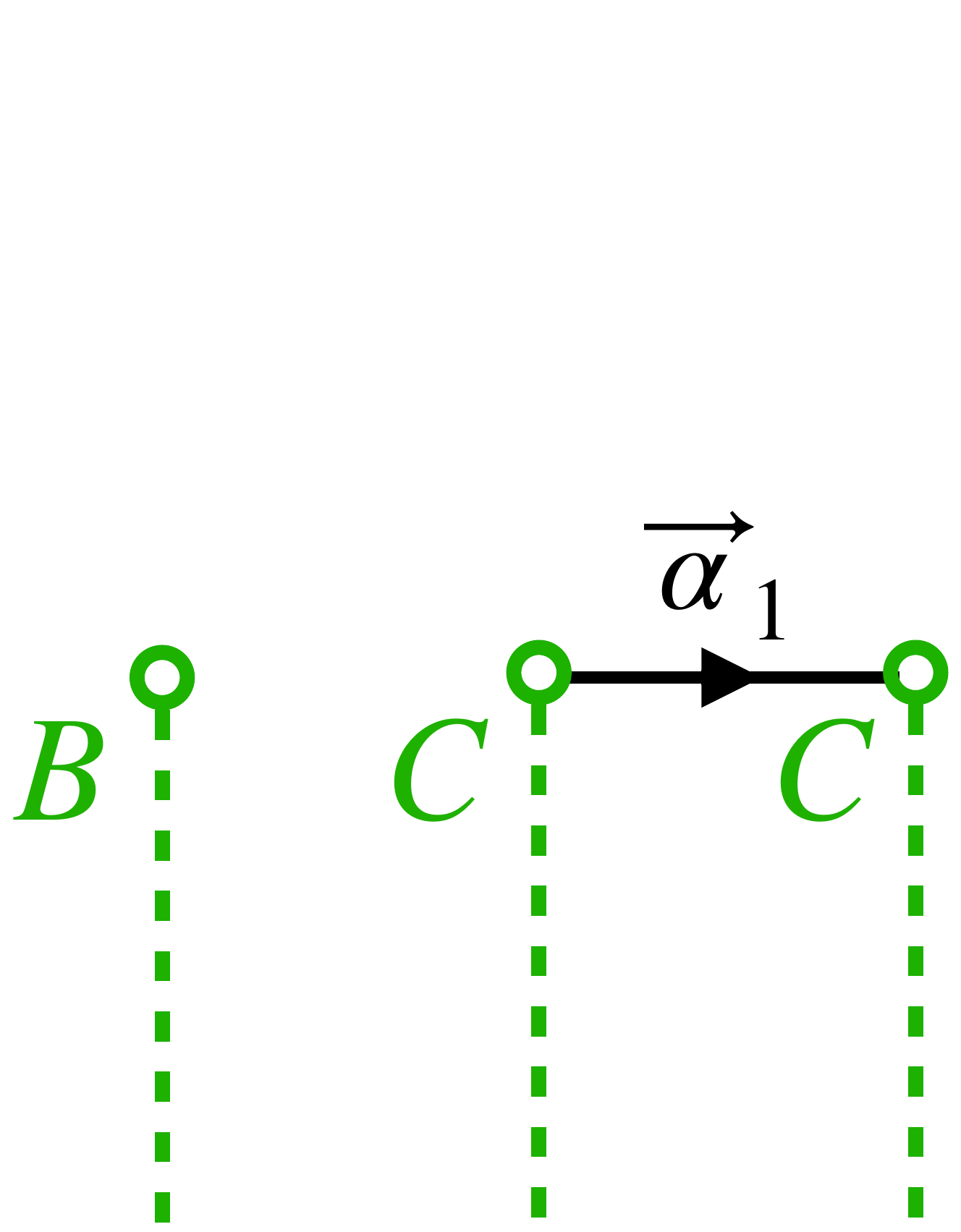}
			\label{fig:su(2)}
		}
		\hspace{2.5cm}
		\subfigure[]{
			\includegraphics[height=135pt]{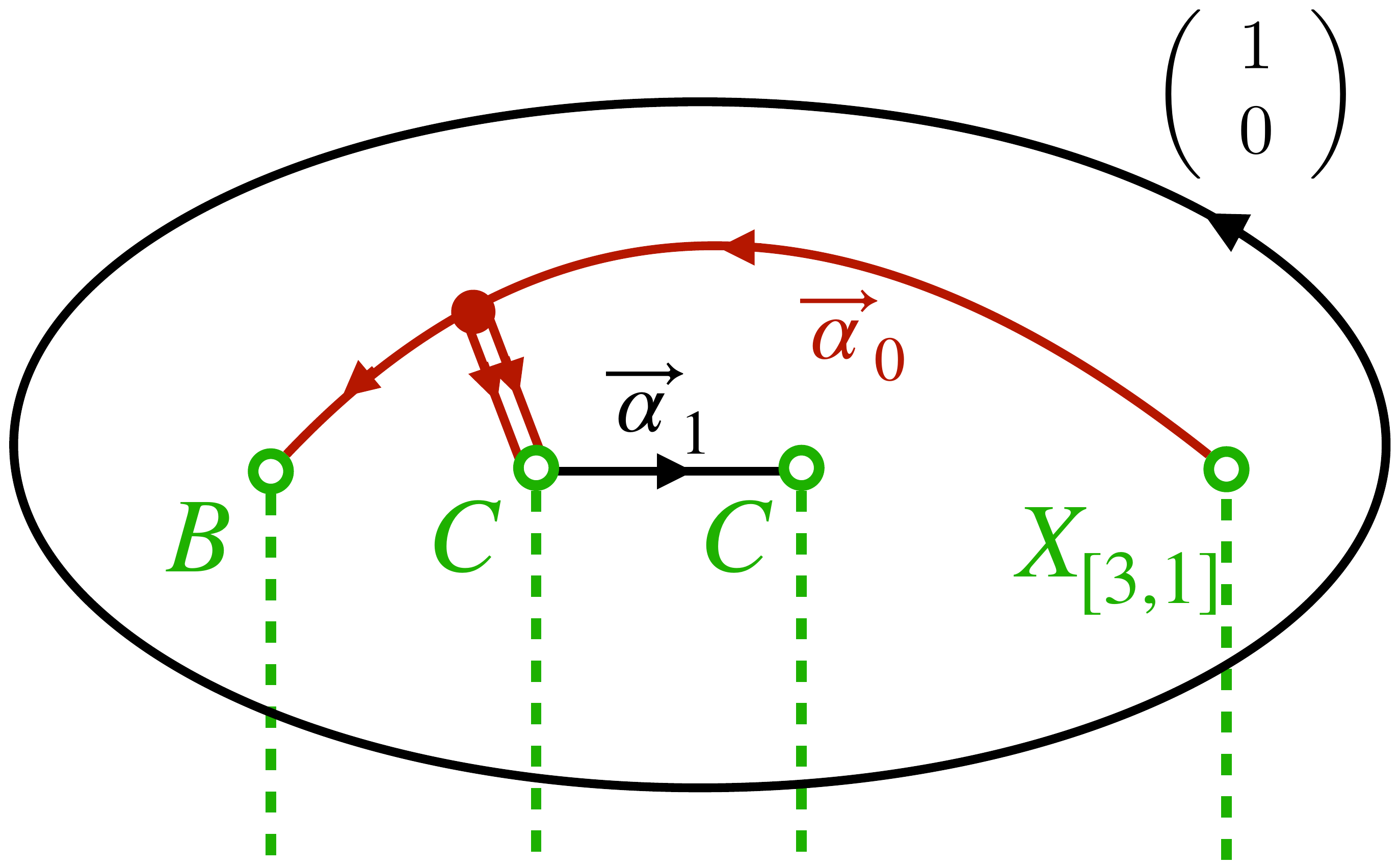}
			\label{fig:su(2)_affine}
		}
		\caption{\footnotesize \textbf{(a)} Realization of the $A_1$ algebra on a BCC stack of branes, where $\vec{\alpha}_1$ gives the simple root.  \textbf{(b)} Realization of the $\widehat{A}_1$ algebra by adding a $X_{[3,1]}$ brane to the BCC stack. The imaginary root, $\vec{\delta}_{(1,0)}$ is given by a $(1,0)$ string loop around the whole configuration, and the affine root (in red) is built as $\vec{\alpha}_0 = \vec{\delta}_{(1,0)}-\vec{\alpha}_1 \,$.}
		\label{fig:su(2)_affinization}
	\end{center}
\end{figure}

Let us consider the case of the affinization of the $A_1$ algebra. Apart from $A^2$, the $A_1=E_1$ algebra can be realized by a configuration of $BCC$ branes. The single simple root, $\vec{\alpha}_1$, which is also the highest root, is given by a junction with charge $\vec{\alpha}_1=(0,1,-1)$ (under the original $BCC$ stack),\footnote{In this simple case this is easy to see, since the requirement of absence of asymptotic charge straightforwardly restricts the possible junctions to those with as many prongs starting in one of the $C$ branes as ending in the remaining $C$ brane. For more general brane configurations these conditions give rise to a more cumbersome set of diophantine equations, studied more systematically in \cite{DeWolfe:1998yf}. } as shown in fig. \ref{fig:su(2)}.
The associated monodromy matrix is given by 
\begin{equation}
M_{BCC}\, = \, \left( \begin{array}{cc}
-2 & -7 \\
-1 & -4
\end{array}\right)\, ,
\label{eq:MonodromyBCC}
\end{equation}
and it cannot support any physical loop junction around it. By adding a $X_{[3,1]}$ brane to the stack, the total monodromy yields
\begin{equation}
M_{BCCX_{[3,1]}}\, = \, \left( \begin{array}{cc}
1 & 8 \\
0 & 1
\end{array}\right)\, ,
\label{eq:MonodromyBCCX31}
\end{equation}
which can support a $(1,0)$ string loop surrounding the four branes. One can easily calculate its charge in the junction lattice by going to the canonical representation, and also build the affine root $\vec{\alpha}_0 = \vec{\delta}_{(1,0)}-\vec{\alpha}_1$, obtaining
\begin{equation}
\vec{\alpha}_1 \, =\, (0,1,-1,0)\, , \qquad \vec{\delta}_{(1,0)}\,=\, (-1,-1,-1,1) \, , \qquad \vec{\alpha}_0\, = \, (-1,-2,0,1) \, .
\end{equation}
By using the pairings given in eqs. \eqref{eq:xpqintequal}-\eqref{eq:abcint}, the intersection matrix for the $BCCX_{[3,1]}$ configuration is
\begin{equation}
(\cdot, \cdot ) \, = \, \left(\begin{array}{cccc}
-1 & 1 & 1 & 2 \\
1 & -1 & 0 & -1 \\
1 & 0 & -1 & -1 \\
2 & -1 & -1 & -1 \\
\end{array}
\right) \, .
\end{equation}
 With this pairing, it can be checked that the imaginary root is indeed orthogonal to the simple roots, namely 
\begin{equation}
(\vec{\delta}_{(1,0)} ,  \vec{\alpha}_1 ) \, = \, (\vec{\delta}_{(1,0)} ,  \vec{\alpha}_0 ) \, = \, 0 \, ,
\end{equation}
and the Gramm matrix between the two simple roots $\vec{\alpha}_0$ and $\vec{\alpha}_1$ takes the form
\begin{equation}
(\vec{\alpha}_i, \vec{\alpha}_j )\, = \,  \left( \begin{array}{cc}
-2 & 2 \\
2 & -2 \\
\end{array}
\right) \,
\end{equation}
which is indeed the negative of the Cartan Matrix for $\widehat{A}_1$.

\section{Heterotic-F theory duality at infinite distance}
\label{Het-Ftheory}
The string junction picture can only give local information about the affinisation process. The types of affine algebras that can actually be realised in F-theory on an elliptically fibered K3 were analysed in detail in \cite{Lee:2021qkx,Lee:2021usk} relying on the global information contained in the degenerations of elliptically fibered K3 lying at infinite distance in the complex structure moduli space. In turn, these can be described and classified in terms of Kulikov models and can be broadly divided into four classes, which we introduce briefly here in relation to the heterotic theory, referring the reader to the Appendix \ref{appendix2} for the detailed geometrical description 
\begin{itemize}
    \item Type II.a: the degenerate K3 surface has non-minimal singularities according to the Kodaira classification. This limit is dual to the full decompactification limit to the $E_8 \times E_8$ Heterotic theory, and realises the algebra $(\widehat{E}_9 \oplus \widehat{E}_9)/ \sim$ as derived in \ref{sec:e9e9}. 
    \item Type II.b: this is an Emergent String limit, where the associated K3 surface has singular fibers already in codimension zero on the base. Since it has no counterpart in the weakly coupled heterotic theory, we will not discuss it further. 
    \item Type III.a: the K3 has non-minimal fibers and it corresponds to a partial decompactification of the heterotic dual from eight to nine dimensions. One example is given in \ref{fth89}.
    \item Type III.b: the degenerate K3 surface has both non minimal singularites and codimension zero singular fibers, and it corresponds to a weak coupling limit accompanied by a full decompactification in the Type IIB string frame. It realises the decompactification to the $SO(32)$ dual heterotic theory in ten dimensions as shown in \ref{sec:so32}. 
\end{itemize}
The states that are dual to the KK towers in the heterotic decompactification limits are partially captured in the M-theory picture as M2 branes wrapping vanishing calibrated 2-tori in the degenerate K3 geometry, as explained in the following paragraphs.

It is important to remark that while in the F-theory setting one can reach all the infinite distance points in moduli space, one can do it in a controlled way only along certain paths, namely the ones compatible with the algebraic nature of the associated Weierstrass models. We elaborate on this in Appendix \ref{appendix:b2}. 

In this Section, we consider the brane configurations studied in \cite{Lee:2021usk} and examine the affine enhancements that arise in the language of string junctions. The geometric idea is that at the infinite distance limits,  the junctions, and therefore all the states realizing the affine algebra, become asymptotically massless. This geometric picture is studied in detail in the references above, so we will not elaborate further here, but instead we will consider the BPS algebra that can arise at each configuration by using their geometric results. 

\subsection{$(\widehat{E}_9 \oplus \widehat{E}_9) / \sim $} \label{sec:e9e9}

The Type II.a Kulikov model is related to the presence of non minimal Kodaira singularities in the K3 degeneration. Following the general procedure, these can be resolved by a blowup of the base that zooms in their brane content. This procedure leads to the so-called stable degeneration limit, at which the geometric picture consists of two del Pezzo surfaces intersecting at a 2-torus, which is known to be dual to the heterotic theory on the torus at the intersection in the limit in which the torus becomes large. We know from the heterotic computation in Section \ref{sec:doubleloop} that the algebra in the full decompactification limit to the ten dimensional heterotic should be $(\widehat{E}_9 \oplus \widehat{E}_9) / \sim $, where the $\sim$ indicates that the imaginary roots enhancing both factors are identified.

In the F-theory setup, the basis degenerates into two components, each of which has twelve branes. There is an $E_8$ factor on each component, which is realized by the brane configuration $A^7 BCC$ made  of ten out of the twelve branes in each basis component. This algebra gets enhanced to $E_9$ by adding a $X_{[3,1]}$ brane, in analogy to the the example in section \ref{ss:JunctionsandAffine}. The total monodromy of this configuration is then
\begin{equation}
M_{A^7BCCX_{[3,1]}}\, = \, \left( \begin{array}{cc}
1 & 1 \\
0 & 1
\end{array}\right)\, ,
\label{eq:MonodromyE9_1}
\end{equation}
which clearly allows for a $\delta_{[1,0]}$ loop junction surrounding the whole configuration yielding the imaginary root for the affine enhancement. Alternatively, by considering the $E_8$ stack together with the extra $A$ brane which is present in the same component of the basis, one obtains the  brane configuration $A^8 BCC$. This brane configuration supports another imaginary root, given by the loop junction $\delta_{[3,1]}$, as can be seen from the invariant eigenspace of the corresponding monodromy matrix
\begin{equation}
M_{A^8BCC}\, = \, \left( \begin{array}{cc}
-2 & 9 \\
-1 & 4
\end{array}\right)\, .
\label{eq:MonodromyE9_2}
\end{equation}
\begin{figure}[t]
	\begin{center}

		\subfigure[]{
			\includegraphics[height=135pt]{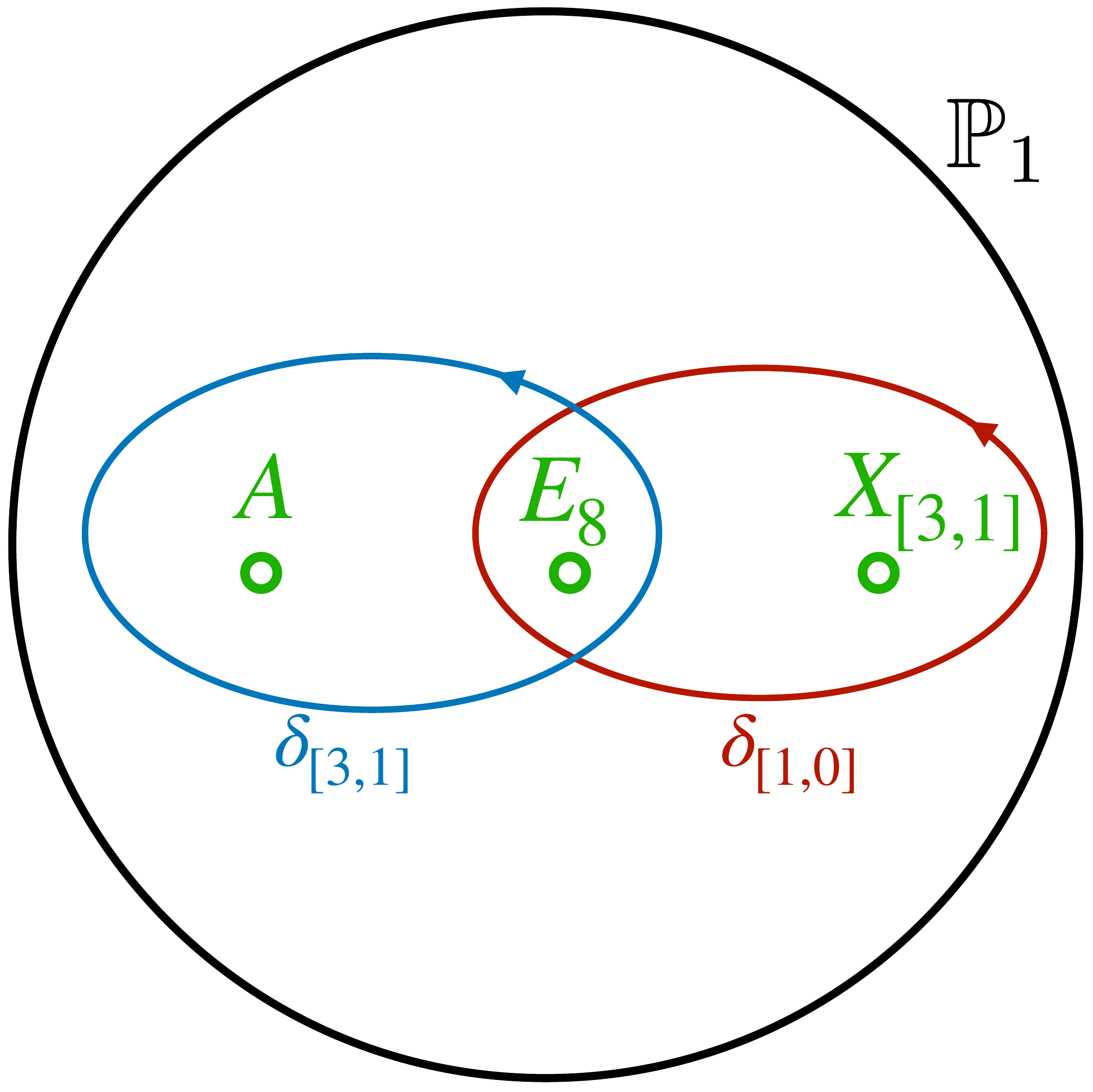}
			\label{fig:hatE9}
		}	\hspace{0.5cm}	
		\subfigure[]{
			\includegraphics[height=135pt]{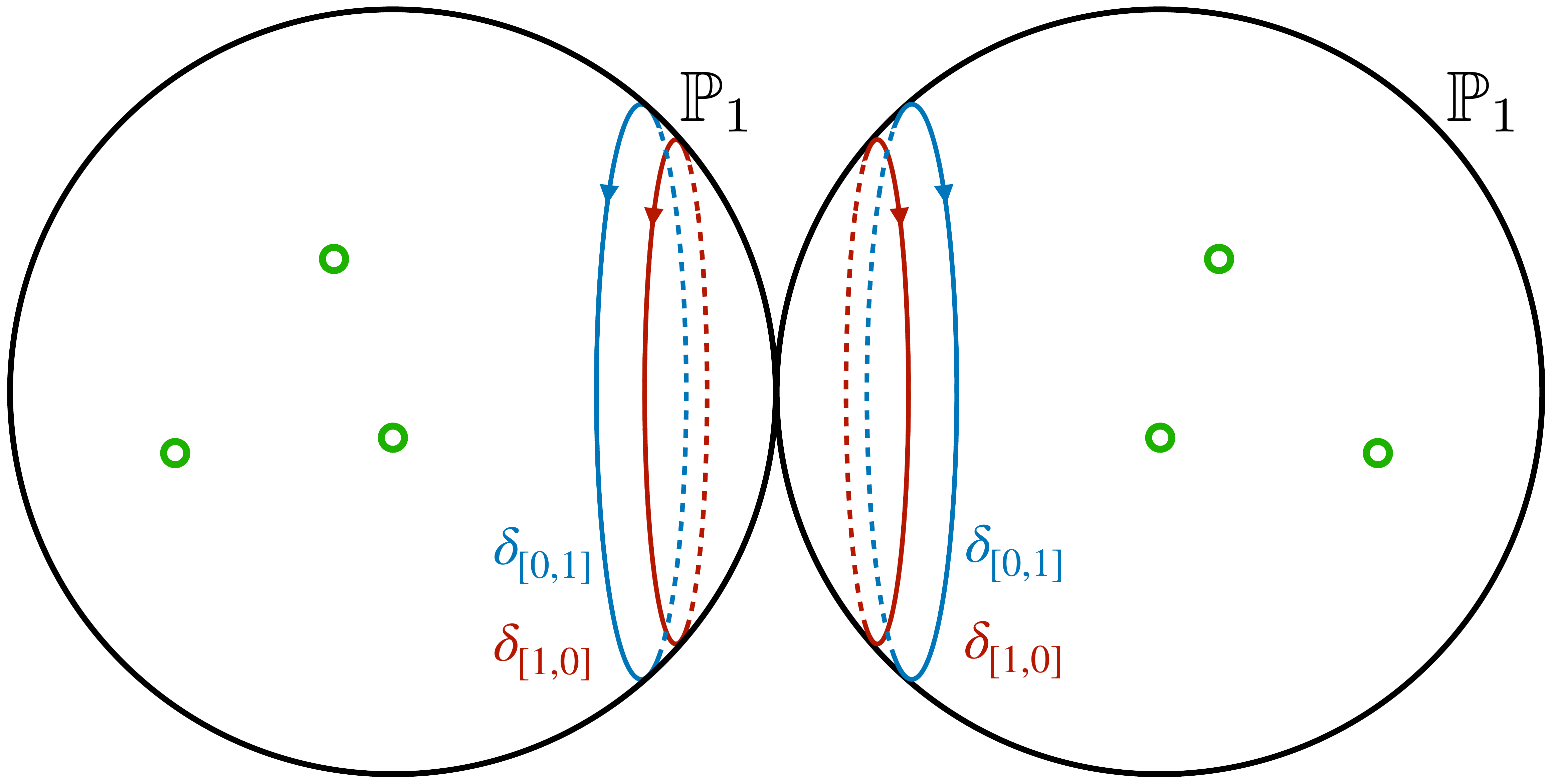}
			\label{fig:hatE9xE9}
		}

		\caption{\footnotesize \textbf{(a)} Brane configuration realizing the loop algebra $\widehat{E}_9$. The two imaginary roots, $\delta_{[1,0]}$ and $\delta_{[3,1]}$, are realized by loop junctions surrounding the $E_8$ stack and one additional brane (note they can also be represented as two independent junctions surrounding the whole configuration). The branes are represented in one of the basis components given by a $\mathbb{P}_1$. \textbf{(b)} Configuration realizing the algebra $(\widehat{E}_9\oplus \widehat{E}_9)/\sim $, with the loop junctions which are identified in the junction lattice represented in both base components. This corresponds to the stable degeneration limit which is dual to the full decompactification of the $E_8 \times E_8$ heterotic string.}			
		    \label{fig:E9s}
	\end{center}
\end{figure} 

By considering the whole configuration, one obtains the loop algebra $\widehat{E}_9$, as can be seen from the fact that the full monodromy matrix is the identity 
\begin{equation}
M_{\widehat{E_9}}\, = \, \left( \begin{array}{cc}
1 & 0 \\
0 & 1
\end{array}\right)\, ,
\label{eq:MonodromyhatE9}
\end{equation}
so that both imaginary roots are present,  as displayed in fig. \ref{fig:hatE9}. Note that both string junctions can also be brought to a form in which they surround the whole configuration, and one can equally choose the basis $\delta_{[1,0]}$ and $\delta_{[0,1]}$ for the string junctions.

The full configuration consists of two components of the base, each of them being a $\mathbb{P}_1$ with a similar brane configuration, as shown in fig. \ref{fig:hatE9xE9}. The two imaginary roots are identified in the junction lattice as they can be seen to be homologous in the M-theory lift. Both loop junctions come  from wrapping an M2-brane in  the $(1,0)$ or $(0,1)$ cycle in the elliptic fiber and the one cycle wrapping the whole brane configuration in each component of the base. As nothing special happens to any of the fiber cycles at the intersection points, the two cycles in both components of the base are homologous and thus they represent the same object in the junction lattice. This is the reason behind the identification of the imaginary roots producing the affinizations in both components. 

In this setting the towers of states are given by the $(n,0)$ and $(0,m)$ strings, which become massless since  they are allowed to shrink to the base intersection point in the blown up version, or to the non minimal singularity from the perspective of the degenerate K3. 

One interesting remark is the following. On the heterotic side, we can decompactify from 8 to 10 dimensions $R_8, \, R_9 \to \infty$ either in one step (corresponding to the case $\frac{R_8}{R_9} = \text{const}$) or in two steps ($\frac{R_8}{R_9} \to 0, \, \infty$) as already noticed in \ref{sec:doubleloop}. In the F-theory picture, this Kulikov model seems to capture only the first possibility.

\subsection{$(\widehat{E}_{8-n} \oplus  \widehat{E}_8 \oplus  \widehat{A}_{n+1} ) / \sim $} \label{fth89}

\begin{figure}[t]
	\begin{center}
			\includegraphics[height=100pt]{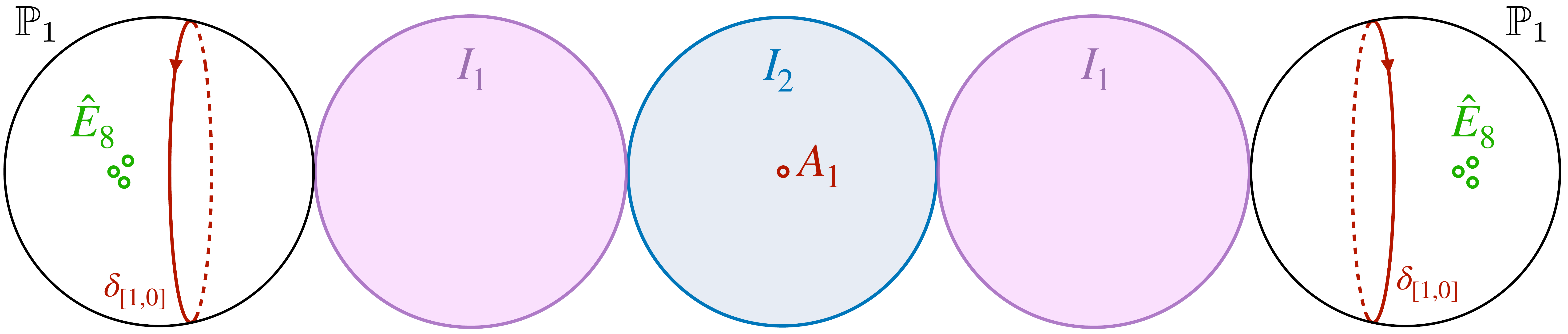}
			\label{fig:hatE8xE8xsu(2)}
		\caption{\footnotesize Schematic configuration of the branes and codimension zero singularities realizing the loop algebra $(\widehat{E}_{8}\oplus \widehat{E}_8\oplus \widehat{A}_{1}) /\sim $,  with the loop junction $\vec{\delta}_{[1,0]}$ giving rise to the imaginary root. For the details on the intermediate components and their corresponding $I_k$ singularities see \cite{Lee:2021usk}.}			
		    \label{fig:E8xE8xsu(2)}
	\end{center}
\end{figure}

Consider now the situation where in the F-theory picture both end components of the base are $dP_9$ surfaces (see section $3.4$ in \cite{Lee:2021usk}), which corresponds to a Type III.a degeneration. In particular, this includes the case where the heterotic dual realizes $E_8\oplus E_8 \oplus A_1$ as explicitly shown in Section \ref{89}, whose F-theory dual is schematically displayed in fig. \ref{fig:E8xE8xsu(2)}. 

The affinization process is similar to the previous one, but with some crucial differences. First of all, the monodromy around the exceptional factors   $\widehat{E}_{8-n}$  in the end components (including the extra $X_{[3,1]}$ brane responsible for the affinization) takes the form 
\begin{equation}
M_{\widehat{E}_{8-n}}\, = \, \left( \begin{array}{cc}
1 & n+1 \\
0 & 1
\end{array}\right)\, =\, M_{I_{n+1}}^{-1} \, ,
\label{eq:MonodromyEnhat}
\end{equation}
so that it only leaves invariant the space of loop junctions generated by $\vec{\delta}_{[1,0]}$, that is, there will be only one tower, as oppossed to the double loop enhancement associated to the two towers before. This can be understood in terms of a dual KK tower from decompactification from eight to nine dimensions, instead of full decompactification to ten dimensions. 

Moreover, the internal components of the base, displayed in different colours for the $(\widehat{E}_{8} \oplus  \widehat{E}_8 \oplus  \widehat{A}_{1})  / \sim$ example in fig. \ref{fig:E8xE8xsu(2)}, have codimension zero singularities of the form $I_k$. Their existence is consistent with the full monodromy composition, which gets more involved in the presence of non-trivial codimension zero singularities. More precisely, from the general blow up procedure \cite{Lee:2021qkx}, given two components intersecting at a point with generic fibers $I_n$ and $I_m$, respectively, the contribution to the monodromy at the component where the generic $I_n$ is located is equivalent to the one from a codimension one singularity (located at the intersection) with monodromy
\begin{equation}\label{eq:codim0monodromycomp}
  M_{I_{m-n}}^{\vphantom{-1}} \, = \, M_{I_n}^{-1}\, M^{\vphantom{-1}}_{I_m}\, .  
\end{equation}
Note that the contribution from the same intersection point, as seen from the component where the codimension zero $I_m$ singularity is located, is given by the inverse monodromy, which can indeed be consistently described by $m \leftrightarrow n$. For the $(\widehat{E}_{8} \oplus  \widehat{E}_8 \oplus  \widehat{A}_{1})  / \sim$ example, this is consistent with the monodromy structure seen by each of the first three base components (starting from the left) in fig. \ref{fig:E8xE8xsu(2)}, namely
\begin{equation}
    M_{\widehat{E}_{8}}\,  M^{\vphantom{-1}}_{I_{1}} \, = \, \mathbb{I}_{2x2}\, , \qquad     M^{\vphantom{-1}}_{I_{-1}}\,  M^{\vphantom{-1}}_{I_{1}} \, = \, \mathbb{I}_{2x2} \, ,\qquad      M_{I_{-1}}\, M^{\vphantom{-1}}_{A_1}\,  M_{I_{-1}} \, = \, \mathbb{I}_{2x2} \, .
\end{equation}
Note that the final two components get the same contributions as the first two.

Finally, the fact that the intermediate components are mutually local is key for the affinization of all the different components of the algebra by one unique imaginary root (given by the loop junction), so that the full algebra is $(\widehat{E}_{8-n} \oplus  \widehat{E}_8 \oplus  \widehat{A}_{n+1})  / \sim $ (for $0\leq n \leq 8$), with the three imaginary roots identified. 

Let us clarify this point, and the reason why one unique $\delta_{[1,0]}$ loop junction, which is invariant under crossing the intermediate component, is now enough for the affinization of the intermediate $A_{n+1}$ factor. Recall that the loop junction given by $\vec{\delta}_{[1,0]}$ is uncharged under the intermediate $A$ branes, and also under the $I_k$ singularities over generic points of the internal components of the base. This means the junction can just be transported from one end component to the other, crossing the $A^{n+2}$ branes forming the stack, as well as the componets with codimension zero singularities, without any effect. Therefore such a junction could never give rise to an affinization of that algebra alone, as it would be trivial because it has no charge in the junction lattice. However, the existence of the exceptional factors in the end components gives this loop junction non-trivial charges in the junction lattice, and thus the $A_{n+1}$ factor can be affinized. In other words, it is the presence of the end components, and the fact that the loop junction has non-trivial charge under them in the charge lattice what allows for the affinization of all factors. In this sense, this affinization is a non-local effect as the imaginary root for all three factors, in the canonical representation, is such that the affine root for the $\widehat{A}_{n+1}$ factor includes string junctions supported in the exceptional factors at the end components. Note that this is precisely what one would expect for an affinization of the $A_{n+1}$ factor, since the imaginary root must not intersect with the rest of the roots, but at the same time it has to be non-trivial in the junction lattice.

\subsection{$\widehat{\widehat{D}}_{16}$} \label{sec:so32}
\begin{figure}[t]
	\begin{center}
		
			\includegraphics[scale=0.20]{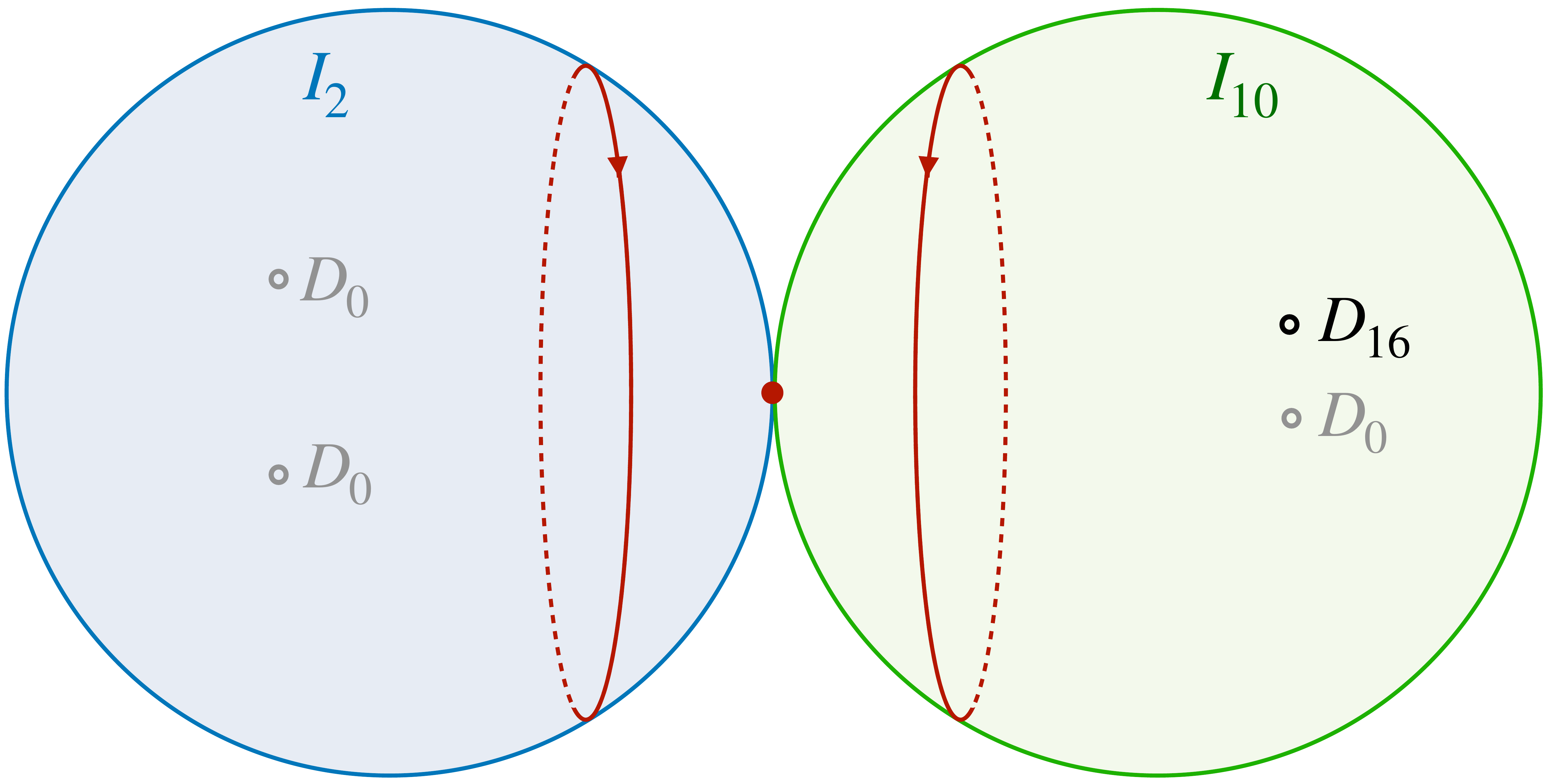} 
		  
		\caption{\footnotesize Configuration associated to the $\widehat{\widehat{D}}_{16}$, with the only loop junction one can detect in the Kulikov model framework in red. The KK tower giving the second imaginary root cannot be realised from this analysis. }     \label{fig:LoopD16}
	\end{center}
\end{figure} 
 The explicit realisation of this algebra, dual to the full decompactification to the $SO(32)$ heterotic theory in \ref{sec:doubleloop}, is shown in Appendix \ref{sec:doubled16} together with the general features of the corresponding Type III.b Kulikov model. We summarise here the main points. As shown in Figure \ref{fig:LoopD16}, the base splits into two $\mathbb{P}^1$ components, $B^0$ and $B^1$, both with non trivial codimension zero singular fibers, respectively of Kodaira type $I_{10}$ and $I_2$. Additionally, there are codimension one singular fibers: on $B^0$ there are one $D_{16}$ and one $D_0$ singularities, while on $B^{1}$ there are two $D_0$ ones. The affinisation of the algebra $D_{16}$ can be seen from the $BC$ system that is approaching the $A^{16}BC$ stack of branes, such that the monodromy of the final configuration is
\begin{equation}
    M_{A^{16}BCBC} = \, \left( \begin{array}{cc}
1 & -8 \\
0 & 1
\end{array}\right)\, .
\end{equation}
This is consistent with the monodromy composition in the case of non-trivial codimension zero singularities $I_n$ and $I_m$ discussed around eq. \eqref{eq:codim0monodromycomp} (see also \cite{Lee:2021qkx}).  In this case the monodromy as seen from each of the two base components reads
\begin{equation}
    M_{BCBC}\, M_{I_{8}}^{\vphantom{-1}}\, = \, \mathbb{I}_{2x2}  \, =\,  M_{A^{16}BCBC}\,  M_{I_8}^{-1} \, .
\end{equation}
The $A^{16}BCBC$ system then supports one loop junction, consisting of a $(1,0)$ string, that is related to the imaginary root making $D_{16}$ affine. From the K3 degeneration perspective which we present in detail in Appendix \ref{appendix:b1}, the imaginary root is given by the M2 brane wrapping the fiber of the vanishing transcendental torus of the Calabi-Yau, which in this case is only one. Thus, apparently the Type III.b degeneration is dual to a partial decompactification limit from eight to nine dimensions, and as such one can indeed realise the affine versions of nine dimensional algebras that are compatible with the weak coupling limit, namely those without E factors. Nevertheless, even though from the geometric perspective only the winding tower is manifest, one should remember the fact that the degeneration limit is taken at fixed K\"ahler moduli, so that in the type IIB string theory the volume of the $T^2$ is constant. This means that the shrinking of one cycle  must be accompanied by the growing of the other one, giving a KK tower that provides the second imaginary root whose presence is geometrically obscure. So, we expect only the algebra $\widehat{\widehat{D}}_{16}$ to be consistently realised in this limit.

\section{Conclusions}
\label{conclusions}
In this work we have studied the infinite distance limits in the moduli space of the heterotic string on $T^d$ correponding to decompactification of one or more dimensions, generalizing the case $d = 1$ studied in \cite{Collazuol:2022jiy}. In these backgrounds, which have 16 supercharges, the heterotic string worldsheet captures neatly all the data relevant to gauge symmetry enhancements and in particular this makes it a suitable frame to study how these symmetries behave at infinite distance. We have seen that in every case, the gauge algebra is asymptotically enhanced to an affine BPS algebra with central extensions equal in number to the decompactified dimensions. This algebra is always the affine version of an allowed gauge algebra in the (partially) decompactified theory, and the affinization works democratically on every one of its simple factors, including the nonabelian factors of type $A$, $D$ and $E$ as well as left- and right-moving Cartans. 

Restricting to the case of eight dimensions, we have studied the dual description given by type IIB string junctions on the sphere and the corresponding F-theory on an elliptic K3 surface. Infinite distance limits in the latter correspond to degenerations of the K3 surface and are exhaustively accounted for in Kulikov models \cite{Lee:2021qkx}. In the case that only one dimension decompactifies in the heterotic dual, i.e. in a Type III.a degeneration, we have shown that the affine algebras emerge in the same way as for the heterotic string. This is a nontrivial statement insofar as only affine algebras of exceptional type had been considered in the string junction literature \cite{DeWolfe:1998yf, DeWolfe:1998pr} and constructed explicitly in F-theory \cite{Lee:2021qkx,Lee:2021usk}. There are in turn two K3 degeneration limits corresponding to decompactification of the heterotic string to ten dimensions. One of them, the Type II.a degeneration, produces the double loop algebra $(\widehat{E}_9\oplus \widehat{E}_9)/\sim$, in perfect agreement with the heterotic string. As we showed, there is also a Type III.b degeneration limit in which double loop algebra $\widehat{\widehat{D}}_{16}$ should appear. However, one of the two associated towers of light states is of KK type and is not realized by string junctions. In fact, the III.b degeneration type admits a priori the realization of various other algebras which we know do not correspond to ten dimensional theories, a symptom of the same underlying problem.  

Moreover, we have seen that from the F-theory point of view there are certain paths to the infinite distance points which cannot be described using Kulikov models. On the contrary, in the heterotic string we can explicitly describe any such path in moduli space. This is nicely illustrated by the case in which both radii go to infinity while their ratio goes to zero.   

Given the simplicity with which the heterotic string allows to study the appearance of affine algebras, it would certainly be interesting to consider other backgrounds with 16 supercharges constructed as asymmetric orbifolds. The simplest case is the CHL string, which in eight dimensions exhibits algebras of type $C_n$, and in lower dimensions also those of type $B_n$ and $F_4$.  The eight dimensional CHL string and its decompactification to nine dimensions has been studied from the point of view of string junctions in \cite{Cvetic:2022uuu}, and it would be interesting to have also the heterotic picture.

\newpage
\noindent {\bf \large Acknowledgments\vspace{0.1in}}\\
We would like to thank Peng Cheng, Gabriele Di Ubaldo, Anamaria Font, Bernardo Fraiman, Carmen Nu\~nez and specially 
Timo Weigand for useful discussions. This work was supported by the ERC Consolidator Grant 772408-Stringlandscape and by CNRS-Imperial College PhD fellowship.

\appendix
\section{Details on the OPEs}
\label{appendix}
In this Appendix, we summarise all the explicit computations of OPEs and commutators.

\subsection{$T^2$ decompactifications}
\label{T2}
From the action normalised as in \eqref{eqn:action}, the relevant OPEs involving the heterotic worldsheet bosons in flat non-compact space-time are
\begin{align}
\label{eqn:XX}
    X^{\mu}(z,\bar{z}) X^{\nu}(w,\bar{w}) \sim& - \frac{1}{2} \eta^{\mu \nu} \log|z-w|^2 \, , \\
\label{eqn:yyope}
    Y^i(z,\bar{z}) Y^j(w,\bar{w}) \sim &-  \frac{1}{2} G^{ij} \log|z-w|^2 \, ,  \\
\label{eqn:oldII}
    X^{I}(z) X^{J}(w) \sim & - \delta^{I J} \log(z-w) \, ,
\end{align}
where $Y^i(z, \bar{z}) = Y^i(z) + \Tilde{Y}^i(\bar{z})\, $, such that
\begin{equation}
    X^{\hat{I}}(z) X^{\hat{J}}(w) \sim - \delta^{\hat{I}\hat{J}} log(z-w)
\end{equation}
are the only non-trivial OPEs among internal fields, the others being either finite or vanishing as $\frac{1}{R_8^2}$.
\subsubsection{$8 \to 9$ dimensions decompactification limit}
\label{appa1}
The vertex operators in the 0 ghost picture associated to the massless vectors \eqref{eqn:8dcartan} and \eqref{eqn:8droots} in the string spectrum are 
\begin{equation}
\label{eqn:vop}
\begin{split}
    &\alpha^{{\hat{I}}}_{-1} \Tilde{\psi}^{\mu}_{-\frac{1}{2}} \ket{0,n_8}_{NS} \rightarrow   \,  i \partial X^{\hat{I}}(z) \Big( i \sqrt{2} \bar{\partial} \Tilde{X}^{\mu}(\bar{z}) + \frac{1}{\sqrt{2}} \kappa \cdot \Tilde{\psi}(\bar{z}) \Tilde{\psi}^{\mu}(\bar{z})\Big)e^{ikX(z, \bar{z})} e^{i n_8 Y^8(z, \bar{z})} \, , \\
    &\alpha^{8}_{-1} \Tilde{\psi}^{\mu}_{-\frac{1}{2}} \ket{0, n_8}_{NS} \rightarrow  \,  i \sqrt{2} \partial Y^8(z) \Big( i \sqrt{2} \bar{\partial} \Tilde{X}^{\mu}(\bar{z}) + \frac{1}{\sqrt{2}} \kappa \cdot \Tilde{\psi}(\bar{z}) \Tilde{\psi}^{\mu}(\bar{z}) \Big)e^{ikX(z, \bar{z})}e^{i n_8 Y^8(z, \bar{z})} \,  ,\\
    &\Tilde{\psi}_{-\frac{1}{2}}^{\mu} \ket{Z_{\mathcal{A}},n_8}_{NS} \rightarrow  \, c_{\alpha} e^{i p_{\alpha; \hat{I}} X^{\hat{I}}(z)}\Big( i \sqrt{2} \bar{\partial} \Tilde{X}^{\mu}(\bar{z}) + \frac{1}{\sqrt{2}} \kappa \cdot \Tilde{\psi}(\bar{z}) \Tilde{\psi}^{\mu}(\bar{z}) \Big)e^{ikX(z, \bar{z})}e^{i (n_8 - E_{89} w^9 - \pi_{\alpha} \cdot A_8)  Y^8(z,\bar{z})} \, ,
\end{split}    
\end{equation}
where $\kappa_M=(k_{\mu}, p_{R,i})$ is the 10 dimensional right momentum. 
In the decompactification limit $G_{88} = R_8^2 \to \infty$, the affine algebra in the presence of a non vanishing $E_{89}=\frac{1}{2}A_8 \cdot A_9$ component is realised by currents generalising the definitions \eqref{eqn:8dcartan} and \eqref{eqn:8droots} of $\{ J^{a}_n \}$ ($a = \hat I, \alpha$), as we will now show. 

Let us first redefine the internal fields as 
\begin{equation}
\label{eqn:newx}
    \mathcal{X}^I(z) = X^I(z) - A^I_8 Y^8(z)
\end{equation}
and $\mathcal{Y}^9(z)$ through the relation 
\begin{equation}
\label{eqn:newy}
    G_{99} \mathcal{Y}^9(z) = G_{99} Y^9(z) + \frac{1}{2} E_{98} Y^8(z) \, .
\end{equation}
In particular, since $G^{99}G_{99}=1$, 
\eqref{eqn:newy} is equivalent to
\begin{equation}
\label{eqn:newy2}
    \mathcal{Y}^9(z) = Y^9(z) + \frac{1}{2} G^{99} E_{98} Y^8(z) \, .
\end{equation}
The leading terms in the OPEs among these redefined internal fields are equivalent to \eqref{eqn:yyope}-\eqref{eqn:oldII} 
\begin{align}
\label{eqn:newyy}
    \mathcal{Y}^9(z) \mathcal{Y}^9(w) &= Y^9(z) Y^9(w) + \mathcal{O}\left( \frac{1}{R_8^2}\right)  \, , \\
    \mathcal{Y}^9(z) \mathcal{X}^I(w) &=  \mathcal{O}\left( \frac{1}{R_8^2}\right) \, , \\
    \mathcal{Y}^9(z) Y^8(w) &= \mathcal{O}\left( \frac{1}{R_8^2}\right) \, , \\
\label{eqn:newij}    
    \mathcal{X}^I(z) \mathcal{X}^J(w) &= X^I(z) X^J(w) + \mathcal{O}\left(\frac{1}{R_8^2}\right) \, .
\end{align}
Finally, let us also define $\mathcal{X}^{\hat{I}}(z)=\left(\sqrt{2} R_9 \mathcal{Y}^9(z), \mathcal{X}^{I} \right)$. The expression for the affine currents in the general case is then
\begin{equation}
\label{eqn:newcurrent}
\begin{split}
    J^{\hat I}_{n_8}(z)  &\equiv i \partial \mathcal{X}^{\hat I}(z) e^{i n_8 Y^8(z)} \, , \\
    J^{\alpha}_{n_8}(z) & \equiv c_{\alpha} e^{i p_{\alpha; \hat{I}} \mathcal{X}^{\hat I}(z)}  e^{in_8 Y^8(z)} \, .
\end{split}
\end{equation}
They are in form equal to \eqref{eqn:8dcartan} and \eqref{eqn:8droots} holding in the case of trivial $A^I_{8}$, and due to \eqref{eqn:newyy}-\eqref{eqn:newij} the two sets of currents have the same OPEs, so that they generate the same algebra.  
The OPEs between currents associated to Cartan vectors read, up to $\mathcal{O}\left( \frac{1}{R_8^2} \right)$ terms
\begin{equation*}
\begin{split}
     J^{\hat{I}}_{n_8}(z) J^{\hat{J}}_{m_8}(w)  &= - :\partial \mathcal{X}^{\hat I}(z) \, e^{i n_8 Y^8(z)}: \, :\partial \mathcal{X}^{\hat J}(w)\, e^{i m_8 Y^8(w)}: \,   \\
    &= (z-w)^{\frac{n_8m_8}{2R_8^2}} :e^{in_8 Y^8(z)} e^{i m_8 Y^8(w)}  \left( - \partial \mathcal{X}^{\hat{I}}(z) \partial \mathcal{X}^{\hat{J}}(w) + \frac{\delta^{\hat{I} \hat J}}{(z-w)^2} \right): \, .
    \end{split}
\end{equation*}
Performing an expansion around $z=w$
\begin{equation}
\begin{split}
 &J^{\hat{I}}_{n_8}(z) J^{\hat{J}}_{m_8}(w) = \\
    &= (z-w)^{\frac{n_8m_8}{2 R_8^2}} :\left( 1 + i n_8 \partial Y^8(w) (z-w) + \ldots \right) e^{i (n_8+m_8) Y^8(w)}\left( - \partial \mathcal{X}^{\hat{I}}(z) \partial \mathcal{X}^{\hat{J}}(w) + \frac{\delta^{\hat{I}\hat{J}}}{(z-w)^2} \right): \,  \\
\label{eqn:survive}
    &\sim (z-w)^{\frac{ n_8m_8}{2 R_8^2}} \left( \delta^{\hat{I}\hat J} \frac{:e^{i(n_8+m_8)Y^8(w)}:}{(z-w)^2} + i \delta^{\hat I \hat J} n_8 \frac{:\partial Y^8(w) e^{i (n_8+m_8)Y^8(w)}: }{z-w} + \mathcal{O}(1) \right) \, ,
    \end{split}
\end{equation}
which is indeed \eqref{eqn:cc}. 

As for the OPE between two affine root currents, up to $\mathcal{O}\left( \frac{1}{R_8^2} \right)$ terms
\begin{equation}
\label{eqn:a8}
\begin{split}
    J^{\alpha}_{n_8}(z) J^{\beta}_{m_8}(w) = &  :c_{\alpha} e^{i p_{\alpha,\hat I} \mathcal{X}^{\hat I }(z)} e^{i n_8 Y^8(z)}: \, :c_{\beta} e^{i p_{\beta;\hat J} \mathcal{X}^{\hat J}(w)} e^{i m
    _8Y^8(w)}:  \\
    =& \,  c_{\alpha} c_{\beta}(z-w)^{ p_{\alpha} \cdot p_{\beta} + \frac{n_8m_8}{2R_8^2}} :e^{i p_{\alpha;\hat I} \mathcal{X}^{\hat I}(z)} e^{i n_8 Y^8(z)}e^{i p_{\beta;\hat J} \mathcal{X}^{\hat J}(w)} e^{i m_8 Y^8(w)}: \\
    = & \, c_{\alpha} c_{\beta}(z-w)^{ p_{\alpha} \cdot p_{\beta} + \frac{n_8m_8}{2R_8^2}} :\big[ 1+i p_{\alpha;\hat I} \partial \mathcal{X}^{\hat I}(w) (z-w)+\ldots)\cdot \\
    & \qquad \qquad \cdot (1+i n_8 \partial Y^8(w) (z-w)+\ldots \big]    e^{i(p_{\alpha;\hat I} + p_{\beta;\hat I})\mathcal{X}^{\hat I}(w)} e^{i (n_8+m_8) Y^8(w)}: \, .
\end{split}
\end{equation}
There are only two non trivial cases for the choice of $p_{\alpha}$ and $p_{\beta}$. Assume that $p_{\alpha}+p_{\beta}=p_{\alpha+\beta}$ is still a root of the algebra $\mathcal{A}$ in 9 dimensions. Then, if the roots are normalised as $|p_{\alpha}|^2=2$, it holds $p_{\alpha} \cdot p_{\beta} = -1$ and \eqref{eqn:a8} reads
\begin{equation}
\begin{split}
        J^{\alpha}_{n_8}(z) J^{\beta}_{m_8}(w) \sim & (z-w)^{  \frac{n_8m_8}{2R_8^2}} \left(   \frac{\epsilon(\alpha, \beta) c_{\alpha+\beta}:e^{i p_{\alpha+\beta;\hat I} \mathcal{X}^{\hat I}(w)} e^{i (n_8+m_8) Y^8(w)} :}{z-w} + \mathcal{O}(1) \right)  \\
        = & (z-w)^{  \frac{n_8m_8}{2R_8^2}} \left(\frac{\epsilon(\alpha, \beta) J^{\alpha+\beta}_{n_8+m_8}(w)}{z-w} + \mathcal{O}(1) \right) \, .
\end{split}
\end{equation}
If instead $p_{\alpha}=-p_{\beta}$, then $p_{\alpha} \cdot p_{\beta} = - |p_{\alpha}|^2=-2$, and so \eqref{eqn:a8} gives
\begin{equation}
\begin{split}
    J^{\alpha}_{n_8}(z) J^{-\alpha}_{m_8}(w) \sim (z-w)^{  \frac{n_8 m_8}{2R_8^2}} :e^{i (n_8+m_8) Y^8(w)} \left( \frac{1}{(z-w)^2} + \frac{i( p_{\alpha;\hat{I}} \partial \mathcal{X}^{\hat I}(w) + n_8 \partial Y^8(w)) }{z-w} + \mathcal{O}(1) \right): \, , \\
    \sim (z-w)^{  \frac{n_8 m_8}{2R_8^2}} \left( \frac{:e^{i (n_8+m_8) Y^8(w)}:}{(z-w)^2} + \frac{p_{\alpha;\hat{I}} J^{\hat I}_{n_8+m_8}(w) + i n_8 :\partial Y^8(w)e^{i (n_8+m_8) Y^8(w)}:}{z-w} + \mathcal{O}(1) \right) \, .
\end{split}
\end{equation}
All the OPEs for other choices of $p_{\alpha}$ and $p_{\beta}$ vanish since in these cases $c_{\alpha} c_{\beta} =0$ 
\begin{equation}
\label{eqn:rrgeneral}
    J^{\alpha}_{n_8}(z) J^{\beta}_{m_8}(w) \sim (z-w)^{\frac{n_8m_8}{2 R_8^2}} \cdot
    \begin{cases}
        \frac{\epsilon(\alpha, \beta) J^{\alpha+\beta}_{n_8+m_8}(w)}{z-w} + \mathcal{O}(1)\qquad \qquad \qquad \qquad \qquad \qquad \qquad \qquad \,    \alpha+\beta \,  {\rm root,} \\
        \frac{:e^{i (n_8+m_8) Y^8(w)}:}{(z-w)^2} + \frac{p_{\alpha;\hat{I}} J^{\hat{I}}_{n_8+m_8}(w)+in_8:\partial Y^8(w) e^{i (n_8+m_8)Y^8(w)}:}{z-w} +\mathcal{O}(1) \, \, \alpha=-\beta\, , \\
        0 \quad \, \, \, \qquad \qquad \qquad \qquad \qquad \qquad \qquad \qquad\qquad \qquad \qquad \quad {\rm \, otherwise,} 
    \end{cases}
\end{equation}
which taking the limit $R_8\to \infty$ reduces indeed to \eqref{eqn:rr1}. 

Finally, considering one Cartan and one root current
\begin{equation}
\label{eqn:ialpha}
\begin{split}
    J^{\hat I}_{n_8}(z) J^{\alpha}_{m_8}(w)= & :i \partial \mathcal{X}^{\hat I}(z) e^{i n_8 Y^8(z)}: \, :c_{\alpha} e^{i  p_{\alpha;\hat J} \mathcal{X}^{\hat J}(w)} e^{i m_8 Y^8(w)}: \\
    = & \, i c_{\alpha} (z-w)^{\frac{n_8m_8}{2 R_8^2}}  \frac{ -i p_{\alpha}^{\hat{I}} :e^{i p_{\alpha;\hat{J}} \mathcal{X}^{\hat J}(w)} e^{in_8 Y^8(z)} e^{i m_8 Y^8(w)}:}{z-w} + \mathcal{O}(1)  \\
    \sim & \,  c_{\alpha} (z-w)^{\frac{n_8m_8}{2 R_8^2}}  \frac{p_{\alpha}^{\hat{I}} :e^{i p_{\alpha;\hat{J}} \mathcal{X}^{\hat J}(w)} e^{i (n_8+m_8) Y^8(w)}: }{z-w} + \mathcal{O}(1) \sim \frac{p_{\alpha}^{\hat I} J^{\alpha}_{n_8+m_8}(w)}{z-w} + \mathcal{O}(1)  
    \end{split}
\end{equation}
as $R_8 \to \infty$, which is \eqref{eqn:cr}. 

The symmetry algebra can be defined as the set of commutation relations between the currents zero modes \eqref{eqn:zeromode} as follows. Let us start considering two Cartan generators, in the limit $G_{88} = R_8^2 \to \infty$ 
\begin{equation}
    [(J^{\hat I}_{n_8})_0, (J^{\hat J}_{m_8})_0] = \oint_{C'} \frac{dz}{2\pi i} \oint_C \frac{dw}{2\pi i}J^{\hat I}_{n_8}(z)  J^{\hat J}_{m_8}(w) - \oint_{C'} \frac{dw}{2\pi i} \oint_C \frac{dz}{2\pi i} J^{\hat I}_{n_8}(z) J^{\hat J}_{m_8}(w)\, 
\end{equation}
where $C'$ is a contour external to $C$. 
\begin{equation}
\label{eqn:cases2}
\begin{split}
    [(J^{\hat I}_{n_8})_0, (J^{\hat J}_{m_8})_0]  = & \oint \frac{dw}{2\pi i} \delta^{\hat I \hat J} \mathrm{Res}_{z \to w} \left[ \frac{:e^{i(n_8+m_8)Y^8(w)}:}{(z-w)^2} + i n_8 \frac{:\partial Y^8(w) e^{i (n_8+m_8) Y^8(w)}: }{z-w} \right] \\
    = & \oint \frac{dw}{2\pi i } in_8 \delta^{\hat I \hat J} \partial Y^8(w) e^{i (n_8+m_8) Y^8(w)} \\
     = &
    \begin{cases}
        \oint \frac{dw}{2\pi i} i n_8 \delta^{\hat I \hat J} \partial y^8(w) = in_8 \delta^{\hat I \hat J} (\partial Y^8)_0 \quad \text{ for } n_8+m_8=0 \, ,\\
         \oint \frac{dw}{2 \pi i } \delta^{\hat I \hat J} \frac{n_8}{n_8+m_8} \partial(e^{i(n_8+m_8)Y^8(w)})=0 \quad \text{ for } n_8+m_8\ne 0 \, 
    \end{cases} \\
    = & \,  i n_8 \delta^{\hat I \hat J} \delta_{n_8+m_8, 0} (\partial Y^8)_0 \, .
    \end{split}
\end{equation}
The commutator between a Cartan and a root generator is
\begin{equation}
\begin{split}
    [(J^{\hat I}_{n_8})_0, (J^{\alpha}_{m_8})_0] = & \oint_{C'} \frac{dz}{2\pi i} \oint_C \frac{dw}{2\pi i}J^{\hat I}_{n_8}(z)  J^{\alpha}_{m_8}(w) - \oint_{C'} \frac{dw}{2\pi i} \oint_C \frac{dz}{2\pi i} J^{\hat I}_{n_8}(z) J^{\alpha}_{m_8}(w) \\
    = & \oint \frac{dw}{2\pi i} \mathrm{Res}_{z \to w} \left[ \frac{p_{\alpha}^{\hat I} c_{\alpha} :e^{i p_{\alpha; \hat{J}}\mathcal{X}^{\hat J}(w)} e^{i  (n_8+m_8) Y^8(w)}:}{z-w} \right] \\
    = &  \oint \frac{dw}{2\pi i} p_{\alpha}^{\hat I} c_{\alpha} :e^{i p_{\alpha;\hat{J}}\mathcal{X}^{\hat J}(w)} e^{i  (n_8+m_8) Y^8(w)}: \\
    =& p_{\alpha}^{\hat I} (J^{\alpha}_{n+m})_0 \, , 
    \end{split}
\end{equation}
and the commutator between two root generators is
\begin{equation}
\begin{split}
   &[(J^{\alpha}_{n_8})_0, (J^{\beta}_{m_8})_0] = \oint_{C'} \frac{dz}{2\pi i} \oint_C \frac{dw}{2\pi i}J^{\alpha}_{n_8}(z)  J^{\beta}_{m_8}(w) - \oint_{C'} \frac{dw}{2\pi i} \oint_C \frac{dz}{2\pi i} J^{\alpha}_{n_8}(z)  J^{\beta}_{m_8}(w)   \\
    &= \oint \frac{dw}{2\pi i} \mathrm{Res}_{z \to w} 
    \begin{cases}
          \frac{\epsilon(\alpha, \beta) c_{\alpha + \beta} e^{i p_{\alpha +\beta;\hat{I}} \mathcal{X}^{\hat I}(w)}e^{i(n_8+m_8)Y^8(w)}}{z-w} + \mathcal{O}(1)\qquad \qquad \qquad \qquad \qquad \qquad  \, \,\,\, \, \alpha+\beta \,  {\rm root,} \\
        \frac{:e^{i (n_8+m_8) Y^8(w)}:}{(z-w)^2} + \frac{:i p_{\alpha;\hat{I}} \partial \mathcal{X}^{\hat I}(w) e^{i (n_8+m_8) Y^8(w)} + i n_8 \partial Y^8(w) e^{i (n_8+m_8) Y^8(w)}:}{z-w} +\mathcal{O}(1)  \, \, \, \alpha=-\beta, \\
        0 \, \, \, \, \qquad \qquad \qquad \qquad \qquad  \qquad\qquad \qquad \qquad \qquad \qquad \qquad \qquad \qquad \, \, {\rm otherwise.} 
    \end{cases}\\
    &\qquad \qquad  \qquad \  = \begin{cases}
          \epsilon(\alpha, \beta) (J^{\alpha + \beta}_{n_8+m_8})_0  \qquad \qquad  \qquad \quad  \quad \,\,\, \,\,  \,  \alpha+\beta \, \, \,  {\rm root} \\
          p_{\alpha;\hat{I}} (J^{\hat I}_{n_8+m_8})_0 + i n_8 \delta_{n_8+m_8, 0} (\partial Y^8)_0 \quad \, \, \alpha=-\beta\\
         0  \qquad \qquad \qquad \qquad \qquad \qquad \qquad \,\,\,\, \, \, \, \, \, {\rm otherwise.} 
    \end{cases}
    \end{split}
\end{equation}

\subsubsection*{Affinisation of the right moving algebra}
Let us now focus on the $U(1)^2_R$ right moving contribution to the algebra and its affinisation pattern. Restricting for simplicity to the case $A_8^{I}=0$, the associated massless states as $R_8 \to \infty$ and their corresponding vertex operators in the 0 ghost picture are
\begin{equation}
    \alpha_{-1}^{\mu} \tilde{\psi}^i_{-\frac{1}{2}} \ket{0,n_8} \to i \sqrt{2} \partial X^{\mu}  \Big( i \sqrt{2} \bar{\partial} \Tilde{Y}^{i}(\bar{z}) + \frac{1}{\sqrt{2}} \kappa \cdot \Tilde{\psi}(\bar{z}) \Tilde{\psi}^{i}(\bar{z})\Big)e^{ikX(z, \bar{z})} e^{i n_8 Y^8(z, \bar{z})} \, , 
\end{equation}
with $i=8,9$, and they also come in massless momentum towers as the eighth direction is decompactified. The associated antiholomorphic currents are 
\begin{equation}
\label{eqn:1}
    \tilde{\Upsilon}^i_n(\bar{z}) = \tilde{\Upsilon}^i(\bar{z}) e^{i n_8 Y^8(\bar{z})} \, ,
\end{equation}
where 
\begin{equation}
    \tilde{\Upsilon}^i(\bar{z}) = \left( i \sqrt{2} \bar{\partial} \tilde{Y}^i(\bar{z}) + \frac{1}{\sqrt{2}} k_{\mu} \tilde{\psi}^{\mu} (\bar{z}) \tilde{\psi}^i(\bar{z}) \right) 
\end{equation}
are the currents of the finite $U(1)^2_R$. Indeed, using \eqref{eqn:yyope} and
\begin{equation}
    \tilde{\psi}^M(\bar{z}) \tilde{\psi}^N(\bar{w}) \sim \frac{\eta^{MN}}{\bar{z}-\bar{w}} \, ,
\end{equation}
and the fact that in the decompactification limit the $T^2$ metric is diagonal, $G^{ij}=\frac{1}{R_i^2} \delta^{ij}$, one finds, up to $\mathcal{O}(1)$ factors 
\begin{equation}
\begin{split}
    \tilde{\Upsilon}^i(\bar{z}) \tilde{\Upsilon}^j(\bar{w}) &= \\
    &= - 2 \wick{  \bar{\partial} \c{  \tilde{ Y}}^i(\bar{z}) \bar{\partial} \c{  \tilde{  Y}}^j}(\bar{w}) - 2   :\bar{\partial} \tilde{ Y}^i(\bar{z}) \bar{\partial}  \tilde{  Y}^j(\bar{w}): - \frac{1}{2} k_{\mu} k_{\nu} (\wick{  \c{ \tilde{\psi}}^{\mu}(\bar{z}) \c{ \tilde{\psi}}^{\nu}(\bar{w})} :\tilde{\psi}^i(\bar{z}) \tilde{\psi}^j(\bar{w}): + \\ 
    &+ :\tilde{\psi}^{\mu}(\bar{z})\tilde{\psi}^{\nu}(\bar{w}): \wick{\c{ \tilde{\psi}^i}(\bar{z}) \c{ \tilde{\psi}^j}(\bar{w})} + \wick{  \c{ \tilde{\psi}}^{\mu}(\bar{z}) \c{ \tilde{\psi}}^{\nu}(\bar{w})} \wick{\c{ \tilde{\psi}^i}(\bar{z}) \c{ \tilde{\psi}^j}(\bar{w})} + :\tilde{\psi}^{\mu}(\bar{z})  \tilde{\psi}^i(\bar{z}) \tilde{\psi}^{\nu}(\bar{w}) \tilde{\psi}^j(\bar{w}):)  \\
    &= \frac{\delta^{ij}}{R_i^2} \frac{1}{(\bar{z}-\bar{w})^2} - \frac{1}{2} k_{\mu} k_{\nu} \left( \frac{\eta^{\mu \nu} :\tilde{\psi}^i(\bar{z}) \tilde{\psi}^j(\bar{w}):}{\bar{z}-\bar{w}} + :\frac{\tilde{\psi}^{\mu}(\bar{z})\tilde{\psi}^{\nu}(\bar{w}): \delta^{ij}}{R_i^2 (\bar{z}-\bar{w})} + \frac{\eta^{\mu \nu} \delta^{ij} }{R_i^2(\bar{z}-\bar{w})^3} \right)  \\
    &= \frac{\delta^{ij}}{R_i^2} \frac{1}{(\bar{z}-\bar{w})^2} \, .
\end{split}    
\end{equation}
The term with $k_{\mu} k_{\nu} \eta^{\mu \nu}$ vanish because $k^2=0$, while the term in $k_{\mu} k_{\nu}\tilde{\psi}^{\mu}(\bar{z})\tilde{\psi}^{\nu}(\bar{w})=0 $ due to symmetry arguments. The absence of the single pole means that the zero modes $(\tilde{\Upsilon}^i(\bar{z}))_0$ commute, so that they are indeed associated to a $U(1)^2_R$ symmetry.

The full asymptotic currents \eqref{eqn:1} satisfy the following OPEs
\begin{equation}
\label{eqn:2}
\begin{split}
    \tilde{\Upsilon}^i_{n_8}(\bar{z}) \tilde{\Upsilon}^j_{m_8}(\bar{w}) &= (\bar{z}-\bar{w})^{\frac{n_8m_8}{2R_8^2}} :e^{in_8\tilde{Y}^8(\bar{z})}  e^{im_8\tilde{Y}^8(\bar{w})} \cdot \\ 
    &\cdot  \left[ \frac{\delta^{ij}}{R_i^2(\bar{z}-\bar{w})^2} - 2i \frac{n_8 \delta^{8j} \bar{\partial}\tilde{Y}^i(\bar{z})-m_8 \delta^{8i} \bar{\partial}\tilde{Y}^j(\bar{w})}{R_8^2(\bar z - \bar w)} -  2 \frac{n_8 m_8 \delta^{8i} \delta^{8j} }{R_8^4(\bar z - \bar w)^2} + \mathcal{O}(1)\right]: \, ,
\end{split}
\end{equation}
and as $R_8 \to \infty$ the only non trivial case is
\begin{equation}
\label{eqn:22}
    \tilde{\Upsilon}^9_{n_8}(\bar{z}) \tilde{\Upsilon}^9_{m_8}(\bar{w}) =   \frac{e^{i(n_8+m_8)\tilde{Y}^8(\bar{z})}  }{R_8^2(\bar{z}-\bar{w})^2} + \frac{in_8 \bar{\partial}\tilde{Y}^8(\bar{w})e^{i(n_8+m_8)\tilde{Y}^8(\bar{z})}}{\bar z -\bar w}+ \mathcal{O}(1) \, ,
\end{equation}
all the other combinations giving only finite terms asymptotically in moduli space. In particular, eq. \eqref{eqn:22} means that the zero modes of the rescaled current $\tilde{\Upsilon}^9_{n_8}(\bar{z}) \to e^9_{\, \, 9} \tilde{\Upsilon}^9_{n_8}(\bar{z})$ obey the OPEs of an $\hat{U}(1)$ algebra, where the central extension is given by $(\bar{\partial} \tilde{Y}^8)_0$.

\subsubsection{$8 \to 10$ dimensions decompactification limit}
\label{appa2}
The vertex operators in the 0 picture associated to the massless vectors \eqref{eqn:kkc}-\eqref{eqn:allroots} for generic finite B field and Wilson lines components are
\begin{equation}
\begin{split}
    \alpha^{i}_{-1} \Tilde{\psi}^{\mu}_{-\frac{1}{2}} \ket{0, n_8, n_9}_{NS} &\to i \sqrt{2} \partial Y^{i}(z) \left( i \sqrt{2} \bar{\partial} \Tilde{X}^{\mu}(\bar{z}) + \frac{1}{\sqrt{2}} \kappa \cdot \Tilde{\psi}(\bar{z}) \Tilde{\psi}^{\mu}(\bar{z})\right)e^{ikX(z, \bar{z})} e^{in_j Y^j(z, \bar{z})} \, , \\
     \alpha^{I}_{-1} \Tilde{\psi}^{\mu}_{-\frac{1}{2}} \ket{0, n_8,n_9}_{NS} &\to i  \partial X^{I}(z) \left( i \sqrt{2} \bar{\partial} \Tilde{X}^{\mu}(\bar{z}) + \frac{1}{\sqrt{2}} \kappa \cdot \Tilde{\psi}(\bar{z}) \Tilde{\psi}^{\mu}(\bar{z})\right)e^{ikX(z, \bar{z})} e^{in_j Y^j(z, \bar{z})} \, , \\
     \Tilde{\psi}_{-\frac{1}{2}}^{\mu} \ket{Z_{E_8 \times E_8}, n_8, n_9}_{NS} &\to c_{\alpha} e^{i\pi_{\alpha;I} X^{I}(z)} \left( i \sqrt{2} \bar{\partial} \Tilde{X}^{\mu}(\bar{z}) + \frac{1}{\sqrt{2}} \kappa \cdot \Tilde{\psi}(\bar{z}) \Tilde{\psi}^{\mu}(\bar{z})\right) e^{i(n_i-\pi_{\alpha;I}A^{I}_i) Y^i(z, \bar{z})} \, ,
\end{split}
\end{equation}
with the same conventions of \ref{appa1}. Let us generalise the definition \eqref{eqn:newx} to
\begin{equation}
    \mathcal{X}^I(z) = X^I(z) - A^I_i Y^i(z) \, .
\end{equation}
In the full decompactification limit, at first order the OPE between two such operators is
\begin{equation}
    \mathcal{X}^I(z) \mathcal{X}^J(w) = X^I(z) X^J(w) 
\end{equation}
The relevant holomorphic currents giving the double loop algebra are
\begin{equation}
\begin{split}
       J^I_{n_8,n_9}(z) &\equiv i \partial \mathcal{X}^I(z) e^{i n_i Y^i(z)} \, , \\
       J^{\alpha}_{n_8,n_9}(z) &\equiv c_{\alpha} e^{i \pi_{\alpha;I} \mathcal{X}^I(z)}e^{i n_i Y^i(z)}\, , 
\end{split}
\end{equation}
related respectively to the Cartan and root sector. They are a generalisation of \eqref{eqn:e8cartanss} and \eqref{eqn:allroots}, the latters holding in the case of vanishing Wilson lines along the two torus directions.

The non trivial OPEs among these currents are 
\begin{equation}
\begin{split}
    J^I_{n_8,n_9}(z) J^J_{m_8,m_9}(w) &= - :\partial \mathcal{X}^I(z) e^{i n_8 Y^8(z)} e^{i n_9 Y^9(z)}: :\partial \mathcal{X}^J(w) e^{i m_8 Y^8(w)} e^{i m_9 Y^9(w)}: 
    \\ 
    &= (z-w)^{\frac{G^{ij}n_i m_j}{2}} :e^{i(n_8+m_8) Y^8(w)}e^{i(n_9+m_9)Y^9(w)}(1+ i n_8 \partial Y^8(w)(z-w)+\ldots) \cdot \\
    & \quad \cdot (1+ i n_9 \partial Y^9(w)(z-w)+\ldots)\left( - \partial \mathcal{X}^{I}(z) \partial \mathcal{X}^{J}(w) + \frac{\delta^{IJ}}{(z-w)^2} \right):  \\
    &= (z-w)^{\frac{G^{ij}n_i m_j}{2}} \left( \frac{\delta^{IJ}:e^{i (n_i+m_i) Y^i(w)}:}{(z-w)^2} + i n_i \frac{:\partial Y^i(w) e^{i (n_j+m_j) Y^j(w)}:}{z-w} + \mathcal{O}(1) \right)
\end{split}
\end{equation}
which helds the following commutator of the zero modes
\begin{equation}
\label{eqn:integral}
\begin{split}
    [(J^{I}_{n_8,n_9})_0, (J^{J}_{m_8,m_9})_0]  = & \oint \frac{dw}{2\pi i} \delta^{IJ} \mathrm{Res}_{z \to w} \left[ \frac{:e^{i(n_i+m_i)Y^i(w)}:}{(z-w)^2} + i n_i \frac{:\partial Y^i(w) e^{i (n_j+m_j) Y^j(w)}: }{z-w} \right] \\
    = & \oint \frac{dw}{2\pi i } i  \delta^{IJ} (n_8 \partial Y^8(w) + n_9 \partial Y^9(w)) e^{i (n_8+m_8) Y^8(w)} e^{i (n_9+m_9) Y^9(w)} \, .
\end{split}    
\end{equation}
This integral can be solved as follows. For any choice of $n_i+m_i, \, i=8, \, 9$ it must hold
\begin{equation}
\label{eqn:sumintegrals}
\begin{split}
    0 &= \frac{1}{i} \oint \frac{dw}{2\pi i} \partial \left( e^{i (n_8+m_8)Y^8(w)} e^{i (n_9+m_9)Y^9(w)} \right)  \\
    &= \oint \frac{dw}{2\pi i }  (n_8+m_8) \partial Y^8(w) e^{i (n_i+m_i) Y^i(w)} + \oint \frac{dw}{2\pi i }  (n_9+m_9) \partial Y^9(w)  e^{i (n_i+m_i) Y^i(w)}  \\
    &= \oint \frac{dw}{2\pi i } (n_8 \partial Y^8(w) + n_9 \partial Y^9(w)) e^{i (n_i+m_i) Y^i(w)}  + \oint \frac{dw}{2\pi i }(m_8 \partial Y^8(w) + m_9 \partial Y^9(w)) e^{i (n_i+m_i) Y^i(w)} \, ,
\end{split}
\end{equation}
where in the last line one can recognize two integrals of the type \eqref{eqn:integral}. Focusing on the second line of \eqref{eqn:sumintegrals} and taking into account the fact that $Y^8(z)$ and $Y^9(z)$ are independent, this means that the $\partial Y^8$ and $\partial Y^9$ contributions in the sum must vanish separately. Let us consider for concreteness 
\begin{equation}
\label{eqn:only8}
    \oint \frac{dw}{2\pi i }  (n_8+m_8) \partial Y^8(w) e^{i (n_8+m_8) Y^8(w)} e^{i (n_9+m_9) Y^9(w)} = 0  \, .
\end{equation}
\begin{itemize}
    \item If $n_8 + m_8 = 0$, \eqref{eqn:only8} vanishes trivially, so that in this case one can have
    \begin{equation}
            \oint \frac{dw}{2\pi i } \partial Y^8(w) e^{i (n_9+m_9) Y^9(w)} \ne 0 \, ,
    \end{equation}
    and from \eqref{eqn:zeromode} it is $(\partial Y^8_{0,n_9+m_9})_0$. 
    \item If $n_8 + m_8 \ne 0$, then it must hold
    \begin{equation}
    \label{eqn:zeroint}
        \oint \frac{dw}{2\pi i }   \partial Y^8(w) e^{i (n_8+m_8) Y^8(w)} e^{i (n_9+m_9) Y^9(w)}  = 0 \, .
    \end{equation}
    This statement is trivial for $n_9 + m_9 =0$, when the integrand in \eqref{eqn:zeroint} reduces to a total derivative, but it must actually hold $\forall (n_9+m_9)$.
\end{itemize}
Summarising
\begin{equation}
\label{eqn:fullintegral}
    \oint \frac{dw}{2\pi i } \partial Y^8(w) e^{i(n_8+m_8)Y^8(w)} e^{i (n_9+m_9) Y^9(w)} = \delta_{n_8+m_8,0} (\partial Y^8_{0,n_9+m_9})_0 \, ,
\end{equation}
and the same holds with $8 \leftrightarrow 9$. Using these results in \eqref{eqn:integral} 
\begin{equation}
    [(J^{I}_{n_8,n_9})_0, (J^{J}_{m_8,m_9})_0] = i  \delta^{IJ} ( n_8 \delta_{n_8+m_8,0}  (\partial Y^8_{0,n_9+m_9})_0 + n_9 \delta_{n_9+m_9,0}  (\partial Y^9_{n_8+m_8,0})_0 ) \, . 
\end{equation}
In the case of a Cartan and a root current
\begin{equation}
\begin{split}
    J^{I}_{n_8,n_9}(z) J^{\alpha}_{m_8,m_9}(w) &= ic_{\alpha}  : \partial \mathcal{X}^I(z) e^{i n_8 Y^8(z)} e^{i n_9 Y^9(z)}: :e^{i \pi_{\alpha;I} \mathcal{X}^I(w)} e^{i m_8 Y^8(w)} e^{i m_9 Y^9(w)}:  \\
    &= (z-w)^{\frac{G^{ij}n_i m_j}{2}} \left( c_{\alpha} \pi_{\alpha}^I \frac{:e^{i \pi_{\alpha;I} \mathcal{X}^I(w)} e^{i(n_8+m_8) Y^8(w)}e^{i(n_9+m_9)Y^9(w)}:}{z-w} + \mathcal{O}(1) \right)  \\
    &= (z-w)^{\frac{G^{ij}n_i m_j}{2}} \left(\frac{\pi_{\alpha}^I J^{\alpha}_{n_8+m_8,n_9+m_9}(w)}{z-w} + \mathcal{O}(1) \right) \, ,
\end{split}
\end{equation}
which gives the following commutation relations
\begin{equation}
\begin{split}
    [(J^{I}_{n_8,n_9})_0, (J^{\alpha}_{m_8,m_9})_0]  = & \oint \frac{dw}{2\pi i} \mathrm{Res}_{z \to w} \left[ \frac{\pi_{\alpha}^I J^{\alpha}_{n_8+m_8,n_9+m_9}(w)}{z-w} \right] \\
    = & \oint \frac{dw}{2\pi i }  \pi_{\alpha}^I J^{\alpha}_{n_8+m_8,n_9+m_9}(w) = \pi^I_{\alpha} (J^{\alpha}_{n_8+m_8,n_9+m_9})_0
\end{split}
\end{equation}
Finally, considering two root currents
\begin{equation}
\begin{split}
    J^{\alpha}_{n_8,n_9}(z) J^{\beta}_{m_8,m_9}(w) &= c_{\alpha} c_{\beta} :e^{i \pi_{\alpha;I} \mathcal{X}^I(z)}e^{i n_8 Y^8(z)}e^{i n_9 Y^9(z)}: :e^{i \pi_{\beta;J} \mathcal{X}^J(w)}e^{i m_8 Y^8(w)}e^{i m_9 Y^9(w)}:  \\
    &= c_{\alpha} c_{\beta} (z-w)^{\pi_{\alpha} \cdot \pi_{\beta} + \frac{G^{ij}n_im_j}{2}} :e^{i (\pi_{\alpha;I} \mathcal{X}^I(z)+ \pi_{\beta;J} \mathcal{X}^J(w))} e^{i (n_i Y^i(z)+m_i Y^i(w))}:  \\
    &= c_{\alpha} c_{\beta} (z-w)^{\pi_{\alpha} \cdot \pi_{\beta} + \frac{G^{ij}n_im_j}{2}} :(1+i\pi_{\alpha;J} \partial \mathcal{X}^J(w)(z-w)+ \ldots) e^{i(\pi_{\alpha;I}+\pi_{\beta;I}) \mathcal{X}^I(w)} \\
    &\quad(1+ in_j \partial Y^j(w) (z-w)+ \ldots ) e^{i (n_k+m_k) Y^k(w)}: \, ,
\end{split}
\end{equation}
which again splits in three different cases. If $\pi_{\alpha}+\pi_{\beta}=\pi_{\alpha+\beta}$, then
\begin{equation}
\begin{split}
    J^{\alpha}_{n_8,n_9}(z) J^{\beta}_{m_8,m_9}(w) &= (z-w)^{\frac{G^{ij}n_im_j}{2}} \left( \frac{\epsilon(\alpha,\beta) c_{\alpha+\beta} :e^{i(\pi_{\alpha;I}+\pi_{\beta;I}) \mathcal{X}^I(w)}e^{i (n_k+m_k) Y^k(w)} : }{z-w} + \mathcal{O}(1) \right) \\
    &= (z-w)^{\frac{G^{ij}n_im_j}{2}} \left( \frac{\epsilon(\alpha,\beta) J^{\alpha+\beta}_{n_8+m_8,n_9+m_9}(w) }{z-w} + \mathcal{O}(1) \right) \, .
\end{split}    
\end{equation}
If $\pi_{\alpha}=-\pi_{\beta}$
\begin{equation}
    J^{\alpha}_{n_8,n_9}(z) J^{\beta}_{m_8,m_9}(w) = (z-w)^{\frac{G^{ij}n_im_j}{2}} :e^{i (n_i+m_i) Y^i(w)}\left( \frac{1}{(z-w)^2} + \frac{i \pi_{\alpha;I} \partial \mathcal{X}^I+i n_j \partial Y ^j(w) }{z-w} + \mathcal{O}(1) \right): \, .
\end{equation}
For all the remaining $\pi_{\alpha}, \, \pi_{\beta}$ instead, $c_{\alpha} c_{\beta}=0$ and so to summarise
\begin{align}
       J^{\alpha}_{n_8, n_9}(z) J^{\beta}_{m_8, m_9}(w)  &=  
    \begin{cases}
        \frac{\epsilon(\alpha, \beta) J^{\alpha+\beta}_{n_8+m_8,n_9+m_9}(w)}{z-w} + \mathcal{O}(1)\qquad \qquad \qquad \qquad \qquad \qquad \qquad \qquad \,  \alpha+\beta \,  {\rm root,} \\
        \frac{:e^{i (n_i+m_i) Y^i(w)}:}{(z-w)^2} + \frac{\pi_{\alpha;I} J^{I}_{n_8+m_8,n_9+m_9}(w)+i n_i:\partial Y^i(w) e^{i (n_j+m_j)Y^j(w)}:}{z-w} +\mathcal{O}(1) \, \, \, \, \alpha=-\beta\, , \\
        0 \qquad \qquad \qquad \qquad \qquad \qquad \qquad \qquad\qquad \qquad \qquad \qquad \qquad \quad {\rm  otherwise,} 
    \end{cases}
\end{align}
The algebra of the zero modes in this case is
\begin{equation}
\begin{split}
   &[(J^{\alpha}_{n_8,n_9})_0, (J^{\beta}_{m_8,m_9})_0] = \oint_{C'} \frac{dz}{2\pi i} \oint_C \frac{dw}{2\pi i}J^{\alpha}_{n_8,n_9}(z)  J^{\beta}_{m_8,m_9}(w) - \oint_{C'} \frac{dw}{2\pi i} \oint_C \frac{dz}{2\pi i} J^{\alpha}_{n_8, n_9}(z)  J^{\beta}_{m_8,m_9}(w)   \\
    &= \oint \frac{dw}{2\pi i} \mathrm{Res}_{z \to w} 
    \begin{cases}
         \frac{\epsilon(\alpha, \beta) J^{\alpha+\beta}_{n_8+m_8,n_9+m_9}(w)}{z-w} + \mathcal{O}(1)\qquad \qquad \qquad \qquad \qquad \qquad \qquad \,\,\, \, \alpha+\beta \,  {\rm root,} \\
        \frac{:e^{i (n_i+m_i) Y^i(w)}:}{(z-w)^2} + \frac{:\pi_{\alpha;I} J^{I}_{n_8+m_8,n_9+m_9} + i n_i \partial Y^i(w) e^{i (n_j+m_j) Y^j(w)}:}{z-w} +\mathcal{O}(1)  \, \, \, \alpha=-\beta, \\
        0 \, \, \,  \qquad \quad \qquad \qquad \qquad  \qquad\qquad \qquad \qquad \qquad \qquad \qquad \qquad \, \, {\rm otherwise.} 
    \end{cases}\\
    & \qquad \qquad \qquad \, = \begin{cases}
          \epsilon(\alpha, \beta) (J^{\alpha + \beta}_{n_8+m_8,n_9+m_9})_0  \qquad \qquad \qquad \qquad  \qquad \qquad \qquad \qquad \,\,\, \, \,    \,  \alpha+\beta \,  {\rm root}, \\
          \pi_{\alpha;I} (J^{I}_{n_8+m_8,n_9+m_9})_0 + \\
          \quad + i (n_8  \delta_{n_8+m_8, 0} (\partial Y^8_{0, n_9+m_9})_0 + n_9  \delta_{n_9+m_9, 0} (\partial Y^9_{n_8+m_8,0})_0 )  \qquad \quad  \, \alpha=-\beta\, ,\\
         0 \qquad \qquad \qquad \qquad \qquad \qquad \qquad \qquad \qquad \qquad \qquad \qquad \qquad \,   {\rm otherwise,} 
    \end{cases}
    \end{split}
\end{equation}

\subsection{$T^d$ decompactification}
\label{appa3}
Let us explicitly derive the commutation relations for the algebra in a generic decompactification limit of the heterotic theory on $T^d$. Since from the $S^1$ and $T^2$ examples we clearly showed how the off-diagonal blocks of metric and B field and the Wilson lines along the directions we decompactify do not play any role in the limit, let us for simplicity set them to 0, the difference otherwise being just a field redefinition which does not affect the OPEs.

The relevant OPEs in this case are
\begin{align}
\label{eqn:XXd}
    X^{\mu}(z,\bar{z}) X^{\nu}(w,\bar{w}) \sim& - \frac{1}{2} \eta^{\mu \nu} \log|z-w|^2 \, , \\
\label{eqn:yyoped}
    Y^{i}(z,\bar{z}) Y^{j}(w,\bar{w}) \sim &-  \frac{1}{2} G^{ij} \log|z-w|^2 \, ,  \\
\label{eqn:oldIId}
    X^{I}(z) X^{J}(w) \sim & - \delta^{I J} \log(z-w) \, ,
\end{align}
where \eqref{eqn:yyoped} for $i=\hat{\imath}$ and \eqref{eqn:oldIId} can be written in terms of the fields $X^{\hat{I}}(z)=(\sqrt{2} e_{\hat{\imath}}^{\hat{a}} Y^{\hat{\imath}}(z),X^I(z))$ defined in Section \ref{Dddecompactification} as
\begin{equation}
    X^{\hat I}(z) X^{\hat J}(w) \sim - \delta^{\hat I \hat J} \log(z-w)
\end{equation}
The vertex operators associated to the massless states \eqref{eqn:masslessd} are
\begin{equation}
\label{eqn:vopd}
\begin{split}
    \alpha^{{\hat{I}}}_{-1} \Tilde{\psi}^{\mu}_{-\frac{1}{2}} \ket{0,n_{\bar{\jmath}}}_{NS} \rightarrow  & \,  i \partial X^{\hat{I}}(z) \Big( i \sqrt{2} \bar{\partial} \Tilde{X}^{\mu}(\bar{z}) + \frac{1}{\sqrt{2}} \kappa \cdot \Tilde{\psi}(\bar{z}) \Tilde{\psi}^{\mu}(\bar{z})\Big)e^{ikX(z, \bar{z})} e^{i n_{\bar{\jmath}} Y^{\bar{\jmath}}(z, \bar{z})} \, , \\
    \alpha^{\bar{\imath}}_{-1} \Tilde{\psi}^{\mu}_{-\frac{1}{2}} \ket{0, n_{\bar{\jmath}}}_{NS} \rightarrow  & \, i \sqrt{2} \partial Y^{\bar{\imath}}(z) \Big( i \sqrt{2} \bar{\partial} \Tilde{X}^{\mu}(\bar{z}) + \frac{1}{\sqrt{2}} \kappa \cdot \Tilde{\psi}(\bar{z}) \Tilde{\psi}^{\mu}(\bar{z}) \Big)e^{ikX(z, \bar{z})}e^{i n_{\bar{\jmath}} Y^{{\bar{\jmath}}}(z, \bar{z})} \,  ,\\
    \Tilde{\psi}_{-\frac{1}{2}}^{\mu} \ket{Z_{\mathcal{A}},n_{\bar{\jmath}}}_{NS} \rightarrow & \, c_{\alpha} e^{i p_{\alpha; \hat{I}} X^{\hat{I}}(z)}\Big( i \sqrt{2} \bar{\partial} \Tilde{X}^{\mu}(\bar{z}) + \frac{1}{\sqrt{2}} \kappa \cdot \Tilde{\psi}(\bar{z}) \Tilde{\psi}^{\mu}(\bar{z}) \Big)e^{ikX(z, \bar{z})} e^{i n_{\bar{\jmath}}Y^{\bar{\jmath}}(z,\bar{z})} \, .
\end{split}    
\end{equation}
The conserved currents related to the affine algebra are
\begin{align}
    J^{\hat{I}}_{\{n_{\bar{\imath}}\}}(z) &\equiv i \partial X^{\hat{I}}(z) e^{i n_{\bar{\imath}}Y^{\bar{\imath}}(z)} \, , \\
    J^{\alpha}_{\{n_{\bar{\imath}}\}}(z) &\equiv c_{\alpha}e^{i p_{\alpha; \hat{I}} X^{\hat{I}}(z)} e^{i n_{\bar{\imath}}Y^{\bar{\imath}}(z)} \, . 
\end{align}
In the following we present their mutual OPEs, up to $\mathcal{O}\left( \frac{1}{R_{\bar{\imath}}^2} \right)$ terms and dropping the prefactor $(z-w)^{\frac{G^{\bar{\imath}\bar{\jmath}}n_{\bar{\imath}}m_{\bar{\jmath}}}{2}} \to 1$ asymptotically, which is common to all of them.
\begin{equation}
\begin{split}
     J^{\hat{I}}_{\{n_{\bar{\imath}}\}}(z) J^{\hat{J}}_{\{m_{\bar{\imath}}\}}(w)  &= - :\partial X^{\hat I}(z) \, e^{i n_{\bar{\imath}} Y^{\bar{\imath}}(z)}: \, :\partial X^{\hat J}(w)\, e^{i m_{\bar{\jmath}} Y^{\bar{\jmath}}(w)}:  \,   \\
    &= :e^{i n_{\bar{\imath}} Y^{\bar{\imath}}(z)} e^{i m_{\bar{\jmath}} Y^{\bar{\jmath}}(w)}  \left( - \partial X^{\hat{I}}(z) \partial X^{\hat{J}}(w) + \frac{\delta^{\hat{I} \hat J}}{(z-w)^2} \right):  \\
    &= :\left( 1 + i n_{\bar{\imath}} \partial Y^{\bar{\imath}}(w) (z-w) + \ldots \right) e^{i (n_{\bar{\jmath}}+m_{\bar{\jmath}}) Y^{\bar{\jmath}}(w)}  \left(\frac{\delta^{\hat{I}\hat{J}}}{(z-w)^2} + \mathcal{O}(1) \right) :   \\
    & \sim \delta^{\hat{I}\hat J} \frac{:e^{i (n_{\bar{\jmath}}+m_{\bar{\jmath}}) Y^{\bar{\jmath}}(w)}:}{(z-w)^2} + i \delta^{\hat I \hat J} n_{\bar{\imath}} \frac{:\partial Y^{\bar{\imath}}(w) e^{i (n_{\bar{\jmath}}+m_{\bar{\jmath}}) Y^{\bar{\jmath}}(w)}: }{z-w} + \mathcal{O}(1)  \, .
    \end{split}
\end{equation}
\begin{equation}
\label{eqn:a8d}
\begin{split}
    J^{\alpha}_{\{n_{\bar{\imath}}\}}(z) J^{\beta}_{\{m_{\bar{\jmath}}\}}(w) = &  :c_{\alpha} e^{i p_{\alpha,\hat I} X^{\hat I }(z)} e^{i n_{\bar{\imath}} Y^{\bar{\imath}}(z)}:=  \, :c_{\beta} e^{i p_{\beta;\hat J} X^{\hat J}(w)} e^{i m_{\bar{\jmath}} Y^{\bar{\jmath}}(w)}:  \\
    =& \,  c_{\alpha} c_{\beta} :e^{i p_{\alpha;\hat I} X^{\hat I}(z)} e^{i n_{\bar{\imath}} Y^{\bar{\imath}}(z)}e^{i p_{\beta;\hat J} X^{\hat J}(w)} e^{i m_{\bar{\jmath}} Y^{\bar{\jmath}}(w)}: \\
    = & \, c_{\alpha} c_{\beta}:\big[ 1+i p_{\alpha;\hat I} \partial X^{\hat I}(w) (z-w)+\ldots)\cdot \\
    & \qquad \qquad \cdot (1+i n_{\bar{\imath}} \partial Y^{\bar{\imath}}(w) (z-w)+\ldots \big]    e^{i(p_{\alpha;\hat I} + p_{\beta;\hat I})X^{\hat I}(w)} e^{i (n_{\bar{\jmath}}+m_{\bar{\jmath}}) Y^{\bar{\jmath}}(w)}: \, .
\end{split}
\end{equation}
Again, one can choose $p_{\alpha}+p_{\beta}=p_{\alpha+\beta}$
\begin{equation}
\begin{split}
        J^{\alpha}_{\{n_{\bar{\imath}}\}}(z) J^{\beta}_{\{m_{\bar{\jmath}}\}}(w)  \sim &    \frac{\epsilon(\alpha, \beta) c_{\alpha+\beta}:e^{i p_{\alpha+\beta;\hat I} X^{\hat I}(w)} e^{i (n_{\bar{\jmath}}+m_{\bar{\jmath}}) Y^{\bar{\jmath}}(w)}:}{z-w} + \mathcal{O}(1)  \\
        = &  \frac{\epsilon(\alpha, \beta) J^{\alpha+\beta}_{\{n_{\bar{\imath}}+m_{\bar{\imath}}\}}(w)}{z-w} + \mathcal{O}(1) \, ,
\end{split}
\end{equation}
or $p_{\alpha}=-p_{\beta}$
\begin{equation}
\begin{split}
     J^{\alpha}_{\{n_{\bar{\imath}}\}}(z) J^{\beta}_{\{m_{\bar{\jmath}}\}}(w)  \sim :e^{i (n_{\bar{\jmath}}+m_{\bar{\jmath}}) Y^{\bar{\jmath}}(w)}\left( \frac{1}{(z-w)^2} + \frac{i( p_{\alpha;\hat{I}} \partial X^{\hat I}(w) + n_{\bar{\imath}} \partial Y^{\bar{\imath}}(w)) }{z-w} + \mathcal{O}(1) \right):  \, , \\
    =\frac{:e^{i (n_{\bar{\jmath}}+m_{\bar{\jmath}}) Y^{\bar{\jmath}}(w)}:}{(z-w)^2} + \frac{p_{\alpha;\hat{I}} J^{\hat I}_{ \{n_{\bar{\imath}}+m_{\bar{\imath}}\}}(w) + i n_{\bar{\imath}} :\partial Y^{\bar{\imath}}(w)e^{i (n_{\bar{\jmath}}+m_{\bar{\jmath}}) Y^{\bar{\jmath}}(w)}:}{z-w} + \mathcal{O}(1) \, .
\end{split}
\end{equation}
All the other choices of $\pi$ give vanishing vanish OPEs.
\begin{equation}
\label{eqn:rrgeneralt}
    J^{\alpha}_{\{n_{\bar{\imath}}\}}(z) J^{\beta}_{\{m_{\bar{\jmath}}\}}(w) \sim 
    \begin{cases}
        \frac{\epsilon(\alpha, \beta) J^{\alpha+\beta}_{ \{n_{\bar{\jmath}}+m_{\bar{\jmath}}\}}}{z-w} + \mathcal{O}(1)\qquad \qquad \qquad \qquad \qquad \qquad \qquad \qquad \qquad \, \, \,    \alpha+\beta \,  {\rm root,} \\
        \frac{:e^{i (n_{\bar{\jmath}}+m_{\bar{\jmath}}) Y^{\bar{\jmath}}(w)}:}{(z-w)^2} + \frac{p_{\alpha;\hat{I}} J^{\hat{I}}_{ \{n_{\bar{\imath}}+m_{\bar{\imath}}\}}(w)+in_{\bar{\imath}}:\partial Y^{\bar{\imath}}(w) e^{i (n_{\bar{\jmath}}+m_{\bar{\jmath}}) Y^{\bar{\jmath}}(w)}:}{z-w} +\mathcal{O}(1) \, \, \qquad \alpha=-\beta\, , \\
        0 \quad \, \, \, \qquad \qquad \qquad \qquad \qquad \qquad \qquad \qquad\qquad \qquad \qquad \qquad \quad {\rm \, otherwise,} 
    \end{cases}
\end{equation}
\begin{equation}
\label{eqn:ialphad}
\begin{split}
    J^{\hat I}_{\{n_{\bar{\imath}}\}}(z) J^{\alpha}_{\{m_{\bar{\imath}}\}}(w)= & :i \partial X^{\hat I}(z) e^{i n_{\bar{\imath}} Y^{\bar{\imath}}(z)}: \, :c_{\alpha} e^{i  p_{\alpha;\hat J} X^{\hat J}(w)} e^{i m_{\bar{\jmath}} Y^{\bar{\jmath}}(w)}: \\
    = & \, i c_{\alpha} \frac{ -i p_{\alpha}^{\hat{I}} :e^{i p_{\alpha;\hat{J}} X^{\hat J}(w)} e^{i (n_{\bar{\jmath}}+m_{\bar{\jmath}}) Y^{\bar{\jmath}}(w)}:}{z-w} + \mathcal{O}(1)  \\
    = & \,  c_{\alpha}   \frac{p_{\alpha}^{\hat{I}} :e^{i p_{\alpha;\hat{J}} X^{\hat J}(w)} e^{i (n_{\bar{\jmath}}+m_{\bar{\jmath}}) Y^{\bar{\jmath}}(w)}: }{z-w} + \mathcal{O}(1) = \frac{p_{\alpha}^{\hat I} J^{\alpha}_{ \{n_{\bar{\imath}}+m_{\bar{\imath}}\}}(w)}{z-w} + \mathcal{O}(1)  
    \end{split}
\end{equation}
The asymptotic algebra among the currents zero modes is
\begin{equation}
    [(J^{\hat I}_{\{n_{\bar{\imath}}\}})_0, J^{\hat J}_{\{m_{\bar{\jmath}}\}})_0] = \oint_{C'} \frac{dz}{2\pi i} \oint_C \frac{dw}{2\pi i}J^{\hat I}_{\{n_{\bar{\imath}}\}}(z)  J^{\hat J}_{\{m_{\bar{\jmath}}\}}(w) - \oint_{C'} \frac{dw}{2\pi i} \oint_C \frac{dz}{2\pi i} J^{\hat I}_{\{n_{\bar{\imath}}\}}(z) J^{\hat J}_{\{m_{\bar{\jmath}}\}}(w)\, 
\end{equation}
where $C'$ is a contour external to $C$. 
\begin{equation}
\label{eqn:cases}
\begin{split}
    [(J^{\hat I}_{\{n_{\bar{\imath}}\}})_0, J^{\hat J}_{\{m_{\bar{\jmath}}\}})_0]  = & \oint \frac{dw}{2\pi i} \delta^{\hat I \hat J} \mathrm{Res}_{z \to w} \left[ \frac{:e^{i (n_{\bar{\jmath}}+m_{\bar{\jmath}}) Y^{\bar{\jmath}}(w)}:}{(z-w)^2} + i n_{\bar{\imath}} \frac{:\partial Y^{\bar{\imath}}(w) e^{i (n_{\bar{\jmath}}+m_{\bar{\jmath}}) Y^{\bar{\jmath}}(w)}: }{z-w}  \right] \\
    = & \oint \frac{dw}{2\pi i } i \delta^{\hat I \hat J} n_{\bar{\imath}} \partial Y^{\bar{\imath}}(w) e^{i (n_{\bar{\jmath}}+m_{\bar{\jmath}}) Y^{\bar{\jmath}}(w)} = i \delta^{\hat I \hat J}  n_{\bar{\imath}} \delta_{n_{\bar{\imath}}+m_{\bar{\imath}},0} (\partial Y^{\bar{\imath}}_{\{ n_{\bar{\jmath}}+m_{\bar{\jmath}} \}})_0  \, ,
    \end{split}
\end{equation}
where a sum over $\bar{\imath}$, the directions we are decompactifying, is understood, and between the second and third line we used the generalised version of \eqref{eqn:fullintegral} 
\begin{equation}
   \oint \frac{dw}{2\pi i } \partial Y^{\bar{\imath}}(w) e^{i(n_{\bar{\jmath}}+m_{\bar{\jmath}})Y^{\bar{\jmath}}(w)}  = \delta_{n_{\bar{\imath}}+m_{\bar{\imath}},0} (\partial Y^{\bar{\imath}}_{\{ n_{\bar{\jmath}}+m_{\bar{\jmath}} \}})_0 \, . 
\end{equation}
Analogously
\begin{equation}
\begin{split}
    [(J^{\hat I}_{\{n_{\bar{\imath}}\}})_0, (J^{\alpha}_{\{m_{\bar{\jmath}}\}})_0] = & \oint \frac{dw}{2\pi i} \mathrm{Res}_{z \to w} \left[ \frac{c_{\alpha}p_{\alpha}^{\hat{I}} :e^{i p_{\alpha;\hat{J}} X^{\hat J}(w)} e^{i (n_{\bar{\jmath}}+m_{\bar{\jmath}}) Y^{\bar{\jmath}}(w)}:}{z-w} \right] \\
    = &  \oint \frac{dw}{2\pi i} p_{\alpha}^{\hat I} c_{\alpha} :e^{i p_{\alpha;\hat{J}} X^{\hat J}(w)} e^{i (n_{\bar{\jmath}}+m_{\bar{\jmath}}) Y^{\bar{\jmath}}(w)}: \\
    =& p_{\alpha}^{\hat I} (J^{\alpha}_{\{ n_{\bar{\jmath}}+m_{\bar{\jmath}} \}})_0 \, , 
    \end{split}
\end{equation}
and 
\begin{equation}
\begin{split}
   &[(J^{\hat I}_{\{n_{\bar{\imath}}\}})_0, (J^{\beta}_{\{m_{\bar{\jmath}}\}})_0] = \\
   &= \oint \frac{dw}{2\pi i} \mathrm{Res}_{z \to w} 
    \begin{cases}
        \frac{\epsilon(\alpha, \beta) J^{\alpha+\beta}_{ \{n_{\bar{\jmath}}+m_{\bar{\jmath}}\}}}{z-w} + \mathcal{O}(1)\qquad \qquad \qquad \qquad \qquad \qquad \qquad \qquad \quad \, \, \,  \alpha+\beta \,  {\rm root,} \\
        \frac{:e^{i (n_{\bar{\jmath}}+m_{\bar{\jmath}}) Y^{\bar{\jmath}}(w)}:}{(z-w)^2} + \frac{p_{\alpha;\hat{I}} J^{\hat{I}}_{ \{n_{\bar{\imath}}+m_{\bar{\imath}}\}}(w)+in_{\bar{\imath}}:\partial Y^{\bar{\imath}}(w) e^{i (n_{\bar{\jmath}}+m_{\bar{\jmath}}) Y^{\bar{\jmath}}(w)}:}{z-w} +\mathcal{O}(1) \, \quad \alpha=-\beta\, , \\
        0 \quad \, \, \, \qquad \qquad \qquad \qquad \qquad \qquad \qquad \qquad\qquad \qquad \qquad \qquad {\rm \, otherwise,} 
    \end{cases}\\
    &\qquad \qquad  \qquad \  = \begin{cases}
          \epsilon(\alpha, \beta) (J^{\alpha + \beta}_{\{ n_{\bar{\jmath}}+m_{\bar{\jmath}} \}})_0  \qquad \qquad  \qquad \qquad  \qquad \,\,\, \,\,  \,  \alpha+\beta \, \, \,  {\rm root} \\
          p_{\alpha;\hat{I}} (J^{\hat I}_{\{ n_{\bar{\jmath}}+m_{\bar{\jmath}} \}})_0 + i n_{\bar{\imath}} \delta_{n_{\bar{\imath}}+m_{\bar{\imath}}, 0} (\partial Y^{\bar{\imath}}_{\{ n_{\bar{\jmath}}+m_{\bar{\jmath}} \}})_0 \quad \, \alpha=-\beta\\
         0  \qquad \qquad \qquad \qquad \qquad \qquad \qquad \qquad \, \, \, \,\,\,\, \, \, \, \, \, {\rm otherwise.} 
    \end{cases}
    \end{split}
\end{equation}
Only as a brief comment on the affinisation of the right moving $U(1)^{d-k}_R$ algebra, the general pattern is that the $U(1)$'s associated to the directions that remain compact are made affine with $k$ central extensions given by the $(\bar{\partial} \tilde{Y}^{\bar{\imath}})$ that correspond to decompactified directions.

\section{Decompactification limits in F-theory}
\label{appendix2}

\subsection{Review of Kulikov models}
\label{appendix:b1}
From the dual F-theory point of view, decompactification limits of the heterotic theory can be analysed geometrically by studying the complex structure degenerations of the elliptically fibered K3, in particular the ones occurring at infinite distance in the complex structure moduli space. In turn, these can be described and classified in terms of Kulikov models \cite{Lee:2021qkx}. 

In order to define them, let us introduce a K3 degeneration as a one-parameter family $\{ X_u \}$ of K3 surfaces fibered over a disk $u \in D = \{ u \in \mathbb{C}, |u|<1 \}$, such that the fiber is smooth everywhere except at $u=0$, the infinite distance limit, where it degenerates. The disk and the K3 surfaces form a threefold $\mathcal{X}$ with base $D$ and fiber $X_{u}$, and by construction it defines a Kulikov model, which is a degeneration with the following properties:
\begin{itemize}
    \item semi-stability: $\mathcal{X}$ is smooth, and the fiber $X_0$ is a reduced variety where all components have multiplicity one and the singularities are of normal crossing type. This is guaranteed by the semi-stable reduction theorem.
    \item Ricci flatness, which can always be achieved for a semi-stable degeneration via base changes ($u \to u^k$) and/or birational transformations.
\end{itemize}
Since the case of interest is the one of elliptically fibered K3's, one can describe this family $\{ X_u\}$ by blowing down each member $X_u$ and associating to the resulting surface a Weierstrass model,  
\begin{equation}
   Y_u: \quad y^2 = x^3 +f_u(s,t) xz^4+ g_u(s,t)z^6 \, ,
\end{equation} 
and the corresponding family $\{ Y_u \}$ is called Kulikov Weierstrass model. Here $[x:y:z]$ are homogeneous coordinates of $\mathbb{P}_{231}$ representing the fiber space and $[s:t]$ are the homogeneous coordinates of the base $\mathbb{P}^1$ of $X_u$. $f_u$ and $g_u$ are homogeneous polynomials in $[s:t]$ of degree 8 and 12 respectively, giving a singular surface $Y_0$ which can be obtained by blowing down $X_0$. 
It can be shown that $Y_0$ is an union of several components
\begin{equation}
    Y_0 = \cup_{i=0}^p Y^i \, 
\end{equation}
where each $Y^i$ is a fibration over a $\mathbb{P}^1$ base component $B_i= \{ e_i=0 \}$ with $e_i$ coordinates on the base space, which arrange in a chain (each $B^i$ intersect two other components at most once), with possibly two kinds of degeneracies that can be read from the vanishing orders of the discriminant
\begin{equation}
    \Delta_0 = 4f_0^3+27g_0^2 \, .
\end{equation}
These can be
\begin{itemize}
    \item non-minimal singularities in the Kodaira classification, namely points on $B^i$ with vanishing orders $ord(f_0,g_0,\Delta_0) = (\geq 4, \geq 6, \geq 12)$.
    
    \item codimension zero singularities, namely singular fibers over any point of the base component $B^i$ of the form $\Delta_0 = e_i^{n_i}\Delta'_0$, where $\Delta'_0$ does not allow for further factorisation in $e_i$. These singularities must be of Kodaira Type $I_{n_i}$ (namely, $f_0$ and $g_0$ do not vanish at these generic points) in order for them  to be of normal crossing type. These singularities can get enhanced at co-dimension one loci to give either type A, or two type D singularities. This is in contrast to the usual smooth case where the codimension zero singularities are of Kodaira Type $I_0$, which can have all types of singular fibers over codimension one loci on the base.     
\end{itemize}

We will not give the details of the classification of all the possible degenerate limits, for which we refer to \cite{Lee:2021qkx}, but just qualitatively introduce the structure of the degenerations which are dual to decompactifications in the heterotic framework. The towers in this approach are recovered - at least partially - from elliptic transcendental 2-cycles with asymptotically vanishing volume on which M2 branes can be wrapped arbitrarily many times. 
All the statements below hold up to base changes and birational transformations.

\subsubsection*{Partial decompactification limits to 9 dimensions}
This case, called `Type III.a', is dual to \ref{89}. The base of the elliptically fibered K3 degenerates into a chain of two or more components. The intermediate components have codimension zero fiber of Kodaira Type $I_{n_i}, \, n_i >0$ and at most codimension one singularities of type A. As for the end components, either they are both rational elliptic surfaces, $dP_9$, or one of them is a $dP_9$ surface and the other one has codimension zero fiber of Type $I_{n>0}$ and two D-type codimension one fibers. All these components intersect at a point, where they share a Type $I_{m}$ singularity.
One example is given explicitly in \ref{fth89}.  
The vanishing transcendental torus is given by fibering the (1,0) vanishing cycle of the fiber at the intersection point over the vanishing 1-cycle of the base, which is denoted by $\delta_{[1,0]}$ in Figure \ref{fig:E8xE8xsu(2)}.

\subsubsection*{Full decompactification limits to 10 dimensions}
In the F-theory setting there are two possible cases in which the dual heterotic theory fully decompactifies. The one that has been explicitly worked out in terms of the Kulikov Weierstrass model is the one corresponding to the algebra $(\widehat{E}_9 \oplus \widehat{E}_9)/\sim$, called `Type II.a' and reviewed in Section \ref{sec:e9e9}, but we can also realise the qualitatively different `Type III.b' limit the algebra $\widehat{\widehat{D}}_{16}$, as we will explicitly show in the following.  

 In the Type III.b limit the base of the K3 surface degenerates into several components $B^i$, all of which have a codimension zero singularity of Kodaira type $I_{n_i}, \, n_i >0$. This corresponds to the weak coupling Sen's orientifold limit. In addition there is a non-minimal singular fiber, which is related to the presence of an affine algebra. Like in the Type III.a case, the Kulikov model captures only one vanishing transcendental elliptic curve, namely the fibration of the vanishing A-cycle in the fiber over the cycle surrounding the point of intersection between two base components. This account for only one of the towers, given in the dual M-theory by  M2 branes wrapping this transcendental torus arbitrarily many times, or alternatively to the winding modes of the weakly coupled type IIB F1 wrapping a vanishing cycle on the $T^2$. Nevertheless, the Type III.b corresponds to a full decompactification limit if one accounts for the fact that the K\"ahler modulus is kept constant in the degeneration, which is possible if the size of the B-cycle goes to infinity, giving the second tower as KK modes along it. 
Let us now explicitly describe the limit corresponding to the algebra $\widehat{\widehat{D}}_{16}$, which is dual to the decompactification to the 10 dimensional $SO(32)$ heterotic theory.

\subsection{Realisation of double loop $D_{16}$ in F-theory} \label{sec:doubled16}
Let us consider the following Weierstrass model in the patch $z=1$
\begin{equation}
    y^2=x^3+f(s,t)x+g(s,t)
\end{equation}
with
\begin{align}
    f(s,t)&=-\frac{1}{3} s^2 h(s,t)^2-ds^8 \, , \\
    g(s,t)&=-\frac{2}{27} s^3 h(s,t)^3-\frac{1}{3}ds^9h(s,t)\, , \\
    h(s,t)&=bt^3+at^2s+s^3 \, ,
\end{align}
$a, \, b, \, d \in \mathbb{C}$. The discriminat is
\begin{equation}
\label{eqn:delta}
\begin{split}
    \Delta(s,t)&= 4 f(s,t)^3+27g(s,t)^2\\ 
    &= -d^2 s^{18} ((1+4d) s^6 + 2a s^4 t^2 + 2b s^3 t^3 +a^2 s^2 t^4  +2abst^5 + b^2t^6 )  \, .    
\end{split}
\end{equation}
There is a $D_{16}$ singularity at $s=0$, with vanishing orders $ord(f,g,\Delta)|_{s=0}=(2,3,18)$, and 6 additional $I_{1}$ singularites, which restricting to the patch $s=1$ are at
\begin{align}
    t_{1, k_{\pm}} &= - \frac{a}{3b} + \frac{\sqrt[3]{2}a^2}{3bk_{\pm}} + \frac{k_{\pm}}{3 \sqrt[3]{2}b} \, , \\
    t_{2, k_{\pm}} &= - \frac{a}{3b} - \frac{(1+ i \sqrt{3})a^2}{3 \sqrt[3]{4} b k_{\pm}} - \frac{(1 - i \sqrt{3} )k_{\pm}}{6 \sqrt[3]{2} b} \, , \\
    t_{3, k_{\pm}} &= - \frac{a}{3b} - \frac{(1- i \sqrt{3})a^2}{3 \sqrt[3]{4} b k_{\pm}} - \frac{(1 + i \sqrt{3} )k_{\pm}}{6 \sqrt[3]{2} b} 
\end{align}
where the two $\pm$ in $t_{2, k_{\pm}, \pm i}$ are not correlated, and we have defined the following quantites
\begin{equation}
    k_{\pm}(a,b,d) \equiv k_{\pm} = \left(-2 a^3 -27 b^2 + \sqrt{-4a^6 +(-2a^3-27b^2\pm 54b^2 \sqrt{-d})^2 \pm 54b^2 \sqrt{-d} }\right)^{\frac{1}{3}} 
\end{equation}
So, this Weierstrass model describes a compactification with algebra $D_{16}$.

For $d=0$ one has $k_{+}=k_{-}$, but also $\Delta$ vanishes identically at order 2. This limit is consistent with a Kulikov model of Type IIb, with $d=d(u)=d_0 u$. We should then work with the redefined quantity
\begin{equation}
    \Delta_0(s,t)=\frac{\Delta(s,t)}{d^2}\bigg|_{d=0} \, ,
\end{equation}
(the subscript $0$ stands for $d=0$), which is regular as $d\to 0$, with
\begin{align}
    f(s,t) |_{d=0}&=-\frac{1}{3} s^2 h(s,t)^2 \, , \\
    g(s,t)|_{d=0}&=-\frac{2}{27} s^3 h(s,t)^3 \, , \\
    \Delta_0(s,t)&= -s^{18} h(s,t)^2 \, .
\end{align}
$\Delta_0$ has zeros at $s=0$, with vanishing orders $ord(f,g,\Delta_0)|_{s=0}=(2,3,18)$ and three additional zeros of vanishing orders $ord(f,g,\Delta_0)=(2,3,2)$, which in the patch $s=1$ are at
\begin{align}
    t_{1} &= - \frac{a}{3b} + \frac{\sqrt[3]{2}a^2}{3bk_0} + \frac{k_0}{3 \sqrt[3]{2}b} \, , \\
    t_{2} &= - \frac{a}{3b} - \frac{(1+ i \sqrt{3})a^2}{3 \sqrt[3]{4} b k_0} - \frac{(1 - i \sqrt{3} )k_0}{6 \sqrt[3]{2} b} \, , \\
    t_{3} &= - \frac{a}{3b} - \frac{(1- i \sqrt{3})a^2}{3 \sqrt[3]{4} b k_0} - \frac{(1 + i \sqrt{3} )k_0}{6 \sqrt[3]{2} b} 
\end{align}
where $k_0 \equiv k_{+} (a,b,0) =  k_{-} (a,b,0)$. Physically, the 6 branes giving six $I_1$ singularities in \eqref{eqn:delta} pair into three $D_0$ singularities.

Additionally, one can take $b \to 0$, so that, introducing $h_0(s,t)=at^2s+s^3=s(at^2+s^2)$ 
\begin{align}
    f(s,t)_{b,d=0}&=-\frac{1}{3} s^2 h_0(s,t)^2 \, , \\
    g(s,t)|_{b,d=0}&=-\frac{2}{27} s^3 h_0(s,t)^3 \, , \\
    \Delta_0(s,t)|_{b=0}&= -s^{18} h_0(s,t)^2 \, .
\end{align}
$\Delta_0(s,t)|_{b=0}$ has zeros at $s=0$, with vanishing orders $ord(f|_{b,d=0},g|_{b,d=0},\Delta_0|_{b=0})\big|_{s=0}=(4,6,20)$, which is a non minimal singularity, 
and two zeros of vanishing orders 
$ord(f|_{b,d=0},g|_{b,d=0},\Delta_0|_{b=0})=(2,3,2)$, which in the patch $s=1$ are at
\begin{equation}
    t'_{1,2}= \pm \frac{i}{\sqrt{a}} \, .
\end{equation}
The singularity becoming non minimal can be interpreted as one of the $D_0$ stacks coming closer to the $D_{16}$ one, the other two having a fixed position far apart from each other and from the $D_{16}+D_0$ system.

The fact that in this case $\Delta$ vanishes at a generic point on the base at order two and there is a non minimal singularity makes this limit consistent with a Kulikov model of Type III.b, with $b=b(u)=b_0 u$. To resolve the non minimal singularity we should consider this Kulikov model explicitly, namely 
\begin{equation}
    y^2=x^3+f_u(s,t)x+g_u(s,t)
\end{equation}
with
\begin{align}
    f_u(s,t)&=-\frac{1}{3} s^2 h_u(s,t)^2-d_0 u s^8 \, , \\
    g_u (s,t)&=-\frac{2}{27} s^3 h_u(s,t)^3-\frac{1}{3}d_0 u s^9h_u(s,t) \, , \\
    h_u(s,t)&=b_0 u t^3+at^2s+s^3 \, ,
\end{align}
with $b_0, \, d_0 \ne 0$.
With this parametrisation of the vanishing coefficients it holds
\begin{equation}
\label{eqn:deltau}
\begin{split}
    \Delta_u(s,t)&= 4 f_u(s,t)^3+27g_u(s,t)^2\\ 
    &= -d_0^2 u^2 s^{18} ((1+4d_0 u) s^6 + 2a s^4 t^2 + 2b_0u s^3 t^3 +a^2 s^2 t^4  +2ab_0ust^5 + b_0^2 u^2 t^6 )  \, .    
\end{split}
\end{equation}
Indeed, one sees that at $u=0, \, s=0$ there is a non minimal singularity with
\begin{align}
    f_u(s,t)|_{s=u\to0}&=-\frac{1}{3} s^4  (h'(s,t)^2+3d_0 s^5) \, , \\
    g_u (s,t)|_{s=u\to0}&=-\frac{1}{27} s^6 h'(s,t) \left(2h'(s,t)^2+9 d_0 s^5\right) \, , \\
    \Delta(s,t)|_{s=u\to0} &= - d_0^2 s^{22} (h'(s,t)^2+4d_0s^5)\, .
\end{align}
where
\begin{equation}
    h_u(s,t)|_{s=u\to0}= s (b_0 t^3+at^2+s^2) \equiv s h'(s,t) 
\end{equation}
so that the vanishing orders in the Kulikov model framework correspond to a non minimal singularity in the Kodaira classification, $ord(f|_{s=u\to0},g|_{s=u\to0},\Delta|_{s=u\to0})=(4,6,22)$. On the other hand, at $u=0, \, t=0$ the threefold has an $I_2$ minimal singularity
\begin{align}
    f_u(s,t)|_{t=u\to0}&=-\frac{1}{3} s^2 \left(h_u(s,t)|_{t=u\to0}\right)^2-d_0 t s^8 \, , \\
    g_u (s,t)|_{t=u\to0}&=-\frac{2}{27} s^3 \left(h_u(s,t)|_{t=u\to0}\right)^3-\frac{1}{3}d_0 t s^9\left(h_u(s,t)|_{t=u\to0}\right) \, , \\
    \Delta(s,t)|_{t=u\to0} &= - d_0^2 t^2 s^{18} (s^6+4 d_0 s^6t + 2 a s^4 t^2 +a^2 s^2 t^4 +2b_0s^3t^4 + 2ab_0st^6+b_0^2t^8) \, ,
\end{align}
with 
\begin{equation}
    h_u(s,t)|_{t=u\to0}=b_0 t^4+at^2s+s^3
\end{equation}
which does not display any further factorisation of $t$ factors.

To get rid of the non minimal singularity one performs the base blowup
\begin{equation}
    s \to s e_1 \, , \quad u \to e_0 e_1
\end{equation}
accompanied by the rescalings (preserving the Calabi-Yau condition)
\begin{equation}
\label{eqn:buf}
\begin{split}
    f_{e_0}(s,t,e_1) &\to \frac{f_{e_0}(s,t,e_1)}{e_1^4} \, , \\ 
    g_{e_0}(s,t,e_1) &\to \frac{g_{e_0}(s,t,e_1)}{e_1^6} \, . 
\end{split}
\end{equation}
After these rescalings, it holds
\begin{align}
\label{eqn:budelta}
    \Delta_{e_0}(s,t,e_1) = - d_0^2 e_0^2 e_1^{10} s^{18} \delta_{e_0}(s,t,e_1) 
\end{align}
where $\delta_{e_0}(s,t,e_1)$ cannot be further factorised, and for completeness it reads
\begin{equation}
    \delta_{e_0}(s,t,e_1) = e_1^4 s^6 +4d_0 e_0 e_1^5 s^6 + 2 a e_1^2 s^4 t^2 + 2 b_0 e_0 e_1^2 s^3 t^3 + a_0^2 s^2 t^4 + 2 a b_0 e_0 st^5 + b_0^2 e_0^2 t^6
\end{equation}
The blown up model has a minimal singularity for $e_1=s=0$, which from \eqref{eqn:buf} and \eqref{eqn:budelta} can be seen to be of the type
$ord(f_{e_0}(s,t,e_1)|_{s=e_1\to 0},g_{e_0}(s,t,e_1)|_{s=e_1\to 0},\Delta_{e_0}(s,t,e_1)|_{s=e_1\to 0})=(2,3,28)$, so the base cannot be further blown up. 

In order to read the physical singularities, namely the ones corresponding to 7 branes, one should define the $K3$ discriminant
\begin{equation}
    \Delta'_{e_0}(s,t,e_1) = \frac{\Delta_{e_0}(s,t,e_1)}{e_0^2 e_1^{10}} 
\end{equation}
\eqref{eqn:buf} and \eqref{eqn:budelta} describe a degenerate $K3$ surface. The base degenerates in a chain of two intersecting surfaces
\begin{itemize}
    \item $\{ e_0 = 0\}$ with generic $I_2$ fiber. Restricting to this component we can set $s=1$ since $s$ and $e_0$ cannot vanish simultaneously and
    \begin{equation}
    \begin{split}
        f_{0}(1,t,e_1) &= - \frac{1}{3} (e_1^2 + a t^2)^2 \, , \\
        g_{0}(1,t,e_1) &= - \frac{2}{27} (e_1^2 + a t^2)^3 \, , \\
        \Delta'_{0}(1,t,e_1) &= - d_0^2 (e_1^2 + a t^2)^2 \, .
    \end{split}    
    \end{equation}
    In this component there are two minimal singularities with vanishing orders given by  $ord(f_{0}(1,t,e_1),g_{0}(1,t,e_1),\Delta'_{e_0}(1,t,e_1))=(2,3,2)$ corresponding to two $D_0$ singularities, separated one from the other.
    \item $\{ e_1 = 0\}$ with generic $I_{10}$ fiber. Restricting to this component we can set $t=1$ and
    \begin{equation}
    \begin{split}
        f_{e_0}(s,1,0) &= - \frac{1}{3} s^2 (b_0 e_0 + a s)^2 \, , \\
        g_{e_0}(s,1,0) &= - \frac{2}{27} s^3 (b_0 e_0 + a s)^3 \, , \\
        \Delta'_{e_0}(s,1,0) &= - d_0^2 s^{18} (b_0 e_0 + a s)^2 \, ,
    \end{split}    
    \end{equation}
    so that in this component there is one $D_{16}$ singularity $ord(f_{e_0}(s,s,0),g_{e_0}(s,1,0),\Delta'_{e_0}(s,1,0))=(2,3,18)$ and one $D_0$ singualrity as above.
\end{itemize}

The monodromy of the $D_{16} + D_{0}$ system is
\begin{equation}
    M_{D_{16}/D_{0}} = \, \left( \begin{array}{cc}
1 & -8 \\
0 & 1
\end{array}\right)\, ,
\end{equation}
so indeed it supports only one invariant $(1,0)$ string surrounding the brane configuration, corresponding to only one imaginary root extending the algebra to $\widehat{D}_{16}$. As already mentioned, the second one cannot be detected in this framework. 

\subsection{Heterotic/F-theory duality map}
\label{appendix:b2}
Let us now examine how the F-theory and heterotic frames differ in the way they capture the decompactification limits. More concretely, as we will show, certain decompactification limits which can be described naturally from the point of view of the heterotic string cannot be encoded in the framework of Kulikov Weierstrass models, as these involve functions that are not rational. 

We work with an elliptic K3 surface with two generic $E_8$ singular fibers, described by the Weierstrass model
\begin{equation}
    y = x^3 + as^4t^4x+ t^5 s^5 (t^2+bst+ds^2) \, ,
\end{equation}
where $a, \, b, \, d \in \mathbb{C}$. The dual $E_8 \times E_8$ heterotic string background is simply characterized by a vanishing Wilson line, with complex structure $\tau$ and complexified Kähler modulus $\rho$ arbitrary. The exact relation between the geometric and heterotic moduli is then \cite{LopesCardoso:1996hq} 
\begin{equation}\label{map}
    j(\tau)j(\rho) = -1728^2 \frac{a^3}{27d}\,, ~~~~~ (j(\tau)-1728)(j(\rho)-1728) = 1728^2 \frac{b^2}{4d}\,,
\end{equation}
where $j(z)$ is the j-invariant modular function. 

Consider a decompactification limit from eight to ten dimensions with square torus and vanishing B-field. We can parametrize
\begin{equation}
    \tau = i\frac{R_8}{R_9} = \frac{i}{u^n}\,, ~~~~~ ~~~~~ ~~~~~ \rho = iR_8 R_9 = \frac{i}{u^m}\,,
\end{equation}
with $u \in \mathbb{R}$ and $m > n > 0$, so that
\begin{equation}\label{radiipar}
    -\tau \rho = R_8^2 =  \frac{1}{u^{m+n}}\,, ~~~~~ ~~~~~ ~~~~~  \frac{\rho}{\tau} = R_9^2 =  \frac{1}{u^{m-n}}\,.
\end{equation}
It follows that as $u$ goes to zero, $R_8$ diverges faster than $R_9$, describing a decompactification from eight to ten dimensions at which the two radii grow at different asymptotic rates. We consider other values for $m$ and $n$ below.

Now we ask how this  decompactification with different rates is described in F-theory. From the map \eqref{map} we get the equations
\begin{equation}\label{alg1}
\begin{split}
    \frac{a^3}{d}+\frac{b^2}{d} &\sim j(i/u^n) + j(i/u^m)\\
    &\sim e^{\frac{2\pi}{u^{n}} } + e^{\frac{2\pi}{u^{m}}}\,
\end{split}
\end{equation}
and
\begin{equation}\label{alg2}
    \frac{b^2}{a^3} \sim 1 - 1728 \frac{e^{\frac{2\pi}{u^n}}+e^{\frac{2\pi}{u^m}}}{e^{\frac{2\pi}{u^n}+\frac{2\pi}{u^m}}}\,, 
\end{equation}
where we have used the expansion $q(z) = e^{-2\pi i z} + 744 + \cdots$ and kept only the leading orders as $u \to 0$. By redefining the degeneration parameter $u \mapsto u' = f(u)$, any of these equations can be made algebraic. However, this cannot be done for both equations at once, and so the parameters $a,b,c$ cannot be expressed as rational functions of any degeneration parameter; this makes it impossible to write down a Kulikov Weierstrass model describing the associated decompactification limit. In other words, a path in moduli space that approaches the limit $R_9 \to \infty$ only after $R_8$ is already infinitely large cannot be described using the tools of algebraic geometry. 

If we set $m = n$ instead, from eq. \eqref{radiipar} we see that only $R_8$ goes to infinity, $R_9$ remaining constant. In turn, equations \eqref{alg1} and \eqref{alg2} can be written as
\begin{equation}
    \frac{a^3}{d}+\frac{b^2}{d} \sim 2 e^{\frac{2\pi}{u^n}}\,, ~~~~~ \frac{b^2}{a^3} \sim 1 - 1728 \frac{2}{e^{\frac{2\pi}{u^n}}}\,,
\end{equation}
hence reparametrizing $u \to u' = \exp(2\pi/u^n)$ makes both equations algebraic and so, as required for consistency, a Kulikov Weirstrass model description exists. A similar situation arises if we set $n = 0$, which corresponds to the two radii diverging at the same rate. The remaining choices for $m$ and $n$ come from exchanging $m \leftrightarrow n$ in the cases already studied and correspond to T-dual frames in which one of the radii goes to zero instead. The map \eqref{map} is T-duality invariant and so the conclusions are the same.

\bibliographystyle{JHEP}
\bibliography{bibliography}

\end{document}